\documentclass[useAMS,usenatbib]{mn2e}
\def\aap{AA}
\def\prl{Phys. Rev. Lett.}

\def\apjl{ApJL}
\def\na{NewA}

\def\mnras{MNRAS}
\def\apj{ApJ}
\def\apjs{ApJS}
\def\aj{AJ}

\def\pasj{PASJ}

\def\prd{PhysRevD}
\def\jcap{JCAP}
\def\physrep{Phys. Rept.}
\def\araa{Annual Review of Astronomy and Astrophysics}

\def\dgemail{gilmanda@ucla.edu}

\usepackage[titletoc]{appendix}
\usepackage{graphicx}
\usepackage{float}
\usepackage{amssymb}
\usepackage{amsfonts}
\usepackage{amsmath} 
\usepackage{color}
\usepackage{wrapfig}
\usepackage{hyperref}
\usepackage{algorithm}
\usepackage{algpseudocode}
\usepackage{physics}

\usepackage{natbib}

\def\msun{{M_{\odot}}}

\def\dlos{{\delta_{\rm{los}}}}
\def\mhm{{m_{\rm{hm}}}}

\def\data{{\bf{d}_{\rm{n}}}}
\def\datasim{{\bf{d}_{\rm{n}}^{\prime}}}
\def\msub{{\bf{m}_{\rm{sub}}}}
\def\qsub{{\bf{q}_{\rm{s}}}}
\def\fsub{{{f}_{\rm{sub}}}}
\def\fsubmean{{\bar{f}_{\rm{sub}}}}
\def\qmac{{\bf{M}}}
\def\qm{{}{\bf{M}}}

\def\vspacing{\\[0.15cm]}

\title[strong lensing constraints on dark matter warmth]{Warm dark matter chills out: constraints on the halo mass function and the free-streaming length of dark matter with 8 quadruple-image strong gravitational lenses}
\author[Gilman et al.]{\parbox{\textwidth}{
		Daniel Gilman$^{1}$\thanks{\dgemail}, 
		Simon Birrer$^{1}$, 
		Anna Nierenberg$^{2}$,
		Tommaso Treu$^{1}$,
		Xiaolong Du$^{3}$,
		Andrew Benson$^{3}$\\
	}
	\\
	\parbox{\textwidth}{
		$^{1}$Department of Physics and Astronomy, University of California,
		Los Angeles, CA 90095, USA\\
		$^{2}$Jet Propulsion Laboratory, California Institute of Technology, 4800 Oak Grove Dr, Pasadena, CA 91109, USA\\
		$^{3}$Carnegie Observatories, 813 Santa Barbara Street, Pasadena, CA 91101, USA
	}
}

\hypersetup{draft}
\begin{document}
	
	\voffset-.6in
	
	\date{Accepted . Received }
	
	\pagerange{\pageref{firstpage}--\pageref{lastpage}} 
	
	\maketitle	
	\label{firstpage}
	\begin{abstract}
		The free-streaming length of dark matter depends on fundamental dark matter physics, and determines the abundance and concentration of dark matter halos on sub-galactic scales. Using the image positions and flux ratios from eight quadruply-imaged quasars, we constrain the free-streaming length of dark matter and the amplitude of the subhalo mass function (SHMF). We model both main deflector subhalos and halos along the line of sight, and account for warm dark matter (WDM) free-streaming effects on the mass function and mass-concentration relation. By calibrating the scaling of the SHMF with host halo mass and redshift using a suite of simulated halos, we infer a global normalization for the SHMF. We account for finite-size background sources, and marginalize over the mass profile of the main deflector. Parameterizing dark matter free-streaming through the half-mode mass $m_{\rm{hm}}$, we constrain the thermal relic particle mass $m_{\rm{DM}}$ corresponding to $m_{\rm{hm}}$. At $95 \%$ CI: $\mhm < 10^{7.8} \msun$ ($m_{\rm{DM}} > 5.2 \ \rm{keV}$). We disfavor $m_{\rm{DM}} = 4.0 \rm{keV}$ and $ m_{\rm{DM}} = 3.0 \rm{keV}$ with likelihood ratios of 7:1 and 30:1, respectively, relative to the peak of the posterior distribution. Assuming cold dark matter, we constrain the projected mass in substructure between $10^6 - 10^{9} M_{\odot}$ near lensed images. At $68 \%$ CI, we infer $2.0 - 6.1 \times 10^{7} \msun \rm{kpc^{-2}}$, corresponding to mean projected mass fraction $\bar{f}_{\rm{sub}} = 0.035_{-0.017}^{+0.021}$. At $95 \%$ CI, we obtain a lower bound on the projected mass of $0.6 \times 10^{7} \msun \rm{kpc^{-2}}$, corresponding to $\bar{f}_{\rm{sub}} > 0.005$. These results agree with the predictions of cold dark matter.
	\end{abstract}
	
	
	\section{Introduction}
	The theory of cold dark matter (CDM) has withstood numerous tests on scales spanning individual galaxies to the large scale structure of the universe and the cosmic microwave background \citep{Tegmark++04,deBlok++08,WMAP9cosmo}. The next frontier for this highly successful theory lies on sub-galactic scales, where CDM makes two distinct predictions: First, CDM predicts a scale-free halo mass function, possibly down to halo masses comparable to that of a planet \citep{Hofmann++01,Angulo++17}. Second, in CDM models halo concentrations decrease monotonically with halo mass, a result of hierarchical structure formation \citep{Moore++99,AvilaReese++01,Zhao++03,DiemerJoyce18}. A confirmation of these predictions through a measurement of the mass function and halo concentrations on mass scales below $10^9 \msun$ would at once constitute a resounding success for CDM and rule out entire classes of alternative dark matter theories. 
	
	The abundance of small-scale dark matter depends on the matter power spectrum at early times. If the velocity distribution of the dark matter particles causes them to diffuse out of small peaks in the density field, this will prevent the direct collapse of over-densities below a characteristic scale referred to as the free-streaming length \citep{Benson++13,Schneider++13}. The delay in structure formation in these scenarios also suppresses the central densities of the smallest collapsed halos, changing the mass-concentration relation for low-mass objects \citep{AvilaReese++01,Schneider++12,Maccio++13,Bose++16,Ludlow++16}. By definition, free-streaming effects are negligible in CDM, while models with cosmologically relevant free-streaming lengths are collectively referred to as warm dark matter (WDM). As the free-streaming length depends on the dark matter particle(s) mass and formation mechanism, an inference on the small-scale structure of dark matter on mass scales where some halos are expected to be completely dark directly constrains fundamental dark matter physics and the viability of specific WDM particle candidates, including sterile neutrinos \citep{Dodelson++94,ShiFuller99,AbazaijanKusenko19} and keV-mass thermal relics.  
	
	Interest in alternatives to the canonical CDM paradigm, such as WDM, were motivated in part by apparent failures of the CDM model on small scales \citep[see][and references therein]{BullockBK17}. Two challenges in particular dominate scientific discourse, and provide illustrative examples of the complexity associated with testing CDM's predictions on sub-galactic scales. The `missing satellites problem' (MSP), first pointed out by \citet{Moore++99}, refers to the paucity of observed satellite galaxies around the Milky Way, in stark contrast to dark-matter-only N-body simulations that predict hundreds of dark matter subhalos hosting a luminous satellite galaxy. Invoking free-streaming effects in WDM to remove these small subhalos would resolve the problem, and hence WDM models gained traction. A second challenge to the CDM picture emerged with the `too big to fail' (TBTF) problem \citep{Boylan-Kolchin++11}, which points out that the subhalos housing the largest Milky Way satellites are either under-dense or too small. Self-interacting dark matter, which results in lower central densities in dark matter subhalos \citep[see][and references therein]{TulinYu18}, gained traction in part as a resolution to the TBTF problem. 
	
	Today, new astrophysical solutions to the MSP and TBTF problems diminish the immediate threat to CDM, but the resolutions to these issues are riddled with assumptions regarding complicated physical processes on sub-galactic scales. The inclusion of baryonic feedback and tidal stripping in N-body simulations results in the destruction of subhalos, pushing the surviving number down to observed levels \citep{Kim++17}, although recently it has been suggested that the role of tidal stripping in N-body simulations is artificially exaggerated by resolution effects \citep{vandenBosch++18,ErraniPenarrubia19}. The continuous discovery of new dwarf galaxies seems to resolve the MSP, and might even suggest a `too-many-satellites problem' \citep{Kim++17b,Homma++19}, but the number of expected satellite galaxies in CDM itself rests on assumptions regarding the process of star formation in low mass halos, which can introduce uncertainties larger than the differences between CDM and WDM on these scales \citep{Nierenberg++16,Dooley++17,Newton++18}.The inclusion of baryonic feedback from star formation processes and supernova in low-mass halos can reduce halo central densities, and at least partially alleviates the issues associated with the TBTF problem \citep{Tollet++16}. However, the degree to which baryonic feedback resolves the problem depends on the manner in which this feedback is implemented in simulations. 
	
	Regarding constraints on WDM models, analysis of the Lyman-$\alpha$ forest \citep{Viel13,Irsic++17} and the luminosity function of distant galaxies \citep{Menci++16,Castellano++19}, while robust to the systematics associated with examining Milky Way satellites, to some degree rely on luminous matter to trace dark matter structure. Constraints from the Lyman-$\alpha$ forest also invokes certain assumptions for the relevant thermodynamics. The common theme is that disentangling the role of baryons and dark matter physics on sub-galactic scales is difficult and fraught with uncertainty. It would be ideal to test the predictions of matter theories irrespective of baryonic physics. 
	
	Strong gravitational lensing by galaxies provides a means of testing the predictions of dark matter theories directly, without relying on baryons to trace the dark matter. As photons emitted from distant background sources traverse the cosmos, they are subject to deflections by the gravitational potential of dark matter halos along the entire line of sight and by subhalos around the a main lensing galaxy. Each warped image produced by a strong lens contains a wealth of information regarding the dark matter structure in the universe. The aim of this work is to extract that information. 
	
	When the lensed background source is spatially extended -- for example, a galaxy -- the lensed image becomes an arc that partially encircles the main deflector. Dark matter halos near the arc produce small surface brightness distortions, which allows for the localization of the perturbing halo and enables constraints on its mass down to scales somewhere between $10^8 - 10^9 \msun$ \citep{Veg++14,Hezaveh++16}. Analysis of the surface brightness fluctuations over the entirety of the arc can also constrain the abundance of small halos too diminutive to be detected individually, and results in a 2 keV lower bound on the mass of thermal relic WDM \citep{Birrer++17a}. A joint analysis of individual detections and non-detections in a sample of arc-lenses can constrain certain models of dark matter and test the predictions of CDM \citep{Vegetti++18,Ritondale++18}. Recently, several works have proposed measuring the substructure convergence power spectrum in by analyzing surface brightness fluctuations in extended arcs \citep{Hezaveh++16b,Cyr-Racine++18,DiazRivero++18,Brennan++18}, and \citet{Bayer++18} applied this method to a strong lens system. 
	
	We focus on a second kind of lens system, quadruply imaged quasars (quads). Rather than extended arcs, the observables in quads are four image positions and three magnification ratios, or flux ratios (the observable is the flux ratio, not the intrinsic flux, because the intrinsic source brightness is unknown) with unresolved sources. Flux ratios depend on non-linear combinations of second derivatives of the lensing potential near an image, providing localized probes of small-scale structure down to scales of $10^{7} \msun$. These systems have been used in the past to constrain the presence of dark matter halos near lensed images \citep{MetcalfMadau01,Metcalf++02,Amara++06,Nierenberg++14,Nierenberg++17} and measure the subhalo mass function \citep{D+K02}. Recently, \citet{Hsueh++19} improved on previous analyses of quadruply imaged quasars by including halos along the line of sight, which can contribute a significant signal in flux ratio perturbations \citep{Xu++12,Gilman++18}. They found results consistent with CDM, ruling out WDM models to a degree comparable to that of the Lyman-$\alpha$ forest \citep{Viel13,Irsic++17}. 
	
	In the case of quadruple-image lenses, the luminous source is often a compact background object, such as the ionized medium around a background quasar. Broad-line emission from the accretion disk is subject to microlensing by stars, whereas light that scatters off of the more spatially extended narrow-line region is immune to microlensing while retaining sensitivity to the milli-arcsecond scale deflection angles produced by dark matter halos in the range $10^7 -10^{10}\msun$ \citep{MoustakasMetcalf02,Sugai++07,Nierenberg++14, Nierenberg++17}. Likewise, radio emission from the background quasar, while generally expected to be more compact than the narrow-line emission based on certain quasar models \citep{ElitzurSholsman06,Combes++19}, is extended enough to absorb micro-lensing effects. 
	
	We carry out an analysis of eight quads using a forward modeling approach we have tested and verified with mock data sets \citep{Gilman++18,Gilman++19}. The sample of lenses we consider contains six systems with flux ratios measured with narrow-line emission presented in \citet{Nierenberg++19}, and two others with data from \citet{Nierenberg++14} and \citet{Nierenberg++17}. We expect the sample is robust to microlensing effects and yield reliable data with which to constrain dark matter models. None of the quads show evidence for morphological complexity in the form of stellar disks, which require more detailed lens modeling \citep{Hsueh++16,Gilman++17,Hsueh++17}. 
	
	This paper is organized as follows: In Section \ref{sec:inference} we describe our forward modeling analysis method and our implementation of a rejection algorithm in Approximate Bayesian Computing. Section \ref{sec:parameterizations} describes our parameterizations for the dark matter structure in the main lens plane and along the line of sight, and our modeling of free-streaming effects in WDM. Section \ref{sec:data} contains a brief description of the data used in our analysis and the relevant references for each system. In Section \ref{sec:assumptionsandpriors} we describe in detail each physical assumption we make and the modeling choices and prior probabilities attached to these assumptions. In Section \ref{sec:results}, we present our inferences on the free-streaming length of dark matter and the amount of lens plane substructure. We discuss the implications of our results and our general conclusions in Section \ref{sec:discussion}. 
	
	All lensing computations are performed using {\tt{lenstronomy}}\footnote{https://github.com/sibirrer/lenstronomy} \citep{BirrerAmara18}. Cosmological computations involving the halo mass function and the matter power spectrum are performed with {\tt{colossus}} \citep{Diemer17}. We assume a standard cosmology using the parameters from WMAP9 \citep{WMAP9cosmo} ($\Omega_m = 0.28, \sigma_8 = 0.82, h=0.7$).  
	
	\section{Bayesian inference in substructure lensing}
	\label{sec:inference}
\begin{figure}
		\includegraphics[clip,trim=3.5cm 4cm 2.5cm
		4cm,width=.49\textwidth,keepaspectratio]{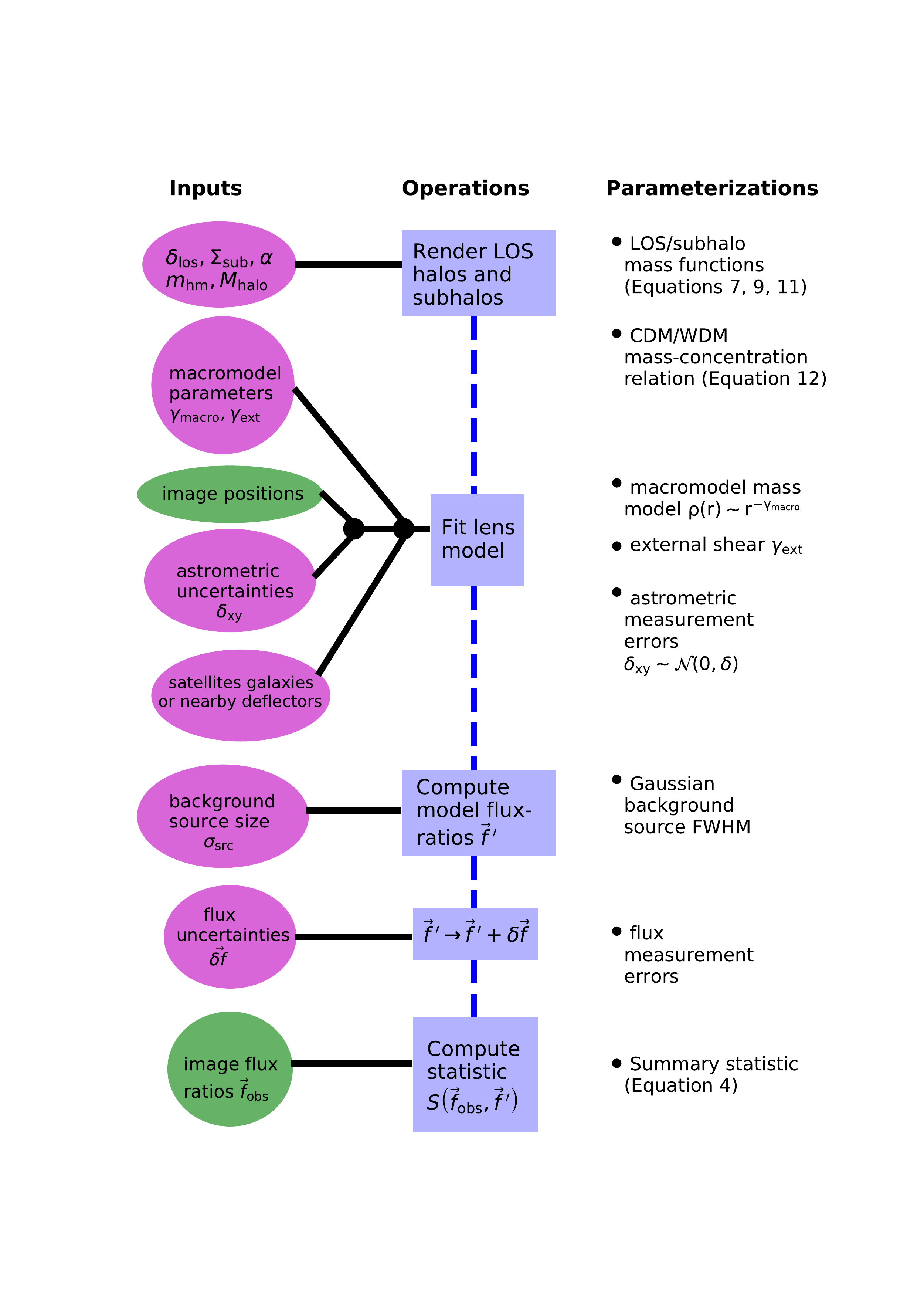}
		\caption{\label{fig:flowchart}  A graphical representation of the forward modeling procedure. Purple colors correspond to the action of sampling from a prior, blue represents an operation performed using the parameters sampled from a prior, and green colors indicate the use of observed information from the lenses. The arrow of time points from top to bottom: The first step is the rendering of dark matter structure, while the use of the information from observed flux ratios happens only at the very end.}
	\end{figure}

	In this section we frame the substructure lensing problem in a Bayesian context, and describe our analysis method which relies on a forward-generative model to sample the target posterior distribution through an implementation of Approximate Bayesian Computing. We have tested this analysis method using simulated data \citep{Gilman++18,Gilman++19}. The full forward modeling procedure we describe in this section is illustrated in Figure \ref{fig:flowchart}, and the relevant parameters are summarized in Table \ref{tab:params}. 
	
	\subsection{The Bayesian inference problem}
	Our goal is to obtain samples from the posterior distribution 
	\begin{equation}
	\label{eqn:posterior}
	p \left(\qsub | \boldsymbol{D} \right) \propto \pi \left(\qsub\right) \prod_{n=1}^{N} \mathcal{L} \left(\data | \qsub \right) 
	\end{equation}
	where $\qsub$ is a set of hyper-parameters describing the subhalo and line of sight halo mass functions, $\boldsymbol{D}$ denotes the set of positions and flux ratios from a set of $N$ lenses with the data from each lens denoted by $\data$, and where $\pi$ represents the prior on $\qsub$. 
	
	A certain dark matter model makes predictions for the parameters in $\qsub$, which includes quantities such as the normalization of the subhalo mass function, the logarithmic slope of the mass function, a free streaming cutoff, etc. For a given $\qsub$, we may generate specific realizations of line of sight halos and main deflector subhalos (including the halo/subhalo masses, positions, concentrations, etc.), that affect lensing observables. We refer to a specific realization of dark matter structure corresponding to a model specified by $\qsub$ as $\msub$. In addition to generating the realizations $\msub$, computing the likelihood function $\mathcal{L} \left(\data | \qsub \right)$ in Equation \ref{eqn:posterior} requires marginalizing over nuisance parameters $\qm$, which include the background source size $\sigma_{\rm{src}}$, and the lens model that describes the main lensing galaxy (hereafter the macromodel). Integrating over the macromodel and the space of possible dark matter realizations $\msub$, the likelihood is given by
	\begin{eqnarray}
	\label{eqn:likelihood}
	\mathcal{L} \left(\data| \qsub \right) = \int p \left(\data | \msub, \qm \right) p \left( \msub, \qm | \ \qsub \right) d \msub \ d \qm.
	\end{eqnarray}
	Note that we write the joint distribution $p\left(\msub, \qmac | \qsub\right)$, and do not assume the parameters in $\qm$ and $\qsub$ are independent. 
	
	Evaluating Equation \ref{eqn:likelihood} is a daunting task. We highlight two main reasons: 
	\begin{itemize}
		\item Exploring the parameter space spanned by $\qsub$ and $\qmac$ through traditional MCMC methods is extremely inefficient. $\qm$ is a high-dimensional space, where the overwhelming majority of volume does not result in model-predicted observables that resemble the data, and in particular does not predict the correct image positions. Thus the overwhelming majority of samples drawn from $\qm$, and the corresponding samples $\qsub$ (even if they described the `true' nature of dark matter) would not contribute to the integral.
		\item The parameters $\qm$ describing the lens macromodel may depend indirectly on the dark matter parameters $\qsub$ through the realizations $\msub$ generated from the model specified by $\qsub$. This necessitates the simultaneous sampling of $\qsub$ and $\qmac$ in the inference. However, it is difficult to impose an informative prior on $\qm$ since the `true' parameters in $\qsub$ are unknown. Recognizing this and using a very uninformative prior on $\qm$, most samples will be rejected since they do not resemble the data, which alludes back to the issue of dimensionality described in the first bullet point. 
	\end{itemize} 
	To address these challenges, we use a statistical method that bypasses the direct computation of the integral in Equation \ref{eqn:likelihood}. 
	
	\subsection{Forward modeling the data}
	Rather than compute the likelihood function, we recognize that by creating simulated observables $\datasim = \datasim \left(\msub, \qm \right)$ from the model $\qsub$, and accepting the proposed $\qsub$ if they satisfy $\datasim = \data$, the accepted $\qsub$ samples will be direct draws from the posterior distribution in Equation \ref{eqn:posterior} \citep{Rubin1984}. In this forward-generative framework, simulating the relevant physics in substructure lensing replaces the task of evaluating the likelihood function in Equation \ref{eqn:likelihood}. We propagate photons from a finite-size background source through lines of sight populated by dark matter halos, a lensing galaxy and its subhalos, and finally into a simulated observation with statistical measurement errors added. Provided the forward model contains all of the relevant physics, the simulated data $\datasim$ will express the same potentially complex covariances present in the observed data.
	
	The `curse of dimensionality' that prohibits direct evaluation of Equation \ref{eqn:likelihood} also afflicts the criterion of exact matching between $\data$ and $\datasim$. In particular, most draws of macromodel parameters $\qmac$ will not yield the observed image positions, and would therefore be rejected from the posterior. To deal with this, our strategy will be to ensure that the macromodel and other nuisance parameters sampled in the forward model, when combined with the full line of sight and subhalo populations specified by $\msub$, yield a lens model that predicts the same image positions as observed in the data. 
	
	Obtaining a lens model that returns the observed image positions amounts to demanding that the the four images seen by the observer on the sky at positions $\boldsymbol{\theta}$ map to the same position on the source plane $\boldsymbol{\beta_K}$. This requires the use of the full multi-plane lens equation describing the path of deflected light rays \citep[e.g.][]{Schnedier1997}, see also \citet{BlandfordNarayan86})
	\begin{equation}
	\label{eqn:raytracing}
	\boldsymbol{\beta_K} = \boldsymbol{\theta} - \frac{1}{D_{\rm{s}}} \sum_{k=1}^{K-1} D_{\rm{ks}}{\boldsymbol{\alpha_{\rm{k}}}} \left(D_{\rm{k}} \boldsymbol{\beta_{\rm{k}}}\right),
	\end{equation} 
	where the quantities $D_{\rm{s}}$, $D_{\rm{k}}$ and $D_{\rm{ks}}$ denote angular diameter distances to the source plane, to the $k$th lens plane, and from the $k$th lens plane to the source plane, respectively. Equation \ref{eqn:raytracing} is a recursive equation for the $\boldsymbol{\beta_{\rm{k}}}$ that couples deflection angles from objects at different redshifts, similar to looking through potentially thousands of magnifying glasses in series. Throughout this process, we account for uncertainties in the measured image positions by sampling astrometric perturbations $\delta_{xy}$, and applying them to the observed image positions during the forward modeling.
	
	To solve for macromodel parameters $\qmac$, for each realization $\msub$ we sample the power-law slope of the main deflector mass profile $\gamma_{\rm{macro}}$ and the external shear strength $\gamma_{\rm{ext}}$. If the lens system in question has satellite galaxies or nearby deflectors, we sample priors for their masses and positions. The remaining parameters describing the lens macromodel\footnote{The full set of macromodel parameters for a power-law ellipsoid are the overall normalization $b_{\rm{macro}}$, the mass centroid $g_x$ and $g_y$, the ellipticity and ellipticity position angle $\epsilon$ and $\theta_{\epsilon}$, the external shear and shear angle $\gamma_{\rm{ext}}$ and $\theta_{\rm{ext}}$, and the power-law slope $\gamma_{\rm{macro}}$. Nearby galaxies are modeled as Singular Isothermal Spheres.} are allowed to vary freely until a lens model that fits the image positions is found\footnote{The four image positions provide $4\times2 = 8$ constraints, and the macromodel parameters that are allowed to vary freely, plus the source position, give $8$ degrees of freedom.}. 
	
	The approach of simultaneously sampling $\qmac$ and $\qsub$ does not involve lens model optimizations with respect to the observed image fluxes, because the information from the observed fluxes is not used at this stage of the analysis. This method therefore avoids potential biases incurred by optimizing the macromodel with respect to the observed fluxes, rather than marginalizing over these parameters. As we will show in Section \ref{ssec:jointinf}, by sampling $\qmac$ and $\qsub$ simultaneously we obtain joint posterior distributions that account for potential covariance between these quantities, recognizing that the addition of substructure may affect the distributions for the macromodel parameters in $\qmac$. 
	
	With a lens model that fits the image positions in hand, we draw a background source size and ray-trace on a finely sampled grid around each image position using Equation \ref{eqn:raytracing} to compute the image fluxes $\boldsymbol{f^{\prime}}$. To incorporate statistical measurement errors in image fluxes, we sample flux uncertainties $\boldsymbol{\delta f}$, and render these perturbations onto the model-predicted fluxes $\boldsymbol{f^{\prime}} \rightarrow\boldsymbol{f^{\prime}} + \boldsymbol{\delta f}$ prior to computing the flux ratios. 
	
	\subsection{Deriving posteriors from the forward model samples}
	For each realization, we compute a summary statistic between the three observed flux ratios $\boldsymbol{f_{\rm{obs}}}$ and those computed in the forward model 
	\begin{equation}
	\label{eqn:summary_stat}
	S_{\rm{lens}} \left(\boldsymbol{f^{\prime}},\boldsymbol{f_{\rm{obs}}} \right) \equiv \sqrt{\sum_{i=1}^{3} \left({f}^{\prime}_{i} - f_{\rm{obs(i)}} \right)^2},
	\end{equation}
	and assign this statistic to the draw of $\qsub$. This summary statistic contains the full information content of the data, as the simultaneous matching of the three ratios requires that the forward model samples that minimize this statistic contain the same correlations present in the data. We repeat this procedure between 300,000 and 1,200,000 times for each quad, depending on with frequency with which the realizations, with the statistical flux uncertainties added, match the observed fluxes to within $1\%$. 
	
	We select the $\qsub$ parameters corresponding to the $800$ lowest summary statistics $S_{\rm{lens}}$. The exact matching criterion $\data = \datasim$, which guarantees that the accepted samples $\qsub$ form the desired posterior, is replaced by selecting the realizations that look most like the data through the summary statistic $S_{\rm{lens}}$. The resulting distribution of $\qsub$ is therefore an approximation to the posterior distribution for each lens, with the approximation converging to the true posterior as the number of forward model samples increases while keeping the number of accepted samples fixed. The quality of the approximation can be quantified through a convergence test, in which we verify that the posteriors are unchanged as one removes realizations from the forward-modeled data while keeping the same number of accepted samples (see Appendix \ref{app:A}). This method is an implementation of a rejection algorithm in Approximate Bayesian Computing \citep{Rubin1984,Marin++11,Lintusaari++17}, a technique applied to problems where it is possible to generate simulated data from the model, but difficult to compute the likelihood \citep[see also][]{Beaumont++02,Akeret++15,Birrer++17a,Hahn++17}.
	
	To obtain the final posterior distribution $p \left(\qsub | \boldsymbol{D}\right)$ (Equation \ref{eqn:posterior}), we multiply together the likelihoods obtained for each lens \footnote{Before taking the product, we use a Gaussian kernel density estimator (KDE) with a first order boundary correction \citep[e.g.][]{Lewis15} to obtain a continuous approximation of the likelihood for each lens. We compute the bandwidth according to Scott's factor \citep{Scott92}, but caution that care should be taken with the choice of bandwidth to avoid over or under smoothing the likelihood.}. This procedure is only possible when using uniform priors in the forward model sampling, as the use of non-uniform priors would effectively move $\pi \left(\qsub\right)$ inside the product in Equation \ref{eqn:posterior} and over-use this information. We may, however, impose any prior we wish a-posteriori by re-weighting the forward model samples accordingly. 
	
	\begin{figure}
		\includegraphics[clip,trim=0cm 0cm 0cm
		0cm,width=.48\textwidth,keepaspectratio]{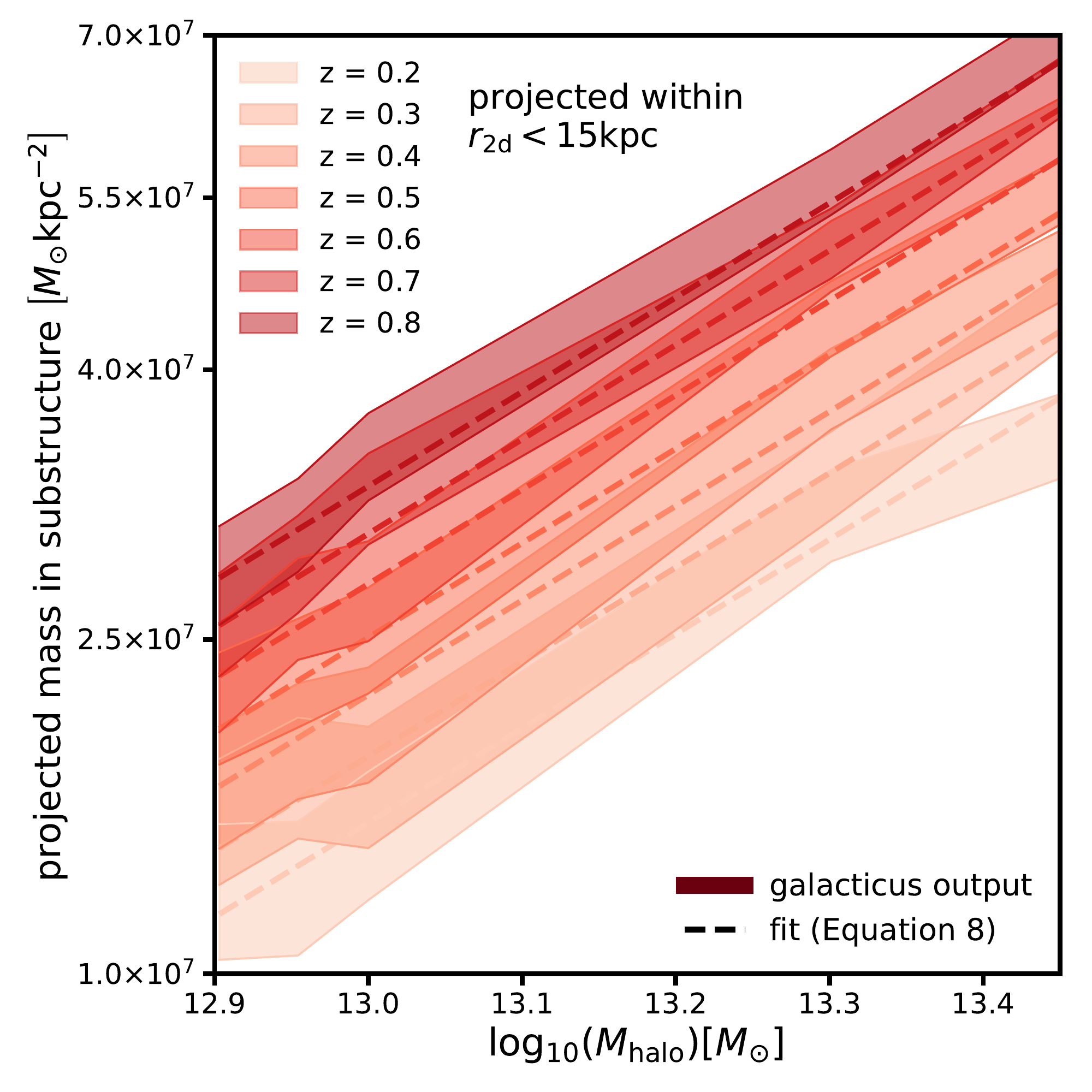}
		\caption{\label{fig:trends}  Output from the {\tt{galacticus}} semi-analytic simulations of substructure within halos used to calibrate the evolution of the subhalo mass function with halo mass and redshift. While on the y-axis we plot the actual projected surface mass density in substructure output by {\tt{galacticus}}, we only use the scaling with halo mass in redshift in our modeling, treating the overall normalization of the subhalo mass function as a free parameter. The projected mass density in substructure on the y-axis corresponds to a mass range $10^6 - 10^{10}\msun$, where we have extrapolated the mass function from the smallest resolved subhalo  ($10^{8}\msun$) to $10^{6}\msun$ to compute the projected mass. }
	\end{figure}	
	\begin{figure}
		\includegraphics[clip,trim=0cm 0cm 0cm
		0cm,width=.48\textwidth,keepaspectratio]{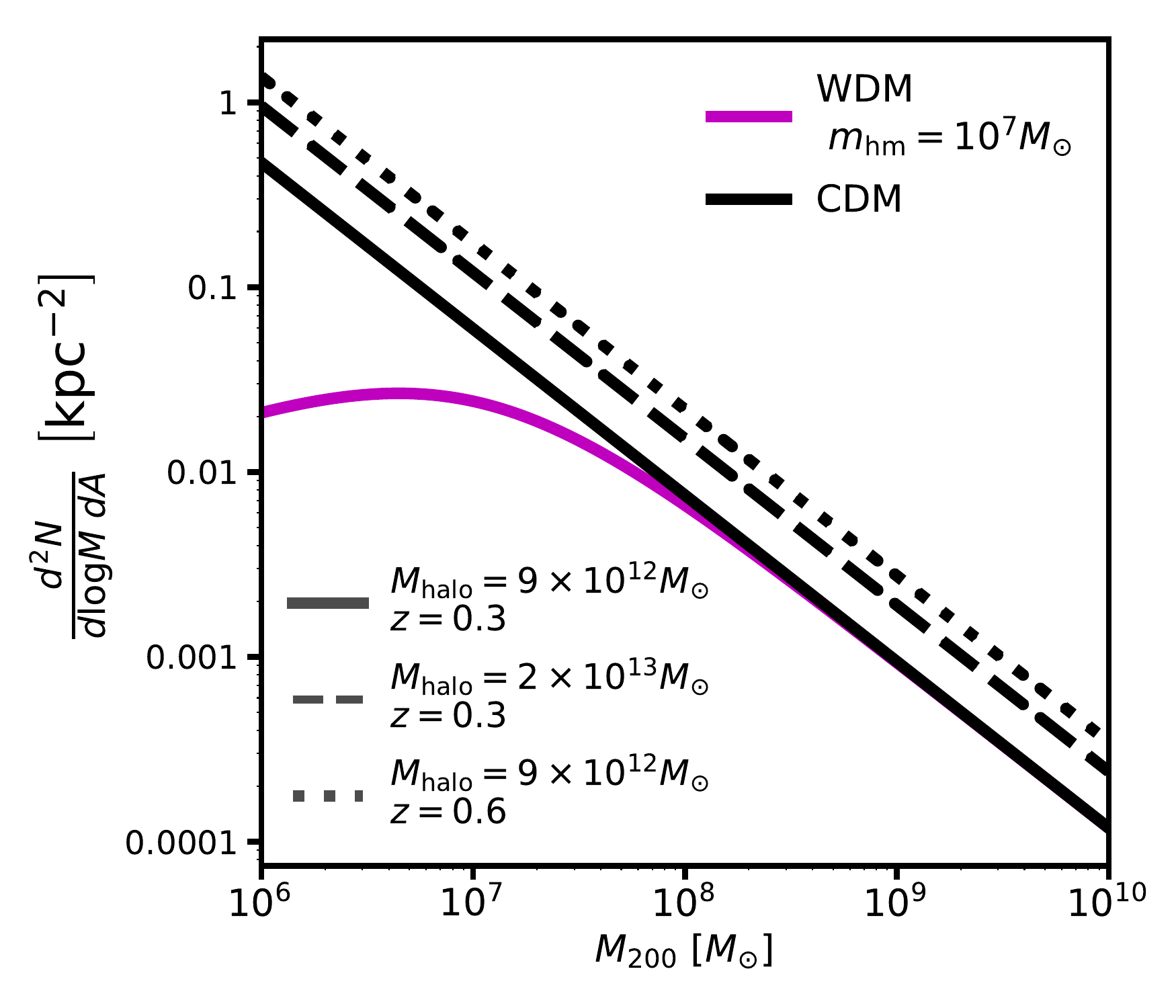}
		\includegraphics[clip,trim=0cm 0cm 0cm
		0cm,width=.48\textwidth,keepaspectratio]{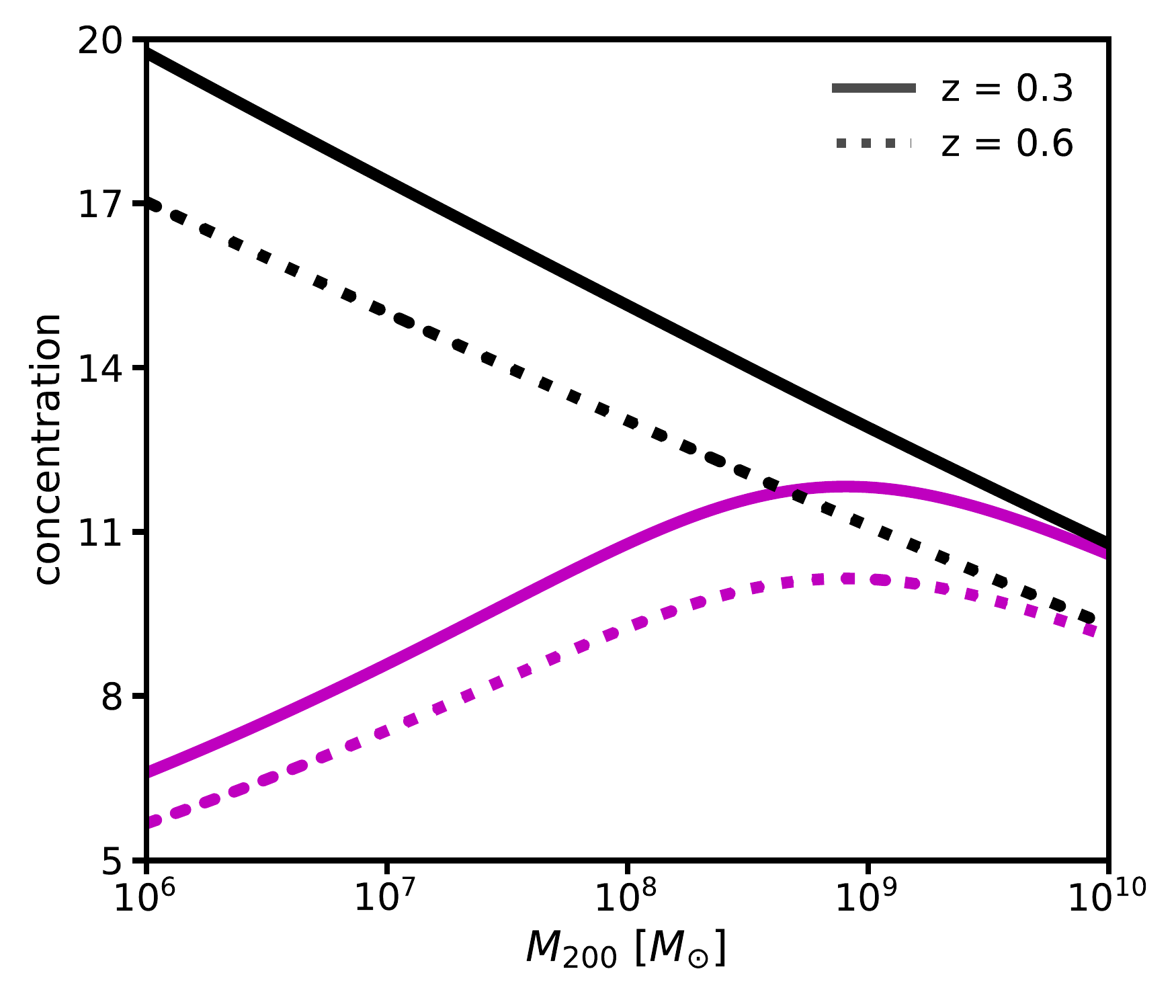}
		\caption{\label{fig:mcrelation} {\bf{Top:}} The subhalo mass function as a function of halo mass, redshift, and the half-mode mass $\mhm = 10^7 \msun$ with $\Sigma_{\rm{sub}} = 0.012 \rm{kpc^{-2}}$. The line of sight halo mass function looks similar, but evolves differently with redshift. {\bf{Bottom:}} The mass-concentration-relation for CDM and the same WDM model with $\mhm = 10^7 \msun$. Free-streaming affects the concentration of halos over one order of magnitude above $\mhm$.}
	\end{figure}	
	\begin{figure*}
		\includegraphics[clip,trim=3.8cm 8cm 2.5cm
		4cm,width=.8\textwidth,keepaspectratio]{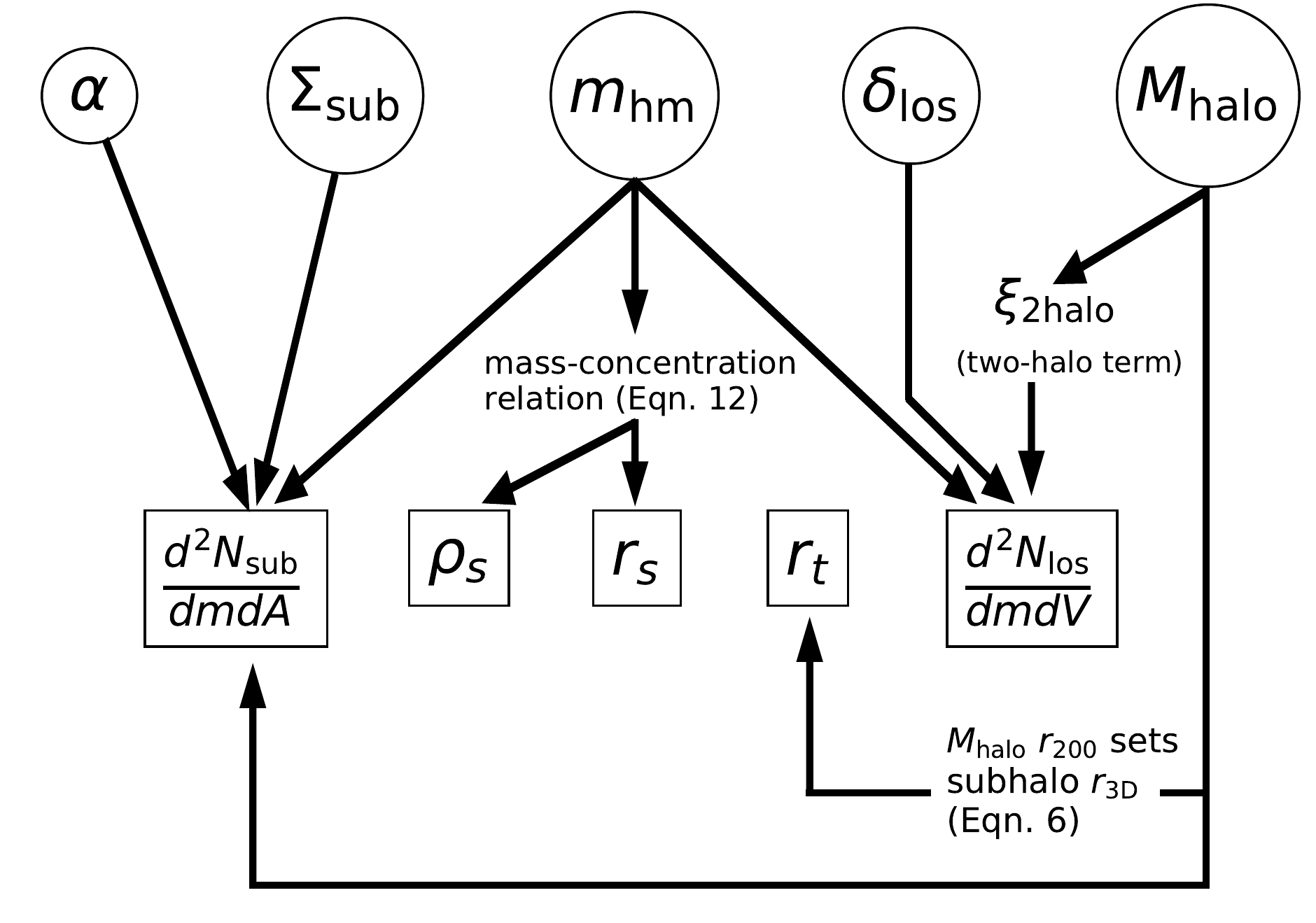}
		\caption{\label{fig:schematic} A graphical representation of the dark matter parameters in $\qsub$: $\alpha$, the logarithmic slope of the subhalo mass function, $\Sigma_{\rm{sub}}$, the overall scaling of the subhalo mass function, $\mhm$, the WDM half-mode mass, $\delta_{\rm{los}}$, the overall factor for the  line of sight halo mass function, and $M_{\rm{halo}}$, the main deflector's parent halo mass. $\xi_{\rm{2halo}}$ is implemented through Equation \ref{eqn:losmfunc} (see Section \ref{ssec:losmfunc}). These parameters are linked to the physical dark matter quantities they affect. From left to right: the subhalo mass function $\frac{d^2 N}{dm dA}$, the normalization $\rho_s$, scale radius $r_s$, and truncation radius $r_t$ of individual halos (see Equation \ref{eqn:massprofile}), and the line of sight halo mass function $\frac{d^2 N}{dm dV}$. The priors for each of these parameters are summarized in Table \ref{tab:lenspriors}, and discussed at length in Section \ref{sec:assumptionsandpriors}.}
	\end{figure*}	
	
	\section{The subhalo and line of sight halo populations}
	\label{sec:parameterizations}
	In this section, we describe the models we implement for the line of sight and subhalo mass functions in cold and warm dark matter that we sample in the forward model. We also describe the density profiles for individual halos, including their truncation radii and their distribution both along the line of sight and in the main lens plane. We begin with the parameterizations used for the halo and subhalo density profiles and the spatial distribution of subhalos in Section \ref{ssec:subhalos}. In Sections \ref{ssec:submfunc} and \ref{ssec:losmfunc} we describe the parameterizations of the subhalo and line of sight halo functions, respectively, and in Section \ref{ssec:modelingwdm} describe how we model WDM free-streaming effects. 
	
	\subsection{Subhalo density profiles and spatial distribution}
	\label{ssec:subhalos}
	We model subhalos as tidally truncated NFW profiles \citep{Baltz++09}
	\begin{equation}
	\label{eqn:massprofile}
	\rho \left(r\right) = \frac{\rho_s}{x \left(1+x\right)^2} \frac{\tau^2}{x^2 + \tau^2}
	\end{equation}
	where $x = \frac{r}{r_s}$, $\tau = \frac{r_t}{r_s}$, and $r_t$ is a truncation radius and $r_s$ is the NFW profile scale radius. We use the mass definition of $M_{200}$ computed with respect to the critical density at $z = 0$, and a concentration mass relation that accounts for free-streaming effects in WDM as is specifically designed to accurately predict the concentrations of low-mass halos (see Section \ref{ssec:modelingwdm}). 
	
	In the main lens plane, we truncate halos according to their three-dimensional position inside the host halo $r_{\rm{3D}}$ through a Roche-limit approximation that assumes a roughly isothermal global mass profile. The relevant scaling is $r_t \propto \left(M_{200} r_{\rm{3D}}^{2}\right)^{\frac{1}{3}}$  \citep{Tormen++98,Cyr-Racine++16}, which we implement as 
	\begin{equation} 
	\label{eqn:truncrad2}
	r_t = 1.4 \left(\frac{M_{200}}{10^7 \msun}\right)^{\frac{1}{3}} \left(\frac{r_{\rm{3D}}}{50 \rm{kpc}}\right)^{\frac{2}{3}} \left[\rm{kpc}\right].
	\end{equation}
	This results in truncation radii of $\sim 4-10 r_s$. We note that the truncation radius depends implicitly on the host halo mass $M_{\rm{halo}}$ through $r_{\rm{3D}}$, which depends on the scale radius and the virial radius of the host halo at the lens redshift (see Figure \ref{fig:schematic}). We note that the definition of $r_t$ in Equation \ref{eqn:truncrad2} does not depend on the structural parameters of the subhalo, which are altered in WDM models (see Section \ref{ssec:modelingwdm}). Incorporating these modeling details requires prescriptions for the tidal evolution of subhalos in the host halo as a function of the physical properties of the subhalo at infall \citep[e.g.][]{GreenvandenBosch19}.
	
	We render subhalos out to a maximum projected radius $3 R_{\rm{Ein}}$ and assign a three-dimensional $z$-coordinate between $-r_{200}$ and $r_{200}$, where $r_{200}$ is the virial radius of the host. Inside this volume, we distribute the subhalos assuming the spatial distribution follows the mass profile of the host dark matter halo outside an inner tidal radius, which we fix to half the scale radius of the host. Inside this radius, we distribute subhalos with a uniform distribution in three dimensions. This choice is motivated by simulations that predict tidal disruption of subhalos near the lensing galaxy, resulting in an approximately uniform number of subhalos per unit volume in the inner regions of the halo \citep{JiangvdB17}. The spatial distribution of subhalos that results from this procedure is approximately uniform in projection, which agrees with the predictions from N-body simulations \citep{Xu++15}.
	
	\begin{table*}
		\centering
		\caption{Free parameters sampled in the forward model. Notation $\mathcal{N} \left(\mu, \sigma\right)$ indicates a Gaussian prior with mean $\mu$ and variance $\sigma$, and $\mathcal{U} \left(u_1, u_2\right)$ indicates a uniform prior between $u_1$ and $u_2$. Lens-specific priors are summarized in Table \ref{tab:lenspriors}. }
		\label{tab:params}
		\begin{tabular}{lccr} 
			\hline
			parameter & definition & prior\\
			\hline 
			$\log_{10} \left(M_{\rm{halo}}\right) \left[\msun\right]$ & main lens parent halo mass &  (lens specific) \\
			\\
			$\Sigma_{\rm{sub}} \left[\rm{kpc}^{-2}\right]$ & normalization of subhalo mass function (Equation \ref{eqn:subhalomfunc})&  $\mathcal{U} \left(0, 0.1\right)$ \\&(rendered between $10^6-10^{10} \msun$) & \\
			\\
			$\alpha$ & logarithmic slope of the subhalo mass function & $\mathcal{U} \left(-1.95, -1.85\right)$\\
			\\
			$\log_{10} \left(\mhm\right) \left[M_{\odot}\right]$ & half-mode mass (Equations \ref{eqn:mfuncwdm} and \ref{eqn:cmrelation})& $\mathcal{U} \left(4.8, 10\right) $ \\
			&$\propto$ to free streaming length and thermal relic mass $m_{\rm{DM}}$ &\\
			\\
			$\delta_{\rm{los}}$ & rescaling factor for the line of sight Sheth-Tormen & $\mathcal{U} \left(0.8, 1.2\right)$ \\
			&mass function (Equation \ref{eqn:losmfunc}, rendered between $10^6-10^{10} \msun$)&\\
			\\
			$\sigma_{\rm{src}} \left[\rm{pc}\right]$ & source size& $\mathcal{U} \left(25, 60\right)$\\
			& parameterized as FWHM of a Gaussian & \\
			\\
			$\gamma_{\rm{macro}}$ & logarithmic slope of main deflector mass model  & $\mathcal{U} \left(1.95, 2.2\right)$ \\
			\\
			$\gamma_{\rm{ext}}$ & external shear in the main lens plane & (lens specific) \\
			\\
			$\delta_{xy} \left[\rm{m.a.s.}\right]$ & image position uncertainties& (lens specific)\\
			\\
			$\delta f$ & image flux uncertainties & (lens specific)\\
			\hline		
		\end{tabular}
	\end{table*}
	\subsection{The CDM subhalo mass function}
	\label{ssec:submfunc}
	In principle, the projected mass in subhalos near the Einstein radius can depend on the host halo mass, redshift, and the severity of tidal stripping by the main lensing galaxy. We will ultimately combine the inferences from multiple lenses at different redshifts and with different host halo masses, so we parameterize the subhalo mass function in such a way that a single parameter $\Sigma_{\rm{sub}}$ can be used to simultaneously describe the projected mass density in substructure for each quad, regardless of halo mass or redshift. 
	
	We use the functional form for the subhalo mass function
	\begin{equation}
	\label{eqn:subhalomfunc}
	\frac{d^2 N_{\rm{sub}}}{dm dA} =  \frac{\Sigma_{\rm{sub}}}{m_0} \left(\frac{m}{m_0}\right)^{\alpha} \mathcal{F} \left(M_{\rm{halo}}, z\right),
	\end{equation}
	where scaling function $\mathcal{F} \left(M_{\rm{halo}}, z\right)$ encodes the differential evolution of the projected number density with redshift and host halo mass, such that $\Sigma_{\rm{sub}}$ can be interpreted as a common parameter for all the lenses. We choose the normalization such that $\mathcal{F} \left(M_{\rm{halo}} = 10^{13} \msun, z = 0.5\right) = 1$, anchoring $\Sigma_{\rm{sub}}$ at $z=0.5$ with a halo mass of $10^{13}\msun$. We use a pivot mass $m_0 = 10^8 \msun$. We will marginalize over $\Sigma_{\rm{sub}}$ and $\alpha$ when quoting constraints on dark matter warmth to account for tidal stripping of subhalos and halo-to-halo scatter.  
	
	To determine the scaling function $\mathcal{F} \left(M_{\rm{halo}}, z\right)$, we run a suite of simulations using the semi-analytic modeling code {\tt{galacticus}}\footnote{Code version 7175:2bd6b8d84a39} \citep{Benson12,Pullen++14}, simulating host halos and their substructure in the redshift range $0.2 < z < 0.8$ and mass range $0.8 - 3 \times 10^{13} \msun$, with a subhalo mass resolution of $10^8 \msun$. In each redshift and mass bin we simulate 24 halos, resulting in 840 halos with $M_{\rm{halo}} \sim 10^{13} \msun$ in total\footnote{The entire simulation suite using {\tt{galacticus}} completed in 1,000 CPU hours.}. We average over the projected number densities along each principle axis inside a 15 kpc aperture to obtain trends in the projected number density with host halo mass and redshift in the vicinity of the Einstein radius, where lensed images appear. The {\tt{galacticus}} simulations include tidal destruction of subhalos by the global dark matter mass profile, which affects the evolution of the projected mass density with host halo redshift: at early times, subhalos are more concentrated in the host, while at later times tidal stripping from the host depletes the population of subhalos at small radii and the projected number density near the Einstein radius decreases. In addition, the physical size of the host halo at higher redshifts is smaller by a factor of $\left(1+z\right)^{-1}$, so the number of subhalos per square physical kpc is higher. We also note that early-type galaxy host halos simulated by \citet{Fiacconi++16} also show significant evolution with redshift in the projected number density of subhalos by about a factor of two, very similar to the {\tt{galacticus}} predictions. 
	
	We fit the evolution with halo mass and redshift predicted by {\tt{galacticus}} with the relation
	
	\begin{equation}
	\label{eqn:scaling}
	\log_{10} \left(\mathcal{F}\right) = k_1 \log_{10} \left(\frac{M_{\rm{halo}}}{10^{13} \msun}\right) + k_2 \log_{10}\left(z+0.5\right)
	\end{equation}
	with $k_1 = 0.88$ and $k_2 = 1.7$. The {\tt{galacticus}} output and the fit from Equation \ref{eqn:scaling} are shown in Figure \ref{fig:trends}. We only extract information regarding the scaling of projected mass density with halo mass and redshift from the {\tt{galacticus}} simulations, and treat the overall normalization of the number density as a free-parameter that absorbs the effects of tidal destruction of subhalos by the main lens galaxy. We discuss our modeling assumptions in more detail in Section \ref{ssec:submfuncassumptions}.  
	
	\subsection{The line of sight halo mass function}
	\label{ssec:losmfunc}
	We model line of sight structure by drawing halo masses from the Sheth-Tormen halo mass function \citep{ST99}, with two modifications. First, we introduce an overall rescaling factor $\delta_{\rm{los}}$ which accounts for theoretical uncertainty in the predicted amplitude of the halo mass function \citep[see e.g.][]{Despali++16}. The factor $\delta_{\rm{los}}$ accounts for the possibility of a selection bias in the quads towards systematically over or under-dense lines of sight. The second modification we add is a contribution from the two-halo term $\xi_{\rm{2halo}} \left(M_{\rm{halo}}, z\right)$, which accounts for the presence of correlated structure in the vicinity of main deflector parent dark matter halo \footnote{In Appendix A of \citet{Gilman++19}, we describe how this effect is implemented and show that this term contributes a $\sim 4\%$ increase in the frequency of flux ratio perturbations induced by objects outside the virial radius of the main deflector.}. With these modifications the line of sight halo mass function takes the form
	
	\begin{equation}
	\label{eqn:losmfunc}
	\frac{d^2N_{\rm{los}}}{dm  dV} = \dlos \big(1+ \xi_{\rm{2halo}}\left(M_{\rm{halo}}, z\right)\big) \frac{d^2N}{dm  dV} \big \vert_{\rm{ShethTormen}}.
	\end{equation}
	Halos along the line of sight are rendered in a double-cone geometry with opening angle $3 R_{\rm{Ein}}$, where $R_{\rm{Ein}}$ is the Einstein radius of the main deflector, and a closing angle behind the main deflector such that the cone closes at the source redshift. Finally, we add negative convergence sheets to subtract the mean expected convergence from line of sight halos at each line of sight plane. Without this numerical procedure, lines of sight are systematically over-dense relative to the expected matter density of the universe, akin to lensing in a universe with positive curvature \citep{Birrer++17b}. This may bias results as the macromodel will attempt to compensate for the artificial focusing of light rays in this scenario.  
	
	\subsection{Modeling free-streaming effects  in WDM}
	\label{ssec:modelingwdm}
	Free-streaming refers to the diffusion of dark matter particles out of small peaks in the matter density field in the early universe. This has the effect of erasing structure on scales below a characteristic free-streaming length which depends on the velocity distribution of the dark matter particles, and hence on their mass and formation mechanism. For a more in-depth discussion, see \citet{Schneider++13}.  
	
	It is convenient to express free-streaming effects in terms of the half-mode mass $\mhm$, which is defined in terms of the length scale where the transfer function between the CDM and WDM power spectra drops to one-half. In the specific case that all of the dark matter exists in the form of thermal relics, a one-to-one mapping between the half-mode mass and the mass of the  candidate particle $m_{\rm{DM}}$ exists, and has the scaling $\mhm \propto m_{\rm{DM}}^{-3.33}$ \citep{Schneider++12}
	\begin{equation}
	\label{eqn:masskev}
	m_{\rm{hm}}\left(m_{\rm{DM}}\right) = 3\times 10^{8} \left(\frac{m_{\rm{DM}}}{\rm{3.3keV}}\right)^{-3.33} \msun.
	\end{equation}
	
	We have run {\tt{galacticus}} models \citep{Benson++13} with WDM mass functions corresponding to $3.3$ and $5$ keV thermal relics to investigate the effects of free-streaming on the trends with host halo mass and redshift of the projected mass in substructure near the Einstein radius, and determine that the fit in Equation \ref{eqn:scaling} is common to both CDM and WDM. We therefore use the same scaling function $\mathcal{F} \left(M_{\rm{halo}}, z\right)$ for WDM subhalo mass functions, and model the effects of free streaming using the fitting formula from \citep{Lovell++14}
	
	\begin{equation}
	\label{eqn:mfuncwdm}
	\frac{dN_{\rm{WDM}}}{dm} = \frac{dN_{\rm{CDM}}}{dm} \left(1+\frac{m_{\rm{hm}}}{m} \right)^{-1.3}.
	\end{equation}
	Since the parameter $\mhm$ is related to the WDM transfer function, it should affect the subhalo and field halo mass functions in a similar manner. We therefore apply the same suppression factor in Equation \ref{eqn:mfuncwdm} to both the subhalo mass function and the line of sight halo mass function in Equations \ref{eqn:subhalomfunc}  and \ref{eqn:losmfunc}, respectively. Lacking a theoretical prediction for the evolution of the turnover with redshift, we do not evolve the shape or position of the free-streaming cutoff in the mass function at higher redshifts. 
	
	In WDM scenarios, the delayed onset of structure formation affects the assembly history of dark matter halos and suppresses their concentrations $c \equiv \frac{r_{\rm{vir}}}{r_s}$\footnote{We define $r_{\rm{vir}}$ with respect to the matter density contrast $200 \rho_{\rm{crit}}$. } on mass scales that extend above $\mhm$ \citep{Schneider++12, Bose++16}. We use the functional form proposed by \citep{Bose++16}, and write the WDM concentration-mass relation as
	
	\begin{equation}
	\label{eqn:cmrelation}
	\frac{c_{\rm{WDM}}\left(m, z\right)}{c_{\rm{CDM}}\left(m, z\right)} =  \left(1+z\right) ^{\beta\left(z\right)} \left(1+60\frac{m_{\rm{hm}}}{m}\right)^{-0.17}
	\end{equation}
	with $\beta \left(z\right) = 0.026z - 0.04$, using the CDM mass-concentration model of \citet{DiemerJoyce18} and a scatter of 0.1 dex \citep{Dutton++14}. The WDM suppression factor for the mass-concentration relation we use was calibrated for halos on mass scales below $M_{200} \sim 10^9 \msun$, and is accurate in the redshift range $z = 0 - 3$. We note that since flux ratios are particularly sensitive to the central density of perturbing halos, the suppression of halo concentrations far above $\mhm$ (because of the factor of 60 in Equation \ref{eqn:cmrelation}) is possibly the dominant effect of dark matter free-streaming on lensing observables. We plot the subhalo mass function and the halo mass-concentration-redshift relation in Figure \ref{fig:mcrelation}. 
	
	\section{The data}
	\label{sec:data}
	We apply the forward-modeling methodology outlined in Section \ref{sec:inference} using the physical model described in Section \ref{sec:parameterizations} to eight quadruply imaged quasars. In this Section, we describe the sample selection, and how the data for these eight systems was obtained. In Table \ref{tab:datasummary} in Appendix \ref{app:C}, we summarize the data used in the analysis and provide the relevant references.
	
	\subsection{The narrow-line systems}
	The quads in our sample have image fluxes measured using the narrow-line emission from the background quasar. Six of these (WGD 2038, WFI 2033, RX J0911, PS J1606, WGD J0405, and WFI 2026) have flux and astrometry presented by \citet{Nierenberg++19}, while the data for B1422 and HE0435 are taken from \citet{Nierenberg++14} and \citet{Nierenberg++17}, respectively. The flux uncertainties for the narrow-line lenses are estimated from the forward-modeling method used to fit the narrow-line spectra. For additional details regarding the measurement methodology for the narrow-line flux ratios, we refer to \citet{Nierenberg++17,Nierenberg++19}. 
	
	\citet{Shajib++18} analyzed several systems in our sample. They measured satellite galaxy location and provided the photometric information for the systems J1606 and WGD J0405, which we used to obtain photometric redshifts (see Appendix \ref{app:B}). 
	
	\subsection{Lenses omitted from our sample}
	
	We apply our analysis to a sample of eight quads, although additional systems exist in the literature with measured flux ratios. We choose only a subset of the total number of possible lenses since the remaining systems either do not have reliable flux measurements, or have complicated deflector morphology that introduces significant uncertainties in the lens modeling. We do not include lenses with fluxes measured using radio emission from the background quasar. Some of these systems may be analyzed in a future work upon revision of our modeling strategy and new flux measurements. 
	
	Specifically, we do not include quads with main lensing galaxies that contain stellar disks, since accurate lens models for these systems require explicit modeling of the disk. This excludes the system J1330 presented by \citet{Nierenberg++19}. We also exclude HS 0810, a system with narrow-line flux measurements presented by \citep{Nierenberg++19} because the flux from the merging images becomes blended together for source sizes larger than 20 pc. This complicates our analysis, as our method for computing image fluxes with extended background sources cannot be applied to merging pairs when the images blur together. 
	
	\begin{table*}
		\centering
		\caption{A summary of deflector $z_d$ and source $z_s$ redshifts, and satellite galaxies included in the lens model for the quads in our sample. Galaxy positions prior marked by $^{*}$ denote observed locations, which may differ from the true physical location due to foreground lensing effects from the lens macromodel. We correct for foreground lensing effects in our inference pipeline (see Section \ref{ssec:satgals}). Satellite galaxy locations are quoted with respect to the light centroid of the main deflector (see Table \ref{tab:datasummary}). All priors on the satellite mass $G2_{\theta_E}$ are positive definite. The raised and lowered numbers around the deflector redshifts for PS J1606, WGD J0405, and WFI 2026 are the $68\%$ confidence intervals on the estimated lens redshifts (see Appendix \ref{app:B}), which we marginalize over.}
		\label{tab:lenspriors}
		\def\arraystretch{1.05}
		\setlength\tabcolsep{0.04in}
		
		\begin{tabular}{lccccccccr} 
			\hline
			lens & $z_d$ & $z_s$ &$ \log_{10} M_{\rm{halo}}$ &$\gamma_{\rm{ext}}$ & $\rm{G2}_x$ & $\rm{G2}_y$ & $\rm{G2}_z$ & $\rm{G2}_{\theta_E}$\\
			\hline
			WGD J0405-3308 & $0.29_{0.25}^{0.32}$ & $1.71$ & $\mathcal{N} \left(13.3, 0.3\right)$  &$\mathcal{U} \left(0.02, 0.1\right)$  & - & - & - & - & \vspacing
			HE0435-1223& 0.45 & 1.69 &  $\mathcal{N} \left(13.2, 0.3\right)$ &$\mathcal{U} \left(0.02, 0.13\right)$  & $ ^{*}\mathcal{N} \left(2.585, 0.05\right)^{*}$ &  $^{*}\mathcal{N}\left(-3.637, 0.05\right)^{*}$  & $z_d + 0.33$ & $\mathcal{N}\left(0.37, 0.03\right)$ \vspacing
			RX J0911+0551 & $0.77$ & $2.76$ & $\mathcal{N} \left(13.1, 0.3\right)$ & see Section \ref{ssec:specificmodeling}  & $\mathcal{N} \left(-0.767,0.05\right)$ & $\mathcal{N} \left(0.657,0.05\right)$  & $z_d$ & $\mathcal{N} \left(0.2,0.2\right)$  \vspacing
			B1422+231 & $0.36$ & $3.67$ & $\mathcal{N} \left(13.3, 0.3\right)$ &$\mathcal{U} \left(0.12, 0.35\right)$ &- & - & - & -\vspacing
			PS J1606-2333 & $0.31_{0.26}^{0.36}$ & $1.70$ & $\mathcal{N} \left(13.3, 0.3\right)$ & $\mathcal{U} \left(0.1, 0.28\right)$  & $\mathcal{N} \left(-0.307, 0.05\right)$  & $\mathcal{N} \left(-1.153, 0.05\right)$ & $z_d$ & $\mathcal{N} \left(0.27, 0.05\right)$ \vspacing
			WFI 2026-4536 & $1.04_{0.9}^{1.12}$ & $2.2$ & $\mathcal{N} \left(13.3, 0.3\right)$  & $\mathcal{U} \left(0.03, 0.16\right)$  & - & - & - & - \vspacing
			WFI 2033-4723 & 0.66 & 1.66 & $\mathcal{N} \left(13.4, 0.3\right)$ &$\mathcal{U} \left(0.13, 0.32\right)$  & $\mathcal{N} \left(0.245, 0.025\right)$ & $\mathcal{N}\left(2.037, 0.025\right)$  & $z_d$ &$\mathcal{N}\left(0.02, 0.005\right)$  \\
			&  &  &  & & $^{*}\mathcal{N}\left(-3.965, 0.025\right)^{*}$ & $^{*}\mathcal{N}\left(-0.025, 0.025\right)^{*}$  & $z_d+0.085$ &$\mathcal{N}\left(0.93, 0.05\right)$ \vspacing
			WGD 2038-4008 & $0.23$ & $0.78$ & $\mathcal{N} \left(13.4, 0.3\right)$ &$\mathcal{U} \left(0.04, 0.12\right)$  &- & - & - & - \vspacing
			\hline		
		\end{tabular}
	\end{table*}
	
	\section{Physical assumptions and priors}
	\label{sec:assumptionsandpriors}
	The parameterizations we introduce in Section \ref{sec:parameterizations} and the priors use in the forward model reflect certain physical assumptions. In this section we describe these assumptions, and the prior probabilities attached to each parameter in the forward model for our sample of quads. 
	
	\subsection{The extended background source}
	\label{ssec:srcassumptions}
	The effect of a dark matter halo of a given mass on the magnification of a lensed image is a function of the background source size \citep{DoblerKeeton02}, see also Figure 14 in \citet{Amara++06} and Figure 8 in \citet{Xu++12}. In general, more extended background sources are less sensitive to dark matter halos (in terms of the image magnifications) on the mass scales relevant for substructure lensing, and the minimum sensitvity threshold for a halo of a given max to produce a measurable flux perturbation is determined by the background source size. 
	
	The lenses in our sample have fluxes measured using emission from the narrow-line region of the background quasar \citep{Nierenberg++17, Nierenberg++19}. The narrow-line region is expected to subtend angular scales larger than a micro-arcsecond, corresponding to physical scales larger than $\sim 1 \rm{pc}$, such that it is immune to microlensing by stars. This physical extent also corresponds to a light-crossing time greater than the typical time delay between lensed images, such that variability in the background quasar should be washed out of the light curves if the source size is indeed large enough to avoid microlensing. 
	
	The size of the narrow-line region typically spans up to $\sim 60 \rm{pc}$ \citep{MullerSanchez++11} defined as the full-width at half maximum (FWHM) of the radially averaged luminosity profile. Upper limits of 50-60 pc may also be obtained by forward modeling the spectrum of the lensed images themselves \citep{Nierenberg++17}. We therefore model the background source as a circular Gaussian and impose a uniform prior on the FWHM between $25 - 60 \rm{pc}$. 
	
	\subsection{Halo and subhalo mass ranges}
	We render halos for both the line of sight and subhalo mass functions in the range $10^6 - 10^{10} \msun$. Halos with masses below $10^{6} \msun$ do not leave imprints on lensing observables for the extended source sizes we consider, which we verify by comparing distributions of image flux ratios with different minimum subhalo masses. The smallest halo masses flux ratios are sensitive to depends on the background source size and the concentration of the halo, but we estimate through ray-tracing simulations that the lower limit lies somewhere between $10^6  - 10^7 \msun$ for the smallest source sizes we model. We include the rare objects more massive than $10^{10} \msun$ by explicitly including them in the lens model, assuming that they host a luminous galaxy, in which case they are detected in the observations of the lenses themselves. This assumption is consistent with current abundance matching techniques \citep{Kim++17b,Nadler++19}. 
	
	\subsection{The line of sight halo mass function}
	\label{ssec:losassumptions}
	We use the Sheth-Tormen \citep{ST99} halo mass function to model structure along the line of sight, with two modifications: First, we introduce a rescaling term $\delta_{\rm{los}}$ to account for a systematic shift in the predicted mean amplitude of the mass function. Second, we include a term $\xi_{\rm{2halo}} \left(M_{\rm{halo}}, z\right)$ that rescales the amplitude of the mass function near the main deflector to account for the presence of correlated structure in the density field near the parent dark matter halo. This results in a $5 - 10 \%$ increase in the number halos near the main deflector. 
	
	Modulo uncertainty in the overall amplitude $\delta_{\rm{los}}$, we assume the halo mass function in the lens cone volume is well-described by the mean halo mass function in the universe. This is a reasonable approximation as lensing volumes span several Gpc, and we expect fluctuations in the dark matter density along the line of sight should average out over large distances. We note, however, that there is some scatter among the predictions from different parameterizations of the halo mass function below $10^{10}\msun$ \citep[e.g.][]{Despali++16} and cosmological model uncertainties, for instance associated with $\sigma_8$ and $\Omega_m$. It is also possible that lenses are selected preferentially in over or under-dense lines of sight. We use a flat prior on $\delta_{\rm{los}}$ between 0.8 and 1.2 to account for these uncertainties.
	
	\subsection{The subhalo mass function}
	\label{ssec:submfuncassumptions}
	Our parameterization of the subhalo mass function is an improvement over previous modeling efforts in predicting strong lensing observables since it explicitly accounts for the evolution of the subhalo mass function with redshift and halo mass, and accounts for the tidal stripping of subhalos by the host dark matter halo. However, since the {\tt{galacticus}} runs do not include a central galaxy\footnote{{\tt{galacticus}} is capable of including the tidal stripping effects from a central galaxy, but we did not include them to minimize computation costs.} we cannot predict the effects of tidal stripping on the projected mass in substructure near the Einstein radius, or the possible redshift and halo mass dependence of this effect. Since tidal destruction of substructures appears to be independent of subhalo mass \citep{GK++17,Graus++18}, we absorb the effects of tidal stripping into the normalization parameter $\Sigma_{\rm{sub}}$ in Equation \ref{eqn:subhalomfunc}. Finally, we note that the prescription for rendering halos outlined in Section \ref{sec:parameterizations} does not couple parameters such as the truncation radius to the concentration of subhalos at infall, and does not model the tidal evolution of subhalos from the time of infall until the time of lensing. These additional degrees of modeling complexity will be implemented in a future analysis that uses a larger sample size of lenses.
	
	To determine reasonable bounds on $\Sigma_{\rm{sub}}$, we compare the predicted surface density in substructure obtained by integrating Equation \ref{eqn:subhalomfunc} over mass with the output from N-body simulations, and from the {\tt{galacticus}} runs. At $z \sim 0.7$, the $\sim 10^{13} \msun$ halos in \citet{Fiacconi++16} have projected substructure mass densities of $10^7 \msun \rm{kpc^{-2}}$ at $0.02 R_{\rm{vir}}$. \citet{Fiacconi++16} show that this value increases when accounting for baryonic contraction of the halo. The {\tt{galacticus}} halos contain more substructure at the same redshift without accounting for baryonic contraction, corresponding to projected mass densities between $2.5 \times 10^{7} \msun \rm{kpc^{-2}}$ and $6 \times 10^{7} \msun \rm{kpc^{-2}}$. Both of these projected mass densities would likely decrease when accounting for tidal stripping. We note, however, that recent works call attention to possible numerical issues that can lead to the artificial fragmentation of subhalos in N-body simulations \citep{vandenBosch++18,ErraniPenarrubia19}. For reference, $\Sigma_{\rm{sub}} = 0.012 \rm{kpc^{-2}}$ corresponds to a projected mass density of $10^{7}\msun \rm{kpc^{-2}}$ at $z=0.5$ in a $10^{13}\msun$ halo, using Equation \ref{eqn:subhalomfunc}. 
	
	With these considerations in mind, we use a wide, flat prior on $\Sigma_{\rm{sub}}$ between 0 and 0.1 $\rm{kpc^{-2}}$ that should encompass the theoretical uncertainties present in the literature. We reiterate that by factoring out the evolution with halo mass and redshift, we intend for the parameter $\Sigma_{\rm{sub}}$ to be common for all the lenses in our sample with scatter from different tidal stripping scenarios and halo-to-halo variance.
	
	The power-law slope $\alpha$ of the subhalo mass function predicted by N-body simulations is consistently in the range $-1.95$ to $-1.85$ \citep{Springel++08,Fiacconi++16}, and because tidal stripping appears independent of mass the presence of a central galaxy should not cause significant deviations from this prediction. We therefore impose a flat prior on $\alpha$ between -1.95 and -1.85. 
	
	\subsection{Free-streaming in WDM}
	\label{ssec:wdmassumptions}
	The prior on $\mhm$ needs to be chosen with care since statements using confidence intervals depend on the choice of prior. We specify the lower bound on the prior for $\mhm$ with the WDM mass-concentration relation (Equation \ref{eqn:cmrelation}) in mind, since the factor of 60 in the denominator of Equation \ref{eqn:cmrelation} results in suppressed halo concentrations nearly two orders of magnitude above the location of the turnover in the mass function (see Figure \ref{fig:mcrelation}). We choose a lower bound for $\mhm$ at $10^{4.8} \msun$ that preserves the CDM-predicted halo concentrations down to $10^7 \msun$. At $10^6 \msun$, even the coldest mass function we model with $\mhm = 10^{4.8} \msun$ result in halo concentrations for $10^{6} \msun$ objects $25\%$ lower than the CDM prediction, but we expect the signal from these very low-mass halos will be sub-dominant given that we model extended background sources which decrease sensitivity to low-mass halos. 
	
	\subsection{The parent dark matter halo mass}
	We use information about the mean population of early-type galaxy lenses, as well as empirical relations between stellar mass, halo mass, and observable quantities such as the image separations and lens/source redshifts, to construct priors for the halo mass of each system.  
	
	First, we estimate the `lensing' velocity dispersion from the Einstein radius and lens/source redshifts using the empirical relation between the stellar mass and velocity dispersion derived by \citet{Auger++10} for a sample of strong lens galaxies. We account for the scatter between spectroscopic velocity dispersion and the `lensing' velocity dispersion \citep{Treu++06}, and uncertainties in the fit by \citet{Auger++10}, and convert the estimated stellar mass into a halo mass using the halo-to-stellar mass ratio $\frac{M_{\rm{halo}}}{M_{*}} = 75_{-27}^{+36}$ inferred by \citet{Lagattuta++10}. The typical uncertainty in the resulting prior for the halo mass is 0.3 dex. 
	
	We use this procedure to construct a prior for the halo mass of each quad, with the exceptions of B1422, PS J1606, and WGD J0405. The stellar velocity dispersions implied by the Einstein radii of these systems is significantly lower than the stellar velocity dispersion in the sample of quads used to calibrate the halo-to-stellar mass ratio in \citet{Lagattuta++10}, and as such the estimate of the halo mass using the above procedure may not be accurate for these systems. For B1422, PS J1606, and WGD J0405, we therefore assume the population mean of $10^{13.3 \pm 0.3} \msun$ inferred by \citet{Lagattuta++10}. We also assume the population mean halo mass for WFI 2026 since the lens redshift used to estimate the central velocity dispersion is very uncertain. 
	
	The system RX J0911 is known to reside near a cluster of galaxies, and thus convergence from the cluster halo contributes to the mass within the Einstein radius. We approximate the contribution from the cluster convergence by noting that it should be approximately equal to the mean external shear we infer of 0.3. We then rescale the Einstein radius by $\sqrt{0.7}$, since the stellar mass scales as $R_{\rm{Ein}}^2$ and where we have used the fact that the mean convergence inside the Einstein radius is approximately equal to one for an isothermal deflector. The priors for the parent halo mass used for each quad are listed in Table \ref{tab:lenspriors}.  
	
	Since we explicitly model the evolution with halo mass, we vary $\Sigma_{\rm{sub}}$ and $M_{\rm{halo}}$ independently. We note however, that $M_{\rm{halo}}$ and $\Sigma_{\rm{sub}}$ are not completely degenerate in our analysis. While the number of lens plane subhalos depends on both parameters, the truncation radius of the subhalos depends on $M_{\rm{halo}}$ through the distribution of subhalo z-coordinates, which in turn depends on the virial radius of the parent halo (see Equation \ref{eqn:truncrad2}), and the 2-halo term appearing in Equation \ref{eqn:losmfunc} depends on the halo mass as a larger halo will have more correlated structure around it. Figure \ref{fig:schematic} provides a visual representation of the link between $M_{\rm{halo}}$, $\Sigma_{\rm{sub}}$, $\alpha$, $\delta_{\rm{los}}$, and $\mhm$. 
	
	\subsection{The main deflector lens model}
	\label{ssec:macromodels}
	The galaxies that dominate the lensing cross-section are typically massive early-types with stellar velocity dispersions $\sigma > 200 \ \rm{km} \ \rm{sec^{-1}}$ \citep{Gavazzi++07,Auger++10,Lagattuta++10}. The mass profiles of these systems are typically inferred to be isothermal, or close to isothermal \citep{Treu++06, Treu++09, Auger++10, Shankar++17}. These observations motivate a simple parameterization for the main deflector lens model, the singular isothermal ellipsoid (SIE) plus external shear. We generalize this model to a power-law ellipsoid with a variable logarithmic slope $\gamma_{\rm{macro}}$ to account for uncertainties associated with the mass profile of the lensing galaxy, and the model-predicted flux ratios. We assume a flat prior on the power-law slope $\gamma_{\rm{macro}}$ between 1.95 and 2.2 for each deflector \citep{Auger++10}. 
	
	In addition to the logarithmic slope of the main deflector mass profile, we sample values for the external shear strength $\gamma_{\rm{ext}}$. The prior for $\gamma_{\rm{ext}}$ is chosen on a lens-by-lens basis by first sampling the macromodel parameter space without subhalos to determine a reasonable starting range for $\gamma_{\rm{ext}}$. The width and center of the prior is adjusted after adding substructure such that the posterior distribution of $\gamma_{\rm{ext}}$ obtained for each lens is contained well within the bounds of the prior. The specific priors used for each system are summarized in Table \ref{tab:lenspriors}. Finally, we use a Gaussian prior for the mass centroid of each quad centered on the main deflector light with a variance of 0.05 arcseconds, a typical modeling uncertainty for quadruple-image systems \citep{Shajib++18, Nierenberg++19}. 
	
	Several studies \citep{EvansWitt03,Hsueh++16,Gilman++17,Hsueh++17,Hsueh++18} explore the role of complicated main deflector morphologies on the model predicted flux ratios. As image magnifications are local probes of the gravitational potential, if there are fluctuations in the surface mass profile on scales comparable to the image separation these structures can affect the image magnifications. In particular, stellar disks, if they go unnoticed, can result in systematically inaccurate lens models. With deep Hubble Space Telescope (HST) images of the narrow-line quads in our sample, we can confirm that they do not contain disks, and indeed are representative of the massive elliptical galaxies with roughly isothermal mass profiles that typically act as strong lenses \citep{Auger++10,Shankar++17}. \citet{Gilman++17} and \citet{Hsueh++18} quantified the systematic uncertainties introduced by modeling early-type galaxy lenses as isothermal ellipsoids with fixed logarithmic slopes $\gamma = 2$. These works found that the resulting systematic uncertainties on image magnifications are typically less than $10\%$. This degree of uncertainty is comparable to the variance in model-predicted image magnifications resulting from marginalizing over a power-law ellipsoid mass model with additional degrees of freedom implemented through a variable logarithmic slope $\gamma$ \citep{Nierenberg++19}. Based on these considerations, we use a power-law ellipsoid with variable logarithmic slopes $\gamma$ to model the main deflector mass profile.
	
	Three quads in our sample do not have measured spectroscopic redshifts. For two of these, we use photometry from \citet{Shajib++18} to compute photometric redshifts probability distributions with the software {\tt{eazy}} \citep{Brammer++08}, and sample the deflector redshift from these distributions in the forward model. For the third system (WFI 2026), which does not have multi-band photometry from \citet{Shajib++18}, we assume a typical velocity dispersion for a massive elliptical galaxy, and derive a probability distribution for the lens redshift from measured quantities such as the source redshift and measured image separation. We give more details regarding this procedure in Appendix \ref{app:B}.    
	
	\subsection{Satellite galaxies and nearby deflectors}
	\label{ssec:satgals}
	We model satellite galaxies and other deflectors near the main lens as Singular Isothermal Spheres (SIS), and assume they lie at the lens redshift unless they have measured redshifts that place them elsewhere. We marginalize over the position and Einstein radius of these objects using Gaussian priors on the positions centered on the light centroid with a variance of $0.05$ arcseconds. We use a Gaussian prior on the Einstein radius which is estimated from lens model fitting, or in some cases by direct measurements on the central velocity dispersion \citep[e.g.][]{Wong++17,Rusu++19}. 
	
	In the cases of HE0435 and WFI 2033, the nearby galaxy lies at a higher redshift than the main lens plane. The light from the galaxy is therefore subject to lensing by the main deflector, and its true physical location differs from its observed position. We estimate the true physical locations of these objects by sampling the macromodel parameter space using the image positions as constraints, and read out the physical position of background satellite given its observed (lensed) position. We then place the satellite at this derived physical location in the forward model sampling with uncertainties of $0.05$ arcseconds. This process significantly speeds up the lensing computations since it does not require the continuous reevaluation of the physical satellite location given its observed position during each lens model computation\footnote{The physical location of the nearby galaxy needs to be continuously reevaluated because it's observed location depends on the foreground lensing effects from the macromodel, and the parameters describing the macromodel are continuously changing while finding a solution to the lens equation (Equation \ref{eqn:raytracing}).}. The boost in speed comes at the cost of decoupling the satellite galaxy position from the dark matter parameters $\qsub$ in the inference, but we expect the covariance between these quantities will be negligible because the satellite galaxies, even when their locations are corrected for foreground lensing effects, are relatively far from the images, introducing convergence at the main deflector light centroid of $< 0.1$ in both cases.\footnote{The default convention in {\tt{lenstronomy}} is to place deflectors at their observed angular locations in the universe, but it is now possible (in code versions 0.8.0+) to specify which objects should be treated using the observed (lensed) position instead. We note that the default convention in {\tt{lensmodel}} \citep{Keeton++97} is to place objects at their observed (lensed) locations during multi-plane ray-tracing.}
	
	In the case of HE0435, we estimate the angular location without foreground lensing of the satellite to be $(-2.37, 2.08)$, while for WFI 2033 we obtain $(-3.63, -0.08)$, for observed (lensed) locations of $(-2.911, 2.339)$ and $(-3.965, -0.022)$, respectively. These coordinates are with respect to the galaxy light centroid (see Table \ref{tab:datasummary}). The angular locations of the lensed background satellites are closer to the mass centroid of the main deflector, just as the physical location of the lensed background quasar is concentric with the mass centroid.  
	
	The lens-specific priors on satellite galaxies are summarized in Table \ref{tab:lenspriors}. 
	\begin{figure*}
		\includegraphics[clip,trim=0.8cm 1cm 0.5cm
		0.5cm,width=.29\textwidth,keepaspectratio]{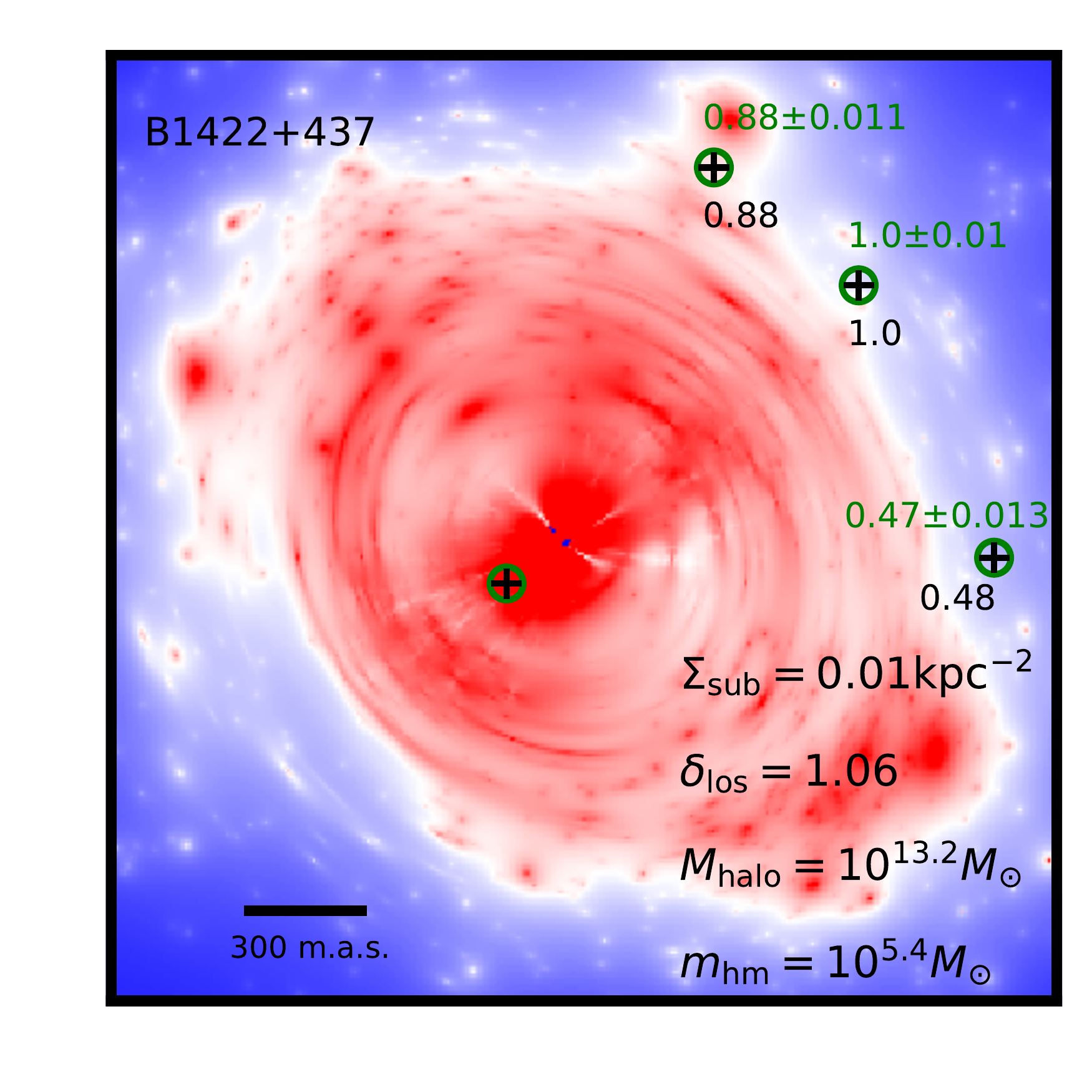}
		\includegraphics[clip,trim=0.8cm 1cm 0.5cm
		0.5cm,,width=.29\textwidth,keepaspectratio]{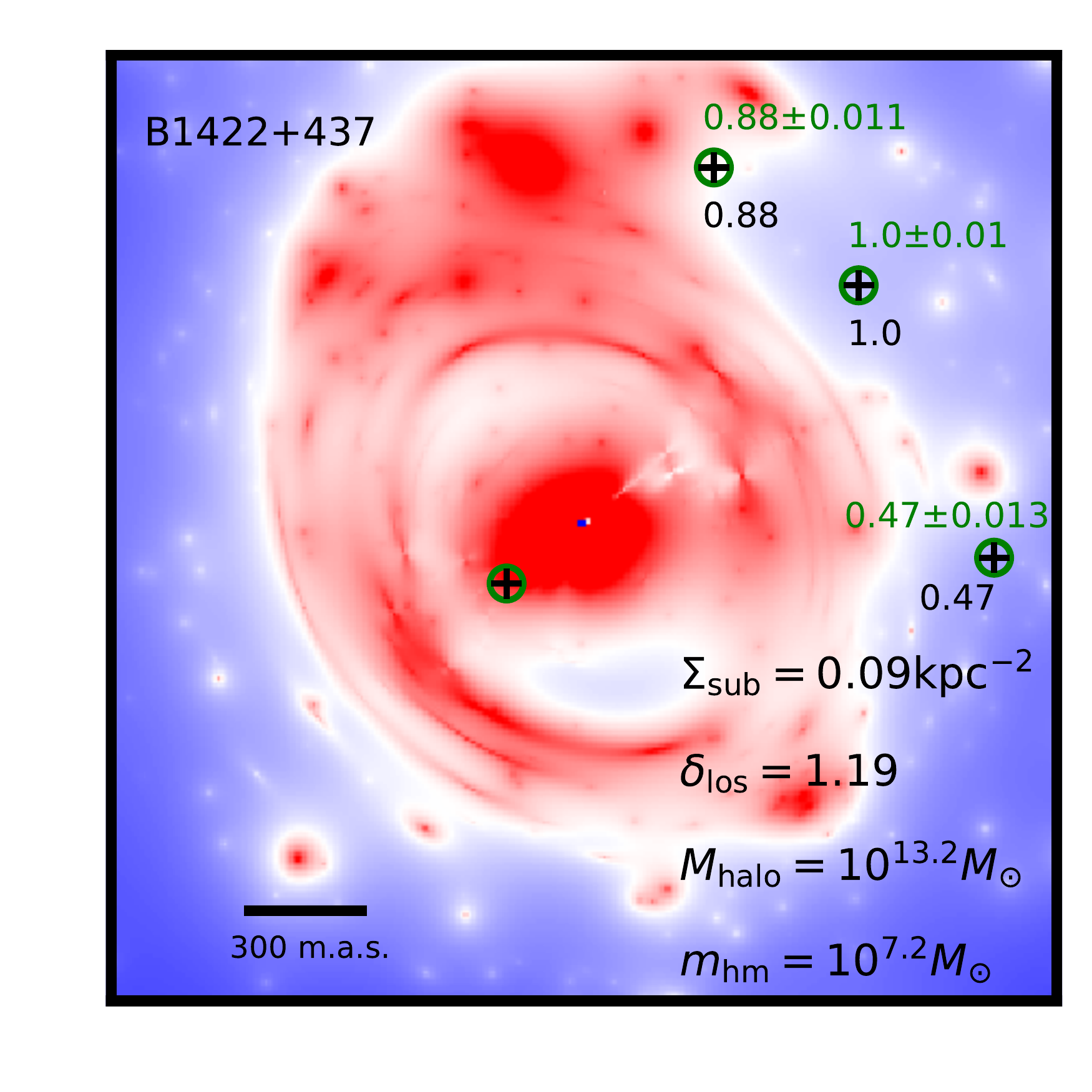}
		\includegraphics[clip,trim=1cm 2.3cm 0.3cm
		2.2cm,width=.34\textwidth,keepaspectratio]{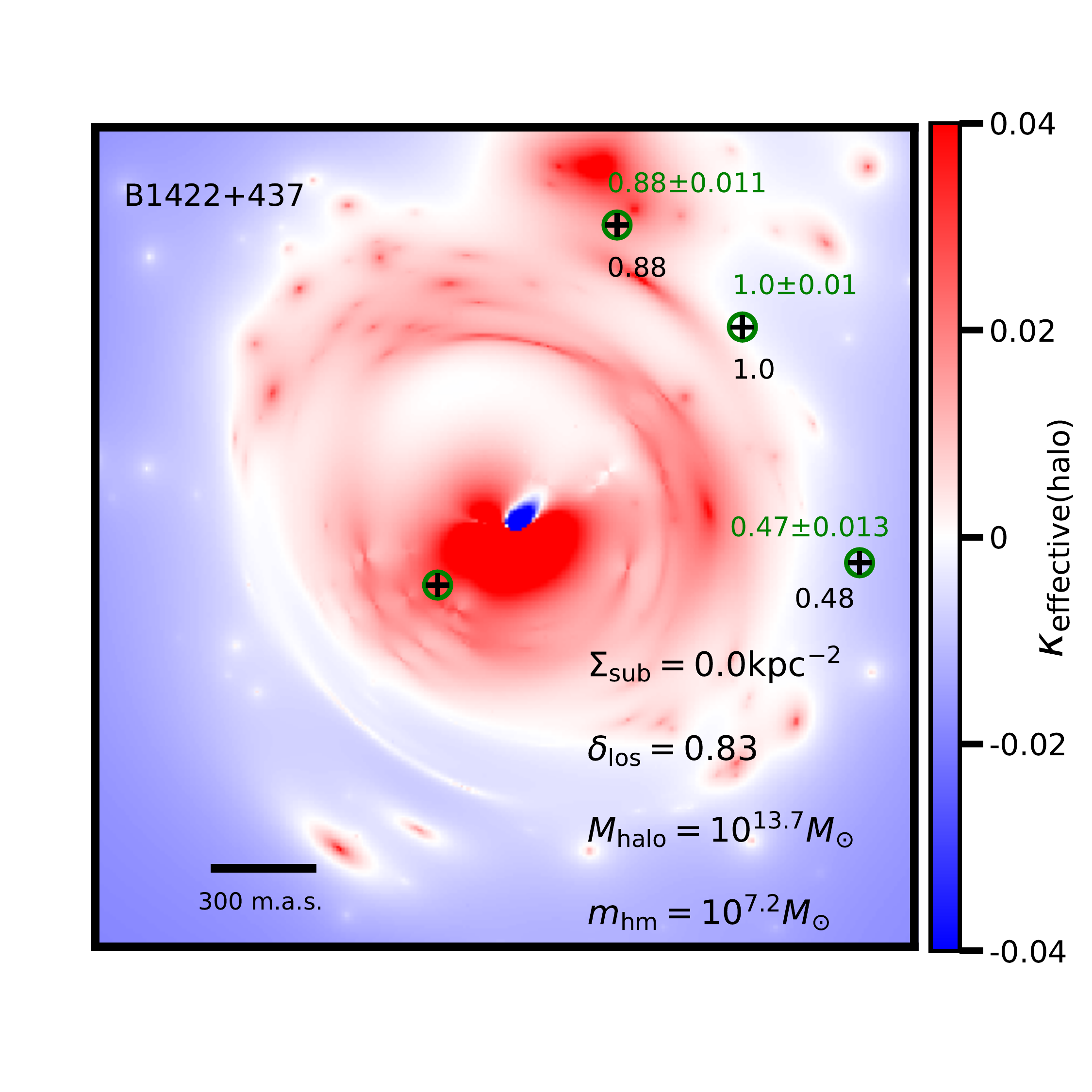}
		\includegraphics[clip,trim=0.8cm 1cm 0.5cm
		0.5cm,width=.29\textwidth,keepaspectratio]{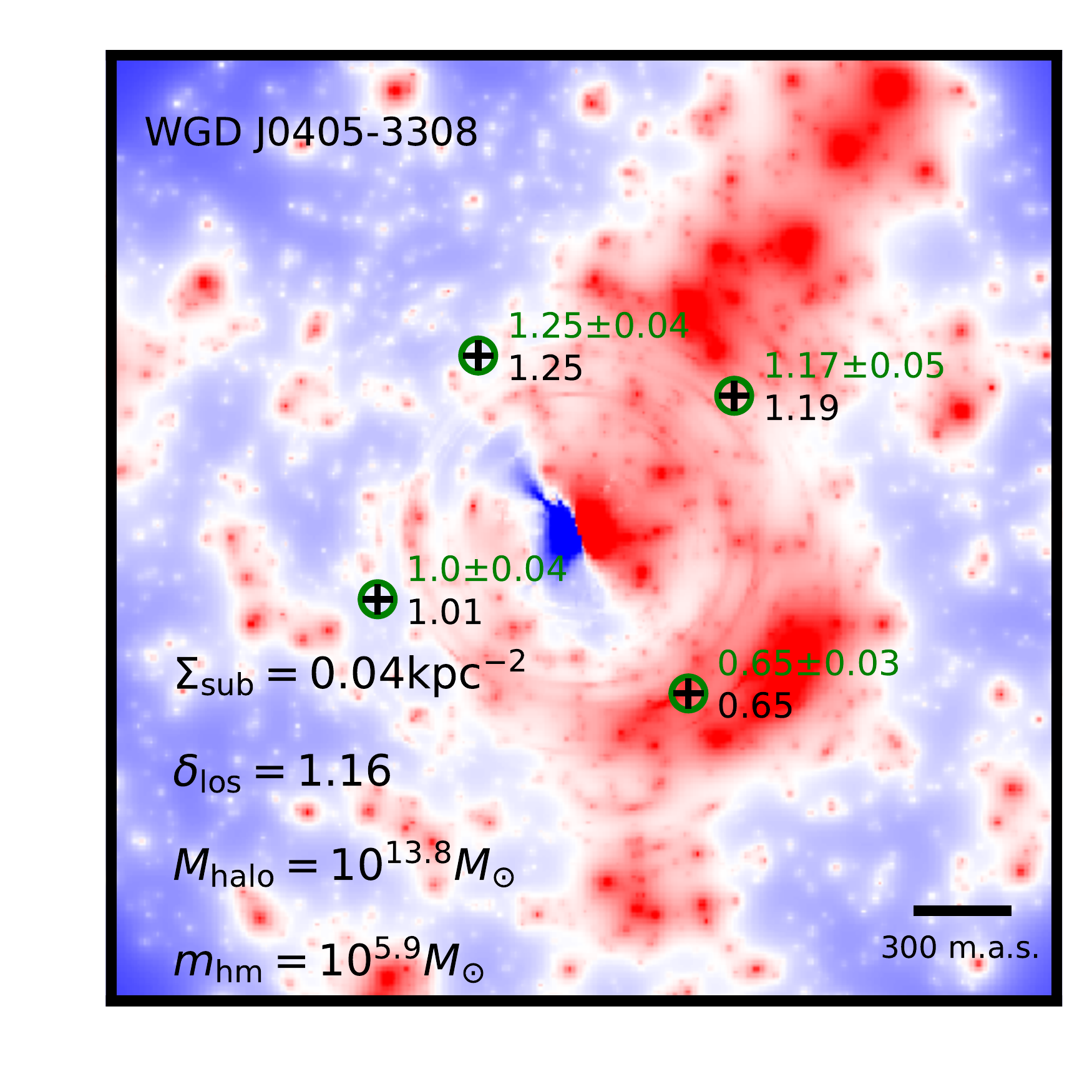}
		\includegraphics[clip,trim=0.8cm 1cm 0.5cm
		0.5cm,,width=.29\textwidth,keepaspectratio]{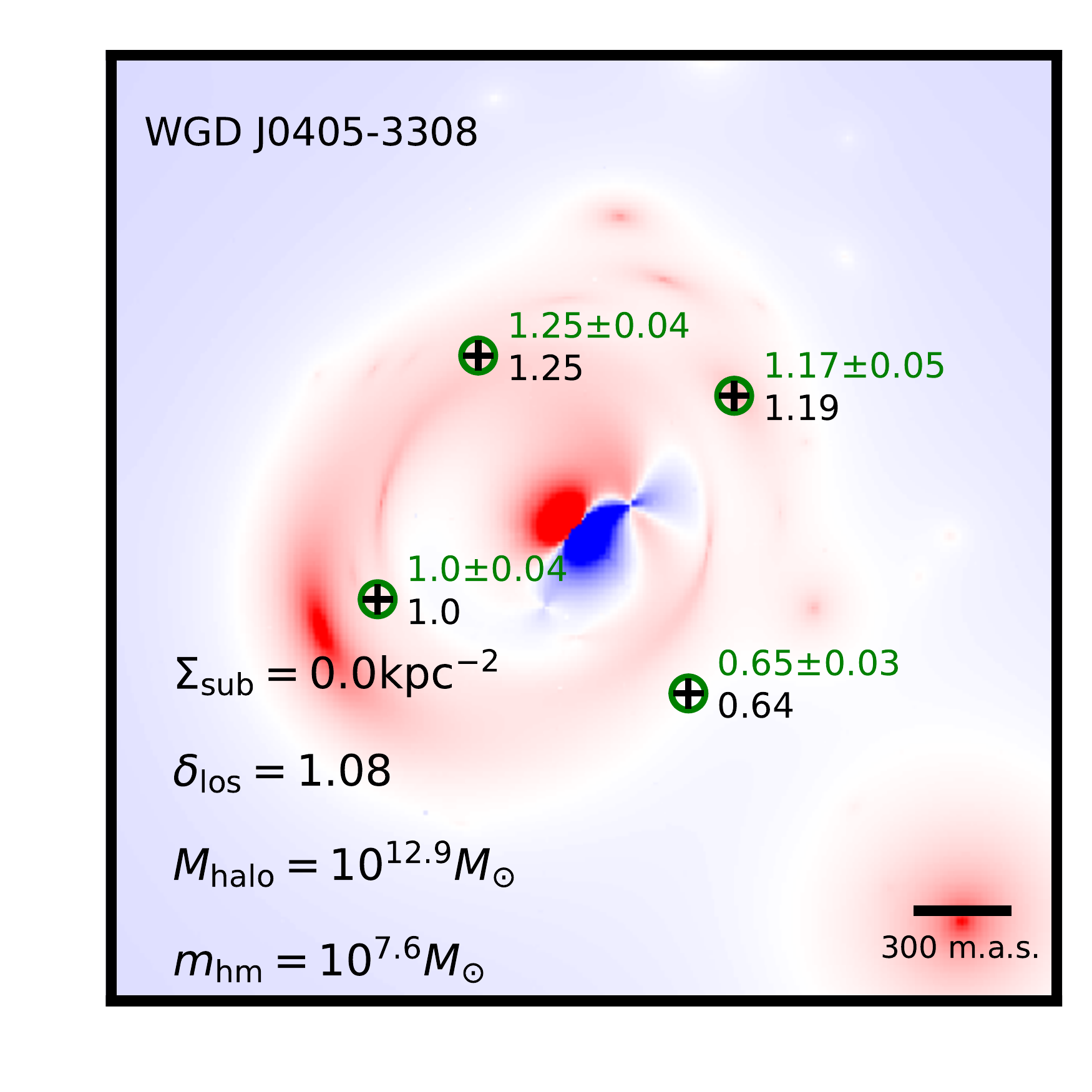}
		\includegraphics[clip,trim=1cm 2.3cm 0.3cm
		2.2cm,width=.34\textwidth,keepaspectratio]{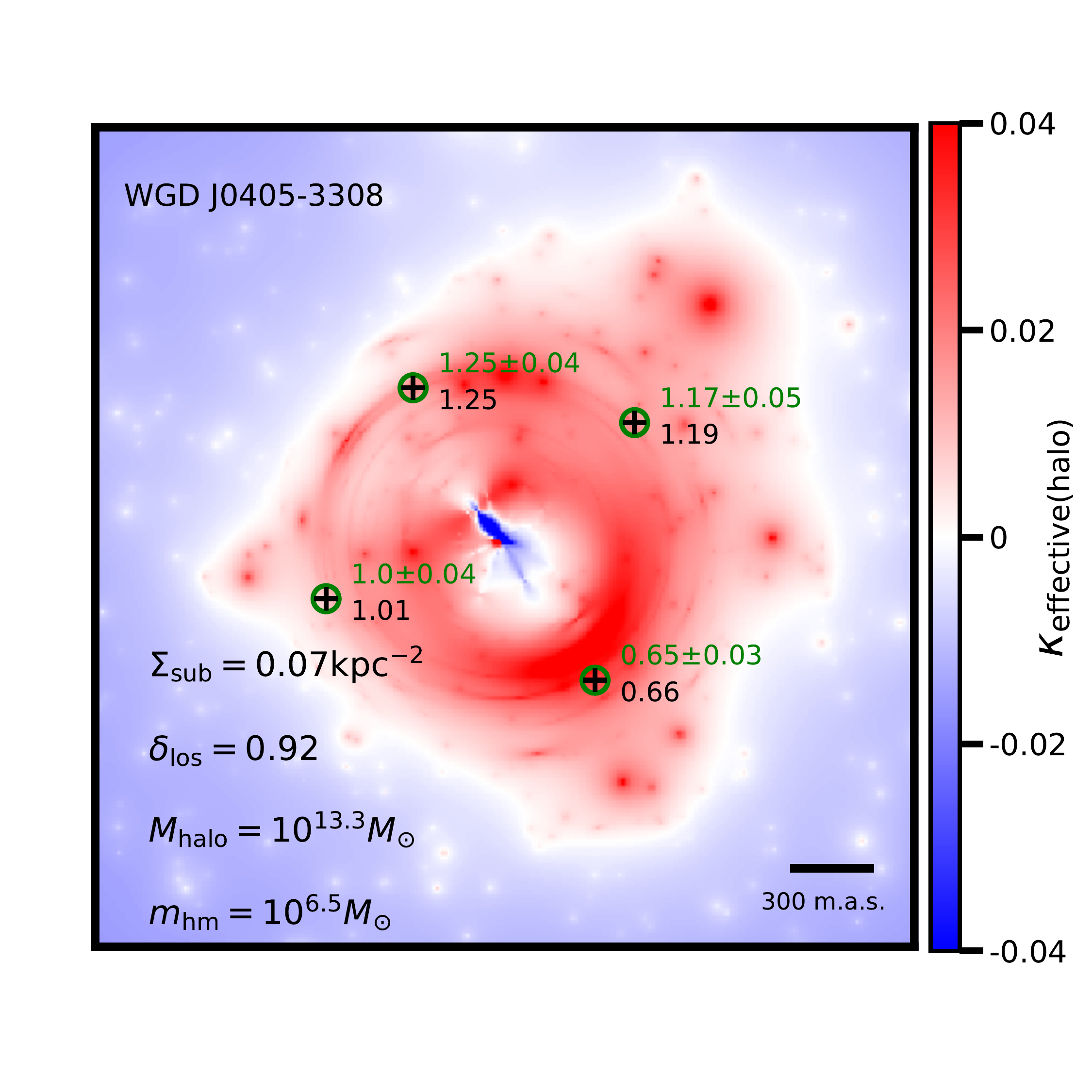}
		\includegraphics[clip,trim=0.8cm 1cm 0.5cm
		0.5cm,width=.29\textwidth,keepaspectratio]{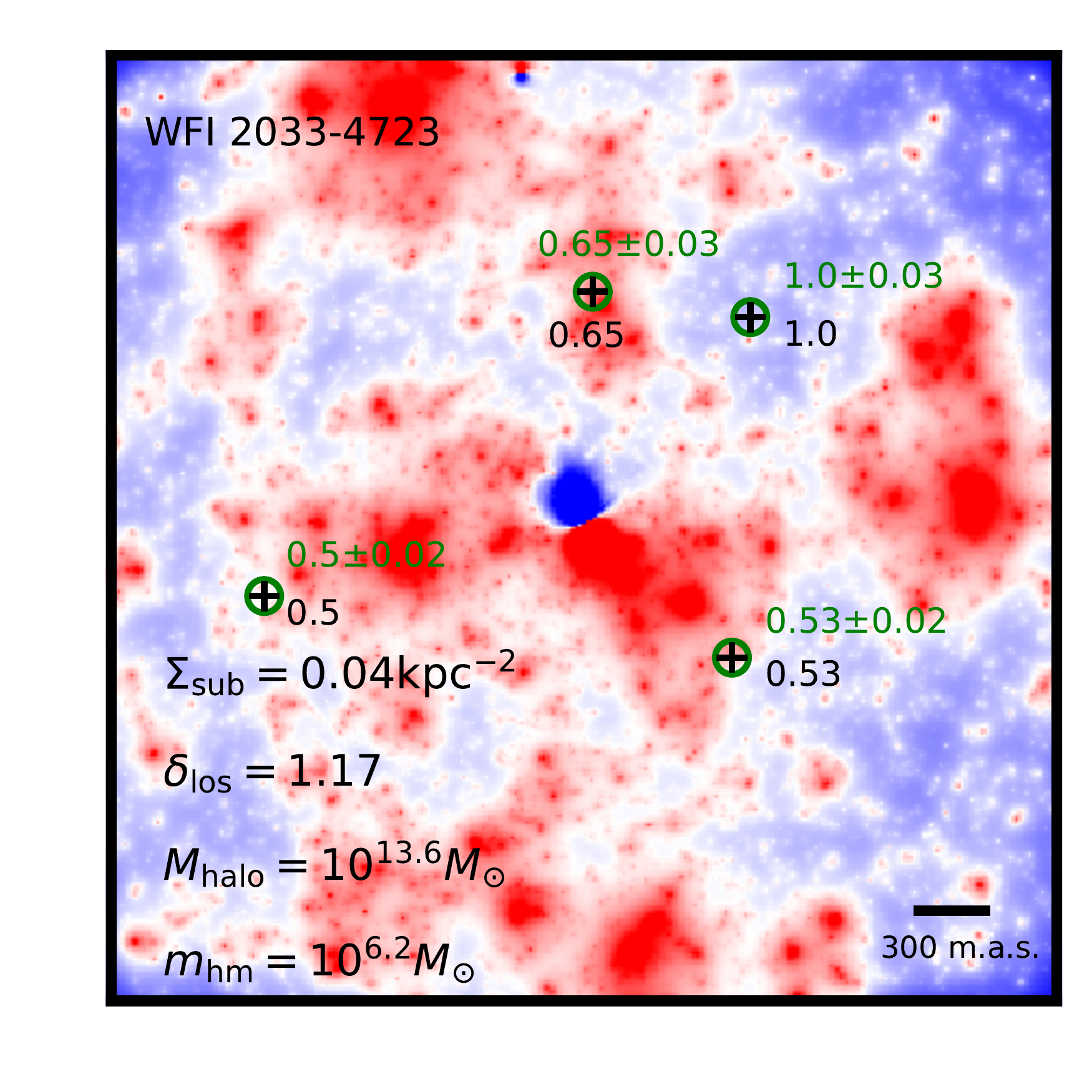}
		\includegraphics[clip,trim=0.8cm 1cm 0.5cm
		0.5cm,,width=.29\textwidth,keepaspectratio]{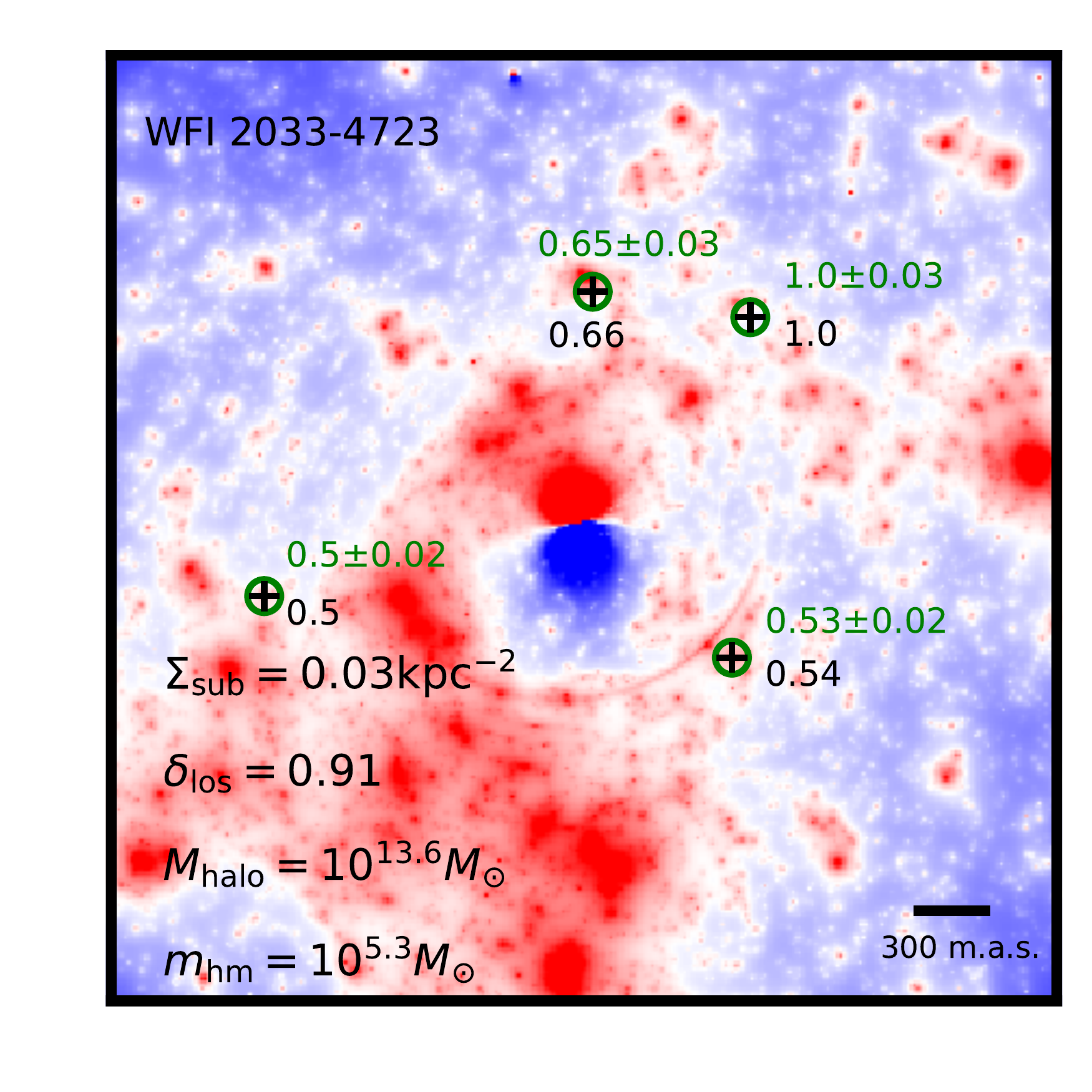}
		\includegraphics[clip,trim=1cm 2.3cm 0.3cm
		2.2cm,width=.34\textwidth,keepaspectratio]{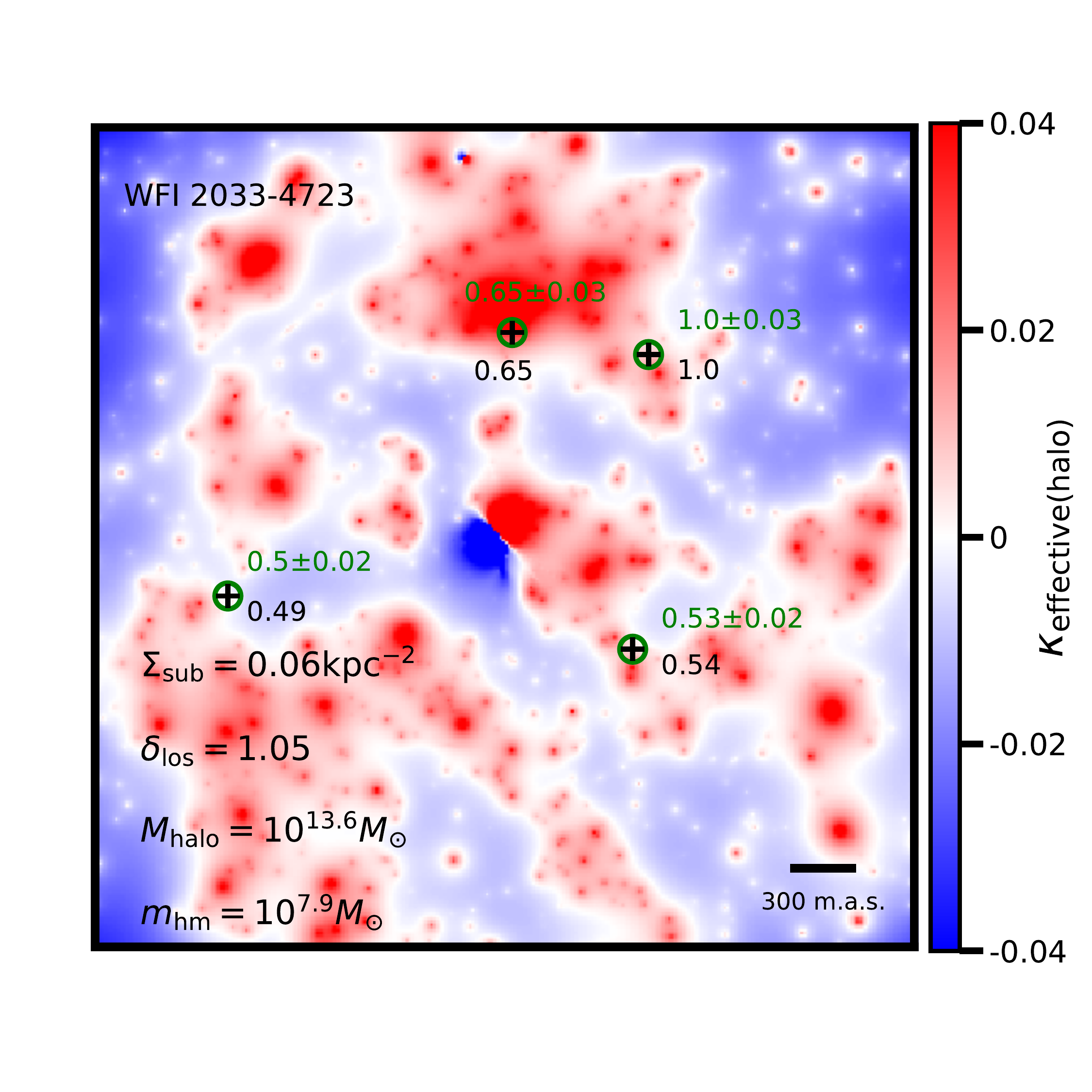}
		\includegraphics[clip,trim=0.8cm 1cm 0.5cm
		0.5cm,width=.29\textwidth,keepaspectratio]{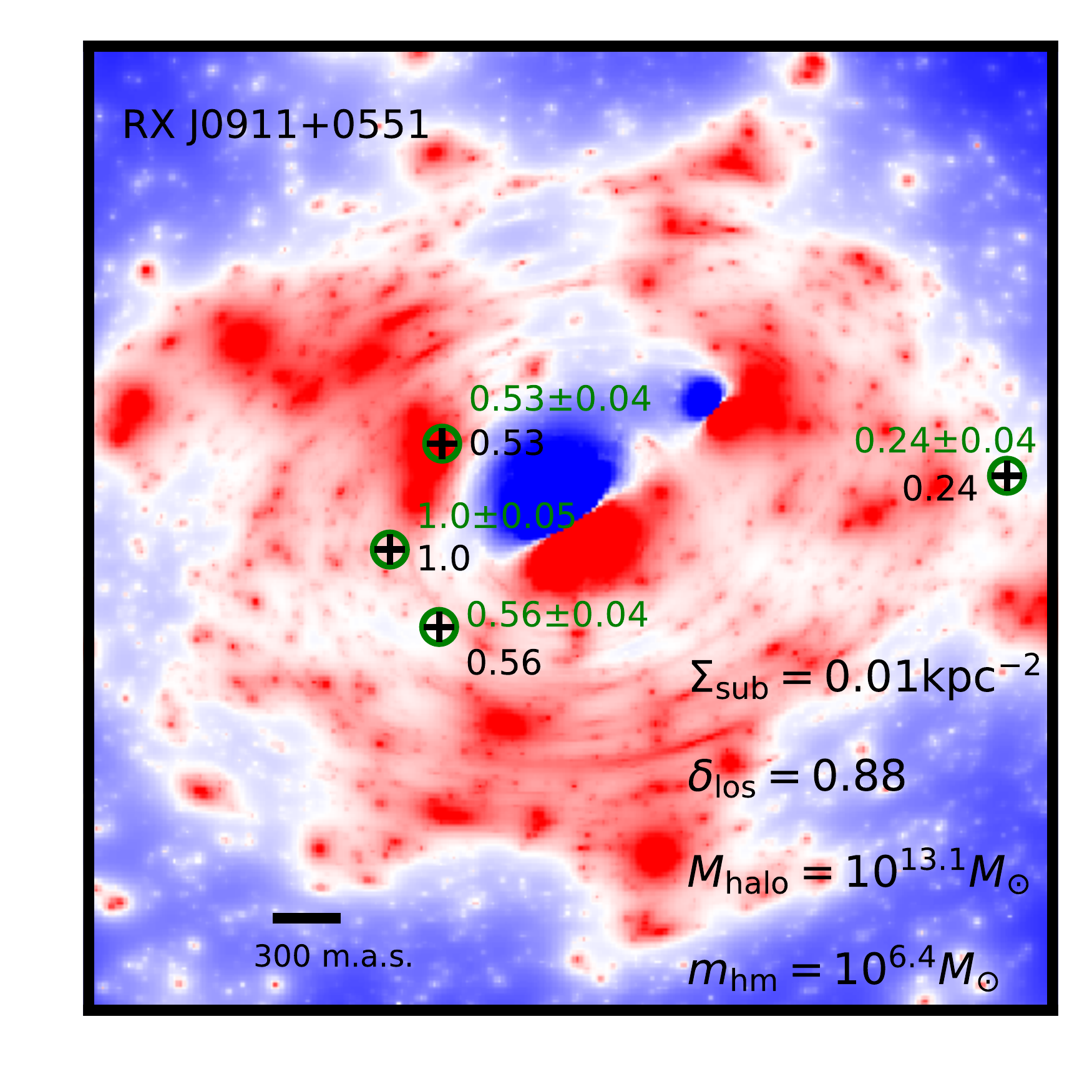}
		\includegraphics[clip,trim=0.8cm 1cm 0.5cm
		0.5cm,,width=.29\textwidth,keepaspectratio]{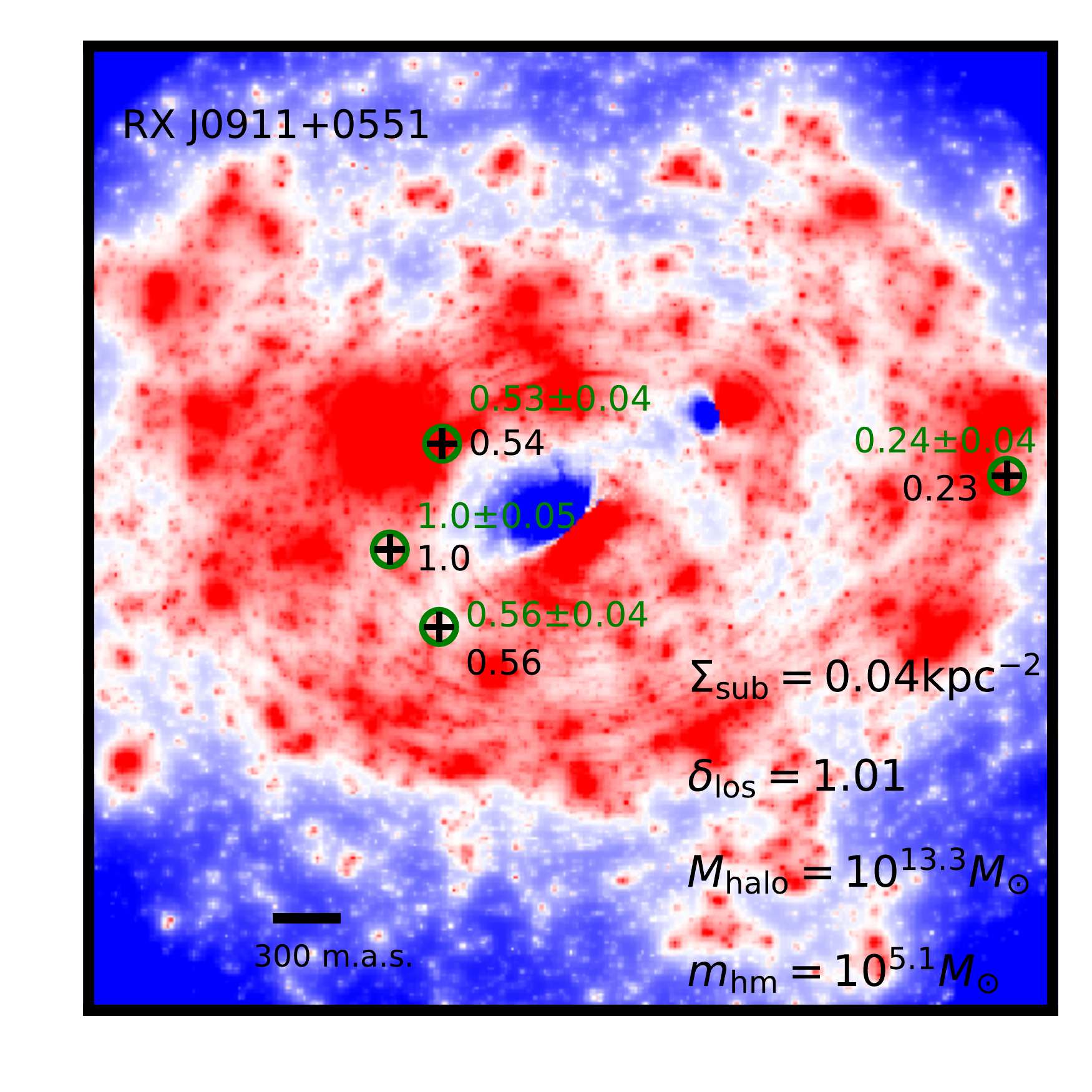}
		\includegraphics[clip,trim=1cm 2.3cm 0.3cm
		2.2cm,width=.34\textwidth,keepaspectratio]{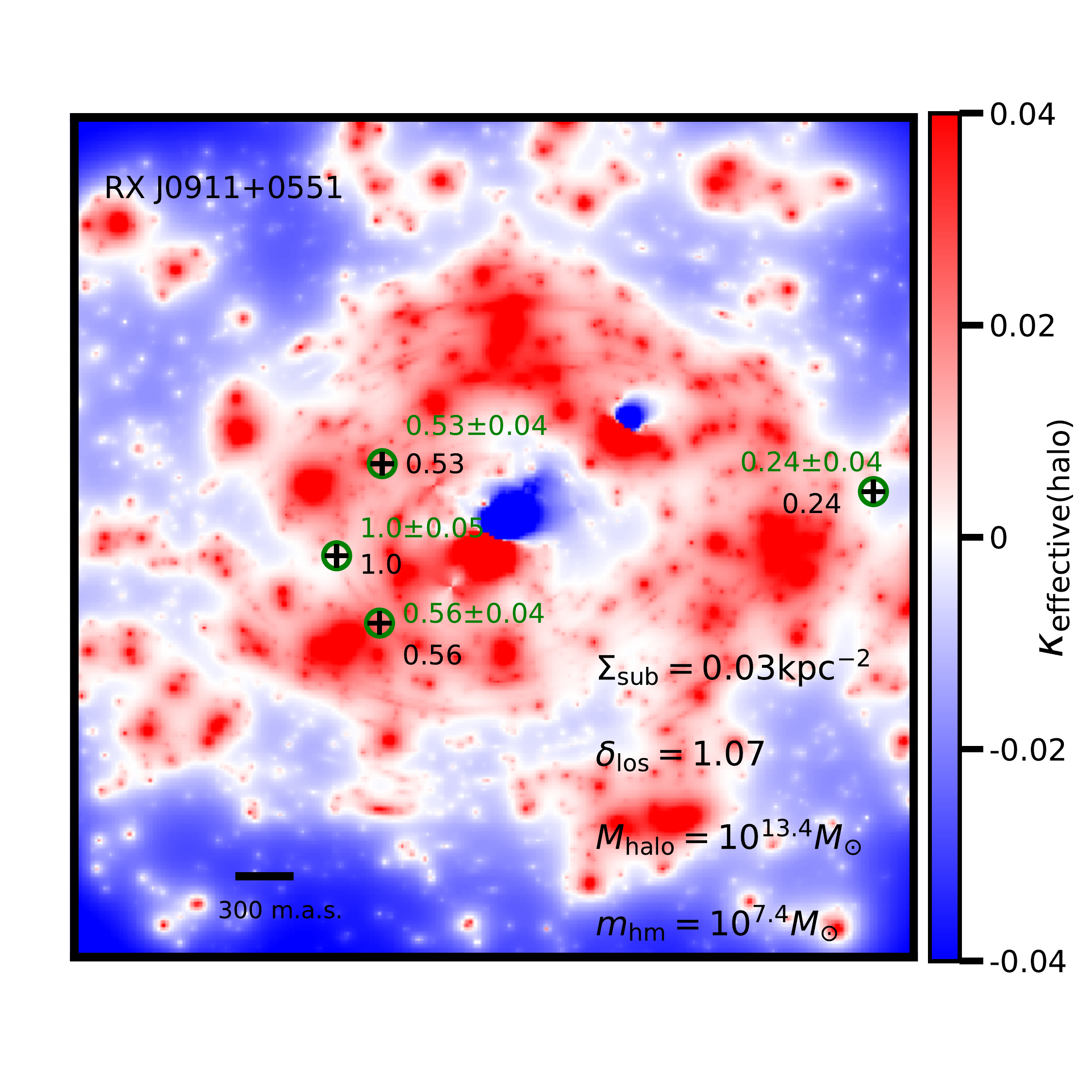}
		\caption{\label{fig:best_realizations} Dark matter halo {\textit{effective multi-plane convergence}} maps for some of the highest-ranked realizations for the subset of quads B1422, WGD J0405, WFI 2033, and RX J0911, each of which has flux ratios inconsistent with smooth lens models. The defintion of the {\textit{effective multi-plane convergence}} takes into account the non-linear effects present in multi-plane lensing, and is defined with respect to the mean dark matter density in the universe such that some regions are underdense (blue), while other regions (specifically, dark matter halos) are over-dense (red). The subhalo mass function normalization, line of sight normalization, halo mass and half-mode mass are displayed for each realization. Green text/circles denote observed image positions and fluxes, while black text/crosses denote the model positions and fluxes. The forward-model data sets fit the image positions and fluxes to within the measurement uncertainties. }
	\end{figure*}	
	\begin{figure*}
		\includegraphics[clip,trim=1.5cm 1.6cm 1cm
		1cm,width=0.95\textwidth,keepaspectratio]{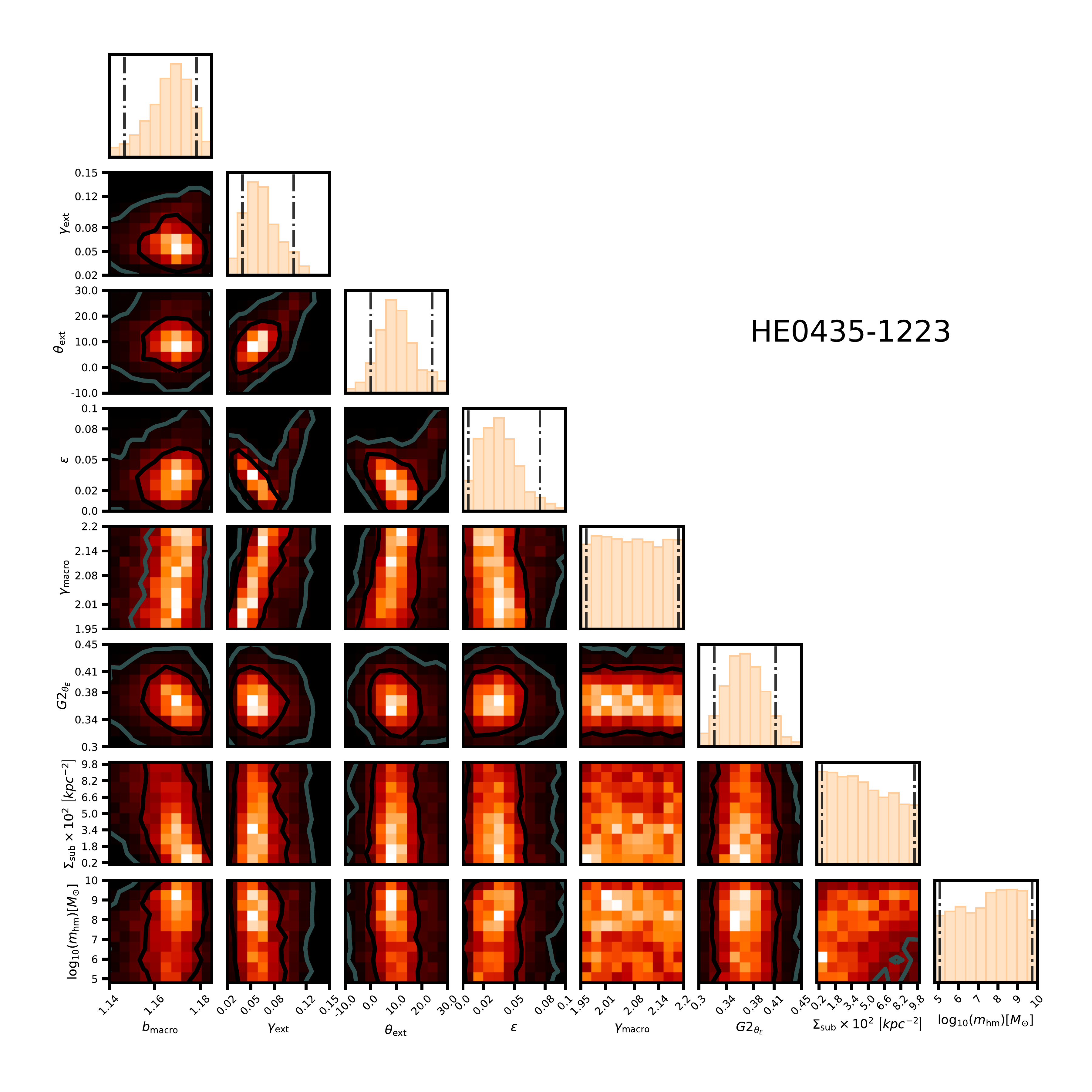}
		\caption{\label{fig:0435inf} Joint posterior distribution for a subset of $\qmac$ and $\qsub$ parameters for the system HE0435. We display the normalization of the main deflector lens model $b_{\rm{macro}}$, the external shear strength and position angle $\gamma_{\rm{ext}}$ and $\theta_{\rm{ext}}$, the deflector ellipticity $\epsilon$, the power-law slope of the main deflector mass profile $\gamma_{\rm{macro}}$, the Einstein radius of the satellite galaxy $G2_{\theta_E}$, the normalization of the subhalo mass function $\Sigma_{\rm{sub}}$, and the half-mode mass $\mhm$. We simultaneously sample the distributions of these parameters to account for covariance between the macromodel and the dark matter hyper-parameters $\qsub$. Vertical lines denote $95\%$ confidence intervals.}
	\end{figure*}	
	\begin{figure*}
		\includegraphics[clip,trim=1.5cm 1.6cm 1cm
		1cm,width=0.95\textwidth,keepaspectratio]{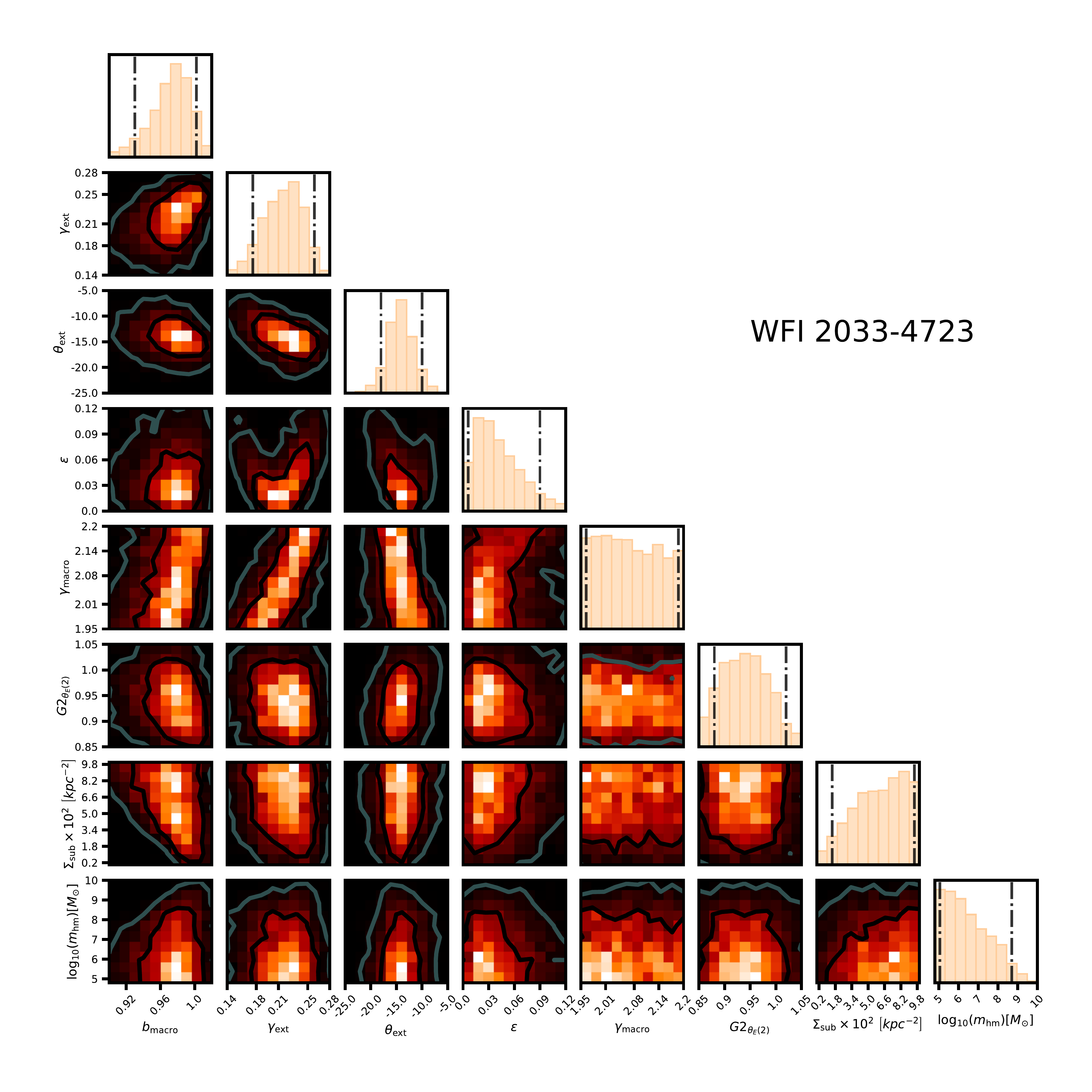}
		\caption{\label{fig:02033inf} Joint posterior distribution for a subset of $\qmac$ and $\qsub$ parameters for the system WFI 2033. The parameters are the same as in Figure \ref{fig:0435inf}. In addition to the main deflector we model two additional nearby galaxies, with Einstein radii $G2_{\theta_E(1)}$ and $G2_{\theta_E(2)}$. We show the distributions of the Einstein radius for the larger nearby galaxy ($G2_{\theta_E(2)}$), whose position we correct for foreground lensing effects (see Section \ref{ssec:satgals}). }
	\end{figure*}	
	
	\begin{figure*}
		\includegraphics[clip,trim=1.5cm 1.2cm 1cm
		1cm,width=0.95\textwidth,keepaspectratio]{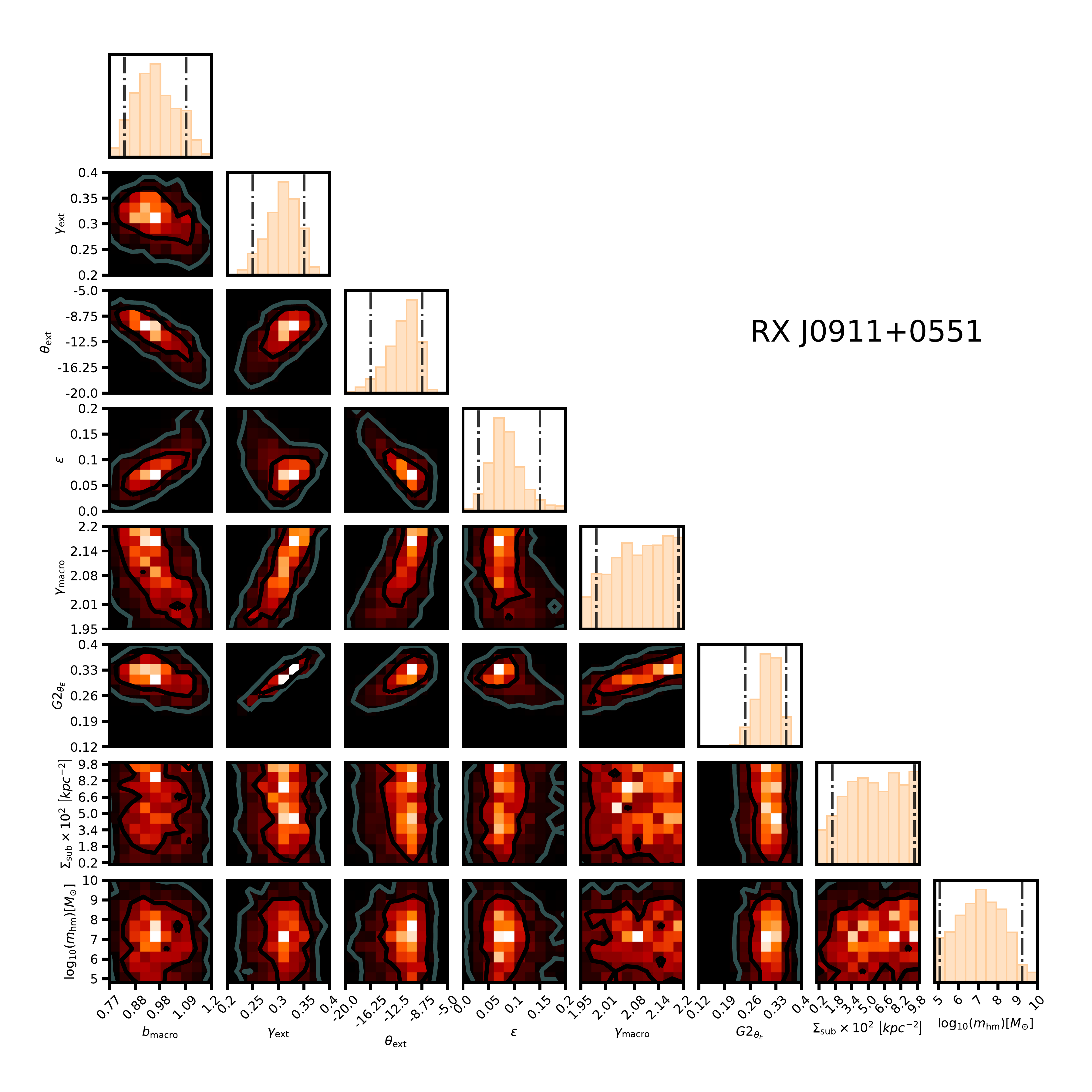}
		\caption{\label{fig:0911inf} Joint posterior distribution for a subset of $\qmac$ and $\qsub$ parameters for the system RX J0911. The parameters are the same as in Figure \ref{fig:0435inf}. }
	\end{figure*}	
	
	\subsection{Lens-specific modeling for RX J0911+0551 and WGD 2038-4008}
	\label{ssec:specificmodeling}
	For system RX J0911, we alter the modeling strategy slightly to increase computational efficiency by allowing the external shear strength $\gamma_{\rm{ext}}$ to vary freely while solving for macromodel parameters that fit the observed image positions. 
	For the system WGD 2038, we widen the prior on the power-law slope of the macromodel as the posterior using the default range for $\gamma_{\rm{macro}}$ between 1.95 - 2.2 is biased towards higher values of $\gamma_{\rm{macro}}$. For WGD 2038, the posterior peaks at $\gamma_{\rm{macro}} \sim 2.25$.  
	
	\section{Results}
	\label{sec:results}
	In this section we present the results of our analysis. We begin in Section \ref{ssec:jointinf} by showing dark matter halo convergence maps for some of the top-ranked realizations drawn in the forward model. We then display the posterior distributions for a few individual lenses, showing the simultaneous inference of parameters describing the macro lens model and the dark matter hyper-parameters. In Section \ref{ssec:mainresults} we present the constraints on the abundance of substructure and dark matter warmth for the full sample of 11 quads.
	
	\subsection{Top-ranked realizations and posteriors for individual lenses}
	\label{ssec:jointinf}
	Minimizing the summary statistic in Equation \ref{eqn:summary_stat} selects realizations that resemble the observed data as closely as possible. This guarantees that the set of accepted dark matter hyper-parameters $\qsub$ yield an accurate approximation of the true posterior distribution for each individual lens with data $\data$: $p\left(\qsub | \data\right)$. For visualization purposes, and to reinforce the fact that the top-ranked realizations look like the data and satisfy $S_{\rm{lens}} \approx 0$ (Equation \ref{eqn:summary_stat}), in Figure \ref{fig:best_realizations} we display the dark matter halo \textit{effective multi-plane convergence} maps for some of the top ranked realizations for a subset of quads in our sample. The \textit{effective multi-plane convergence} is defined as half the divergence of the full deflection field $\boldsymbol{\alpha}$
	\begin{equation}
	\label{eqn:kappaeff}
	\kappa_{\rm{effective}}  \equiv \frac{1}{2} \ \div{ \boldsymbol{\alpha}}.
	\end{equation}
	This definition of the multi-plane convergence accounts for the non-linear effects present in multi-plane lensing, and satisfies the single-plane definition of convergence as second derivatives of a lensing potential in the absence of multiple lens planes. 
	
	To visualize individual realizations of dark matter structure, we define $\kappa_{\rm{effective(halo)}} \equiv \kappa_{\rm{effective}} - \kappa_{\rm{macro}}$, where $\kappa_{\rm{macro}}$ is the convergence from the lens macromodel, including satellite galaxies and nearby deflectors. In the resulting convergence maps, halos located behind the main lens plane appear sheared tangentially around the Einstein radius due to coupling to the large deflections produced by the macromodel. 
	
	In Figure \ref{fig:best_realizations}, we show $\kappa_{\rm{effective(halo)}}$ maps of randomly selected realizations of dark matter structure whose corresponding $\qsub$ parameters were accepted in the final posterior on the basis of their summary statistic $S_{\rm{lens}}$. The specific realizations and the corresponding dark matter parameters $\qsub$ correspond to a diverse set of substructure populations, warm and cold, which yield similarly good fits to the observed flux ratios satisfying $S_{\rm{lens}} \sim 0$. Some models, however, predict flux ratios that match the observed flux ratios more frequently than others. In terms of the Approximate Bayesian Computing algorithm described in Section \ref{sec:inference}, the frequency with which one dark matter model relative to another predicts observables that resemble the data is a surrogate for the relative likelihood of the models. The probability of accepting a proposed $\qsub$ based on the summary statistic in Equation \ref{eqn:summary_stat} is therefore equal to the likelihood $p\left(\data | \qsub\right)$ (Equation \ref{eqn:likelihood}), even though the form of this function is unknown and it is never directly evaluated. 
	
	The top-ranked realizations for B1422 shown in Figure \ref{fig:best_realizations} each have a relatively massive dark matter halo, or several smaller ones, located near the top left merging triplet image with (normalized) flux 0.88. This is in agreement with the analysis by \citet{Nierenberg++14}, who find that a blob of dark matter near this image brings the model-predicted flux ratios into agreement with a smooth lens model. 
	
	Although not obvious from examining Figure \ref{fig:best_realizations}, the underlying macromodels for each accepted realization are unique, with different external shears, power-law slopes, lens ellipticity, etc. We marginalize over different macromodel configurations by simultaneously sampling the macromodel parameters and the dark matter hyper-parameters in the forward model. To illustrate, in Figures \ref{fig:0435inf}, \ref{fig:02033inf}, and \ref{fig:0911inf} we show the posterior distributions for several parameters in the lens macromodel, along with the dark matter hyper-parameters $\Sigma_{\rm{sub}}$ and $\mhm$ for HE0435, WFI 2033, and RX J0911. The system HE0435 generally favors models with low subhalo mass function normalizations (low $\Sigma_{\rm{sub}}$), or a turnover the mass function with higher $\Sigma_{\rm{sub}}$. The system WFI 2033 is the opposite, with a posterior favoring CDM-like mass functions with many lens plane subhalos. The system RX J0911 lies somewhere in between, with a peak in the posterior distribution of $\mhm$ near $10^{7} \msun$. 
	
	For each of these systems, in particular WFI 2033, there is a visibly obvious covariance between the overall normalization of the main deflector mass profile $b_{\rm{macro}}$\footnote{$b_{\rm{macro}}$ has units of convergence, or projected mass density divided by the critical surface mass density for lensing.}, and the parameters $\Sigma_{\rm{sub}}$ and $\mhm$. This covariance is readily understood: To reproduce the observed image positions, the macromodel responds to the addition of mass in the form of subhalos in main lens plane by decreasing the overall normalization of the main deflector mass profile, and hence these quantities are anti-correlated. Similarly, WDM models correspond to macromodels with larger $b_{\rm{macro}}$ because WDM realizations contain fewer subhalos. Interestingly, there is some structure in the posterior distribution for the lens ellipticity $\epsilon$ in WFI 2033, and both $\mhm$ and $\Sigma_{\rm{sub}}$. 
	
	By simultaneously sampling the lens macromodel and dark matter hyper-parameters, we obtain posterior distributions that account for covariance between $\qmac$ and $\qsub$. We do not use lens model priors from more sophisticated lens modeling efforts (e.g. \citet{Wong++17,Shajib++18}) because these analyses did not include substructure in the lens models and therefore do not account for covariances between the macromodel parameters and the dark matter parameters of interest. For the same reason, we do not decouple the lens macromodel parameters from the dark matter hyper-parameters by first sampling the macromodel parameter space that fits the image positions, and using these distributions as priors in the forward modeling. 
	
	\begin{figure*}
		\includegraphics[clip,trim=0.2cm 0cm 0.5cm
		0.5cm,width=0.95\textwidth,keepaspectratio]{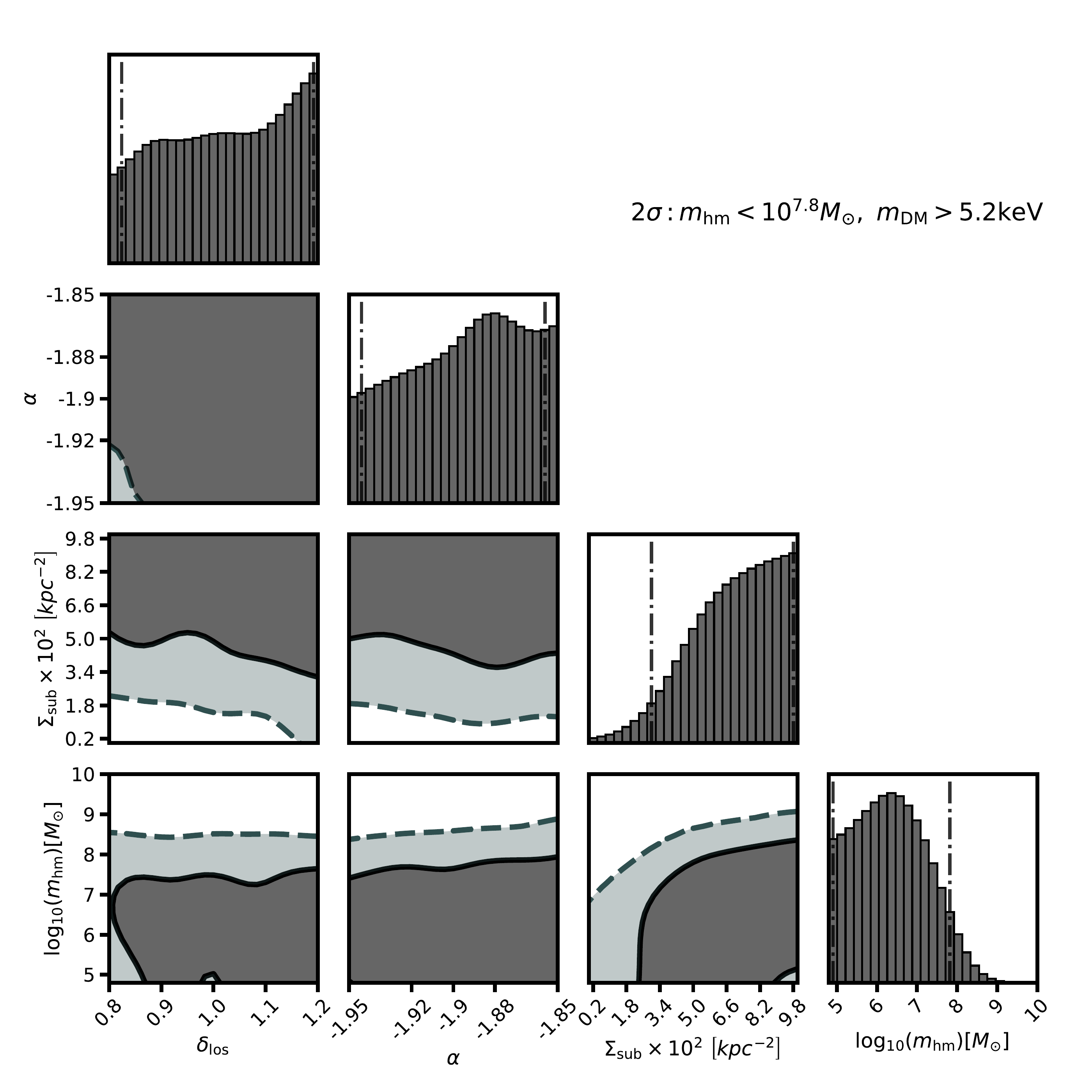}
		\caption{\label{fig:mainresult} Marginal and joint posterior distributions for the dark matter hyper-parameters $\delta_{\rm{los}}$, $\alpha$, $\Sigma_{\rm{sub}}$, and $\mhm$, which represent the overall scaling of the line of sight halo mass function, the logarithmic slope of the subhalo mass function, the global normalization of the subhalo mass function that accounts for evolution with halo mass and redshfit (see Equation \ref{eqn:subhalomfunc}), and the half-mode mass $\mhm$ relevant to WDM models. Contours show $68\%$ and $95\%$ confidence intervals, while the dot-dashed lines on the marginal distributions show the $95\%$ confidence intervals.}
	\end{figure*}	

	\begin{figure}
	\includegraphics[clip,trim=3cm 0.3cm 3cm
	0cm,width=0.45\textwidth,keepaspectratio]{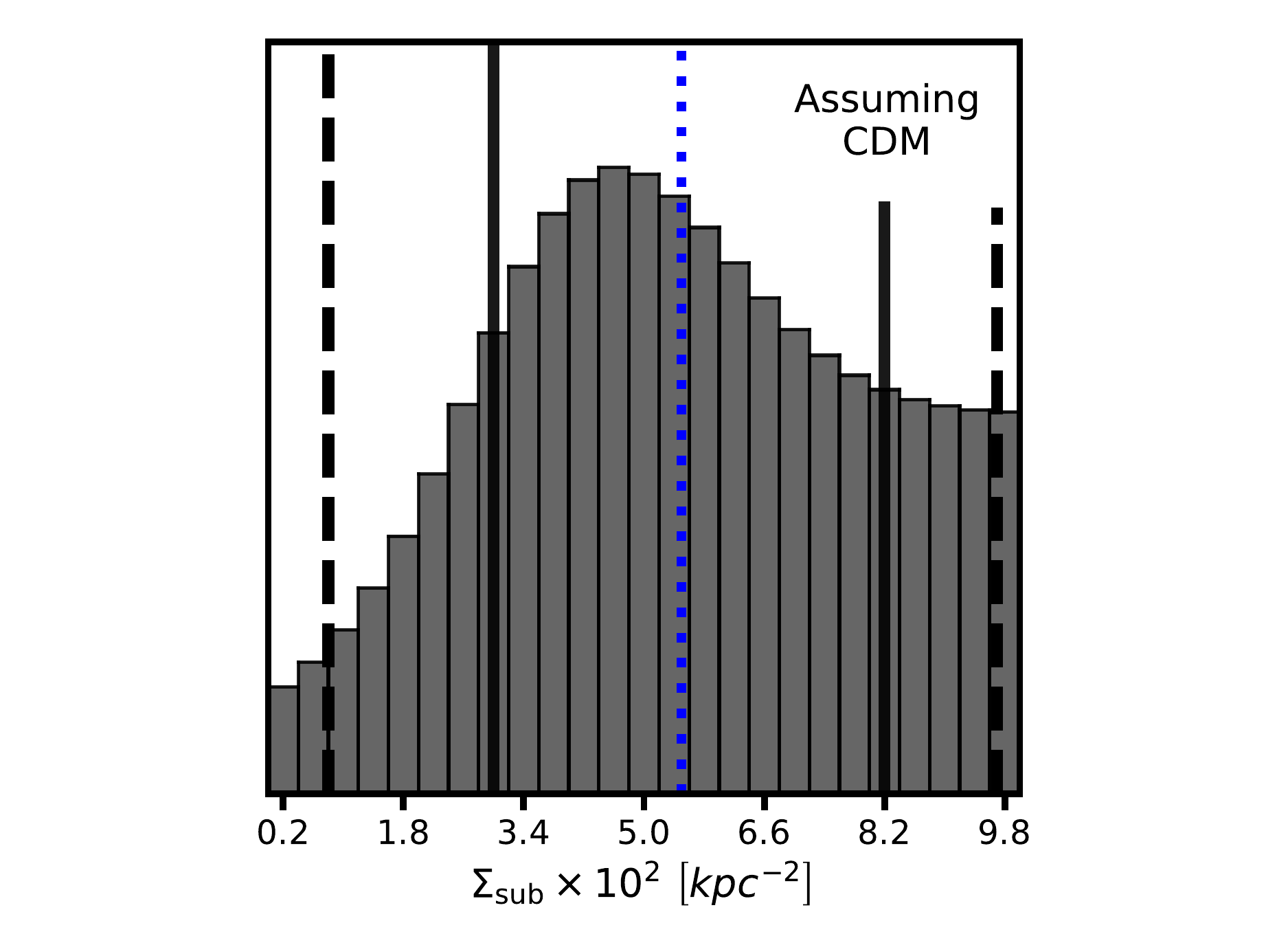}
	\caption{\label{fig:cdmsigmasub} Inference on the global normalization of the subhalo mass function $\Sigma_{\rm{sub}}$ assuming CDM, marginalized over the logarithmic slope $\alpha$ and uncertainty in the overall amplitude of the line of sight halo mass function $\delta_{\rm{los}}$. The blue dashed lines shows the mean of the marginal distribution, while black solid (dashed) lines represent $68\%$ and $95 \%$ confidence intervals. The contours in the joint distribution also represent $68\%$ and $95 \%$ confidence intervals.}
	\end{figure}	
	
	\subsection{Constraints on the free-streaming length of dark matter}
	\label{ssec:mainresults}
	For each quad, we obtain a joint likelihood between the macromodel parameters $\qmac$ and the dark matter-hyper parameters $\qsub$. We marginalize over the parameters in this 20+ dimensional space to obtain the four-dimensional space of $\qsub$ parameters that includes logarithmic slope of the subhalo mass function $\alpha$, the scaling of the line of sight halo mass function $\delta_{\rm{los}}$, the overall scaling of the subhalo mass function $\Sigma_{\rm{sub}}$, and the half-mode mass $\mhm$. We reiterate that these four parameters describe universal properties of dark matter and should therefore be common to all the lenses, while the parameters $\qmac$ and the halo mass $M_{\rm{halo}}$ are lens-specific. After marginalizing, we compute the product of the resulting likelihoods and obtain the desired posterior distribution in Equation \ref{eqn:posterior}, which we display in Figure \ref{fig:mainresult}.
	
	The marginalized constraints on $\mhm$ rule out $\mhm > 10^{7.8} \msun$ at $2 \sigma$, corresponding to thermal relic particle mass of $<5.2 \ \rm{keV}$. It is apparent from Figure \ref{fig:mainresult} that $\mhm$ and $\Sigma_{\rm{sub}}$ are correlated, since halos added by increasing the normalization can be subsequently removed by increasing $\mhm$ such that the total amount of lensing substructure remains relatively constant. As a result, the marginalized distribution for the normalization $\Sigma_{\rm{sub}}$ appears unconstrained from above, as the normalization can be significantly higher in WDM scenarios. With only eight quads we cannot simultaneously measure $\mhm$ and $\Sigma_{\rm{sub}}$, although our previous forecasts indicate this is possible with more lenses \citep{Gilman++18}. 
	
	The constraints on dark matter warmth in terms of confidence intervals depend on the range of allowed values specified by the prior on $\Sigma_{\rm{sub}}$. Similarly, the confidence interval on $\mhm$ depends on the lower bound of this parameter that is set by the prior on $\mhm$. As discussed in Section \ref{ssec:wdmassumptions}, we have chosen the prior on $\mhm$ to encompass the region of parameter space where the data can constrain $\mhm$, keeping in mind that the WDM mass concentration relation affects the central densities of subhalos 60 times above $\mhm$ (Equation \ref{eqn:cmrelation}), and the upper bound of $\Sigma_{\rm{sub}} = 0.1 \rm{kpc^{-2}}$ is a conservative choice as most N-body simulations and the {\tt{galacticus}} runs predict values below $0.05 \rm{kpc^{-2}}$. In light of these complications, we also quote likelihood ratios which do not depend on the choice of prior. Relative to the peak of the $\mhm$ posterior, we obtain likelihood ratios for WDM with $\mhm = 10^{8.2}\msun$ ($\mhm = 10^{8.6}\msun$) of 7:1 (30:1)\footnote{We remind the reader that the relative heights of the peaks in the posterior somewhat depend on the binning method, or in this case the bandwidth estimator of the KDE. In this work, we have applied a KDE with a first order boundary correction and a bandwidth selected according to Scott's factor \citep{Scott92}.}. 
	
	The posterior for $\delta_{\rm{los}}$ indicates the data favors more line of sight structure, but the preference is not statistically significant. The parameters $\delta_{\rm{los}}$ and $\Sigma_{\rm{sub}}$ are anti-correlated, as one would expect as one can, to a certain degree, remove lens plane subhalos and replace them with line of sight halos while keeping the total amount of flux perturbation constant. This is not a perfect degeneracy, however, since lensing efficiency and the relative number of subhalos and line of sight halos changes with redshift. Thus, a larger sample of quads at different redshifts could break the covariance between $\Sigma_{\rm{sub}}$ and $\delta_{\rm{los}}$. 
	
	\subsection{Constraints on the subhalo mass function assuming CDM}
	
	We perform a suite of CDM simulations using the same priors listed in Table \ref{tab:lenspriors}, minus the WDM parameter $m_{\rm{hm}}$, with the aim of inferring $\Sigma_{\rm{sub}}$. We marginalize over $\delta_{\rm{los}}$, and over a theoretical-motivated prior on $\alpha$ (between -1.95 and -1.85) based on predictions from N-body simulations \citep{Springel++08,Fiacconi++16}.
	
	The inference on $\Sigma_{\rm{sub}}$ is shown in Figure \ref{fig:cdmsigmasub}. We infer $\Sigma_{\rm{sub}} = 0.055 \rm{kpc^{-2}}$, with a $1 \sigma$ confidence interval $0.029 < \Sigma_{\rm{sub}} < 0.083 \ \rm{kpc^{-2}}$. At the $2 \sigma$ level we obtain $\Sigma_{\rm{sub}} > 0.008 \rm{kpc^{-2}}$. We do not quote an upper $2\sigma$ bound on $\Sigma_{\rm{sub}}$ as it is prior dominated. To put these numbers in physical units, the mean value of $\Sigma_{\rm{sub}}$ corresponds to a mean projected mass in substructure for the lenses in our sample between $10^6 - 10^{9} \msun$ of $4.0 \times 10^{7} \msun \rm{kpc^{-2}}$, and the $1 \sigma$ confidence interval corresponds to $2.0 - 6.1 \times 10^{7} \msun \rm{kpc^{-2}}$. At $2 \sigma$, the projected mass constraint is $\Sigma_{\rm{sub}} > 0.6 \times 10^{7} \msun \rm{kpc^{-2}}$. To convert into the average projected mass, we have computed the average of the projected masses for each of the eight lenses in our sample, using the scaling of the halo mass function with redshift in Equation \ref{eqn:scaling} while assuming a halo mass of $10^{13} \msun$. 
	
	\section{Discussion and conclusions}
	\label{sec:discussion}
	In this section we review the main results of this work and discuss the implications for cold and warm dark matter. In Section \ref{ssec:mainsummary} we summarize our main results, and in Section \ref{ssec:comparison} we compare our results with those obtained in previous works. In Section \ref{ssec:systematics} we discuss the sources of systematic uncertainty in our analysis, and we conclude in Section \ref{ssec:implications} by discussing the implications of our result for cold and warm dark matter. 
	
	\subsection{Summary of the analysis and main results}
	\label{ssec:mainsummary}
	We have carried out a measurement of the free-streaming length of dark matter and the subhalo mass function using a sample of eight quadruply-imaged quasars. The methodology we use to constrain the dark matter parameters of interest has been tested and verified with simulated data \citep{Gilman++19}. Lenses that show evidence for morphological complexity in the form of stellar disks are excluded from our analysis. We model halos both in the main deflector and along the line of sight, including correlated structure around the main deflector through the two-halo term, and account for evolution of the projected subhalo mass function with redshift and halo mass using a suite of simulations using the semi-analytic modeling code {\tt{galacticus}}. We compute image flux ratios by ray-tracing to finite-size background sources, which correctly accounts for the sensitivity of image flux ratios to perturbing halos. We also marginalize over the macromodel parameters for each system, including the power-law slope of the main deflector, and simultaneously constrain the lens macromodel and dark matter hyper-parameters to account for covariance between these quantities. In addition to the turnover in the halo mass function, we model WDM free-streaming effects on the mass-concentration relation, accounting for the effect of reduced central densities of WDM halos on lensing observables. 
	
	The main results of this analysis are summarized as follows:
	
	\begin{itemize}
		\item We constrain the half-mode mass $\mhm$ (thermal relic dark matter particle mass) to $\mhm < 10^{7.8} \msun$ ($m_{\rm{DM}} > 5.2 \rm{keV}$) at 2$\sigma$. Since the confidence intervals depend on the prior used for both $\mhm$ and $\Sigma_{\rm{sub}}$, we also quote likelihood ratios relative to the peak of the posterior distribution for $\mhm$: we disfavor $\mhm = 10^{8.2} \msun$ ($m_{\rm{DM}} = 4 \rm{keV}$) with a likelihood ratio of 7:1, and with $\mhm = 10^{8.6} \msun$ ($m_{\rm{DM}} = 3.0 \rm{keV}$) the relative likelihood is 30:1. These bounds are marginalized over the amplitude of the subhalo mass function, the amplitude of the line of sight halo mass function, the power-law slope of the subhalo mass function, the parent halo mass, the background source size, and the parameters describing the main deflector mass profile.
		\item Assuming cold dark matter, we infer a value of the global amplitude of the subhalo mass function $\Sigma_{\rm{sub}} = 0.055_{-0.027}^{0.032} \rm{kpc^{-2}}$ at $1 \sigma$, and $\Sigma_{\rm{sub}} > 0.008 \rm{kpc^{-2}}$ at $2 \sigma$. In our lens sample, these values correspond to an average projected mass density in substructure between $10^6 - 10^{9} \msun$ of $4.0_{-2.0}^{+2.1} \times 10^{7} \msun \rm{kpc^{-2}}$ and a lower bound of $0.6  \times 10^{7} \msun \rm{kpc^{-2}}$, respectively. At fixed redshifts, for a $10^{13} \msun$ halo at $z=0.2$ ($z=0.6$) the $1 \sigma$ constraint corresponds to a projected mass in substructure of $1.9_{-0.9}^{+0.9} \times 10^7 \msun \rm{kpc^{-2}}$ ($4.1_{-2.0}^{+2.0} \times 10^7 \msun \rm{kpc^{-2}}$) in the subhalo mass range $10^6-10^{9} \msun$. The $2 \sigma$ constraint corresponds to a projected mass in substructure of greater than $0.3 \times 10^7 \msun \rm{kpc^{-2}}$ ($0.6 \times 10^7 \msun \rm{kpc^{-2}}$) in the same mass range.
	
	\end{itemize}
	
	\subsection{Discussion and comparison with previous work}
	\label{ssec:comparison}
	
	\subsubsection{Constraints on dark matter warmth and the amplitude of the CDM subhalo mass function}
	The first comprehensive analysis of multiply-imaged quasars was carried out by \citet{D+K02} (hereafter DK2), who inferred a projected mass fraction in substructure $\fsubmean$ \footnote{Throughout this section, we will use $\fsubmean$ to refer to the average mass fraction in substructure inferred from a sample of multiple lenses in halos of different masses at different redshifts, and $\fsub$ to refer to the mass fraction in substructure implied by a certain $\Sigma_{\rm{sub}}$ value at a specific redshift and halo mass.} between $0.006 < \fsubmean < 0.07$ at $2 \sigma$ modeling only lens-plane substructure, and assuming CDM. Recently, \citet{Hsueh++19} (hereafter H19) improved on the analysis of DK2 by including the effects of line of sight halos, measuring $0.006 < \fsubmean < 0.018$ at $1 \sigma$ with a mean of $0.011$ assuming CDM, and also constrained the free-streaming length of dark matter to $\mhm < 10^{8.4}$ ($m_{\rm{DM}} > 3.8 \rm{keV}$). 
	
	The $2 \sigma$ bound from H19 of $\mhm < 10^{8.4} \msun$ is weaker than the constraint from this work $\mhm < 10^{7.8}\msun$. One possible reason for this difference is that unlike previous work \citep{Birrer++17a,Gilman++18,Gilman++19} H19 did not model the suppression of the mass-concentration relation in warm dark matter scenarios, which suppresses the lensing signal more than one order of magnitude above the position of the turnover in the mass function. This is of particular relevance for flux ratio studies because the effect of a perturbing dark matter halo depends on its central density profile. Free-streaming effects on the mass-concentration relation therefore increase the relative differences between CDM and WDM on the scales relevant for substructure lensing, which leads to greater constraining power over WDM models. Finally, we note that in a future analysis modeling the tidal evolution of substructures from the time of infall to the time of lensing may introduce additional constraining power over WDM models by coupling the structural parameters of subhalos at the time of lensing to their structural properties, such as concentration, at the time of infall.
	
	To facilitate direct comparison between this analysis and that of DK2 and H19 regarding the constraints on the subhalo mass function assuming CDM, we convert our $\Sigma_{\rm{sub}}$ values into estimates of $\fsubmean$ by computing the projected mass density $\Sigma$, and then using the fact that $\frac{\Sigma}{\Sigma_{\rm{crit}}} = 0.5$ near the Einstein radius, where $\Sigma_{\rm{crit}}$ is the critical surface mass density for lensing. In these conversions, we also assume a halo mass of $10^{13} \msun$, and take care to compute $\fsubmean$ using the same mass range $10^6 -10^9 \msun$ used by H19. Our $2 \sigma$ bounds on $\Sigma_{\rm{sub}}$ correspond to an average mass fraction in substructure $\fsubmean > 0.005$ with a mean of $\fsubmean = 0.035$. At $1 \sigma$ $0.018 < \fsubmean < 0.056$. This result is statistically consistent with the constraints from H19, and also with those of DK2. 
	
	There are several key differences between our analysis and those of H19 and DK2 that pull in opposite directions in terms of constraining power over dark matter models. As mentioned previously we model free-streaming effects on the mass-concentration relation, and include the contribution from the two-halo term to account for correlated structure near the main deflector. These pieces of additional physics add information and increase our constraining power over WDM models. On the other hand, accounting for finite-size background sources decreases the expected magnification signal caused by dark matter halos and subhalos, and we expect to infer a higher normalization of the subhalo mass function in our analysis as more substructure is needed to produce the same degree of flux perturbation. Explicitly, by ray-tracing to finite-size background sources we find that the peak of the magnification cross section for a $5\times 10^7 \msun$ halo is reduced by a factor of two for a $15 \rm{pc}$ background source relative to a $5 \rm{pc}$ background source, and by a factor of three for a $40 \rm{pc}$ source. The simplifying assumption of point-sources for the background quasar invoked by H19 and DK2 introduces signal from low-mass halos whose effects would otherwise be washed out by an extended source. 
	
	The tidal truncation of lens plane subhalos that we model may also reduce the overall impact of subhalos on lensing observables. We also marginalize over the power-law slope of the main deflector and simultaneously sample the macromodel parameters and the dark matter hyper-parameters. These processes introduce additional covariances in the posterior distributions, and should lead to weaker constraints on $\Sigma_{\rm{sub}}$ and $\mhm$.  
	
	Other lensing studies, primarily those using the technique of gravitational imaging, have also sought to measure the subhalo mass function. \citet{Veg++14} inferred $\fsubmean= 0.0064_{-0.0042}^{0.0080}$ at $1 \sigma$ in the mass range $4 \times 10^6 - 4 \times 10^9 \msun$ assuming a prior on the slope of the subhalo mass function centered on $\alpha = -1.9$, while \citet{Hezaveh++16} constrained the normalization of subhalo mass function assuming $\alpha = -1.9$, inferring $\bar{f}_{\rm{fsub}}$ values comparable to the median $\bar{f}_{\rm{sub}} = 0.02$ result from DK2 (and our constraint), but with larger uncertainties. 
	
	To compare with the analysis of \citet{Veg++14}, we assume a halo mass of $10^{13}\msun$ at a lens redshift $z_d = 0.25$ and a source at $z_{\rm{src}} = 0.7$, characteristic values for the lens sample analyzed by \citet{Veg++14}. Using these values with our expression for the subhalo mass function in Equation \ref{eqn:subhalomfunc}, we obtain $\fsub= 0.014_{-0.007}^{+0.008}$ between $4 \times 10^6$ and $6 \times 10^9 \msun$ at $1 \sigma$, in the same mass range used by \citet{Veg++14}. This result is consistent with that of \citet{Veg++14}\footnote{Although \citet{Veg++14} did not model line of sight halos, the low lens/source redshifts their sample lessen the impact of line of sight halos on the inferred subhalo mass fraction such that we may compare our results, which include line of sight halos, with theirs.}. We quote constraints on $f_{\rm{sub}}$ to make comparisons with previous work, but we caution that the conclusions derived from inferences of $f_{\rm{sub}}$ should be interpreted with care. The physical meaning of this parameter depends on specific assumptions regarding the subhalo mass range and the contribution from dark substructure to the convergence near the Einstein radius, which may change with halo mass and redshift. 
	
	Comparing our results with semi-analytic simulations of massive $10^{13} \msun$ hosts, our results in terms of the projected mass in substructure is consistent with the {\tt{galacticus}} simulations used to calibrate the evolution of the subhalo mass function with halo mass and redshift. We stress that our model was not tuned to match the normalization predicted by {\tt{galacticus}}, it only made use of the trends of projected substructure mass density with host halo mass and redshift. 
	
	Our results are also consistent with N-body simulations of $10^{13} \msun$ halos by \citet{Fiacconi++16}, who predict projected substructure mass densities of $2.0-2.8 \times 10^{7} \msun \rm{kpc^{-2}}$ after accounting for baryonic contraction of the halo. We infer roughly triple the predicted mass in substructure than the amount predicted by \citet{Xu++15}, who simulated $10^{13}\msun$ halos by rescaling Milky Way size and cluster size hosts to halo masses of $\sim 10^{13} \msun$. Finally, we note that our results arrive on the heels of several works that examine numerical features of N-body simulations that may result in the artificial fragmentation of subhalos \citep{vandenBosch++18,ErraniPenarrubia19}. Taken at face value, these results suggest that N-body simulations may underpredict substructure abundance in dark matter halos. 
	
	We may also compare our constraints with the projections from  \citet{Gilman++19}. With a sample of ten quads, they projected a $2 \sigma$ bound on $\mhm$ with $\Sigma_{\rm{sub}} = 0.022 \rm{kpc^{-2}}$ of $10^{7.7} \msun$ with $2\%$ uncertainties in image fluxes, and $10^{8.6} \msun$ with $6 \%$ uncertainties. Our constraint of $\mhm < 10^{7.8} \msun$ is broadly consistent with these predictions\footnote{The conversion between the half-mode mass and the mass of the corresponding thermal relic dark matter particle used by \citet{Gilman++19} is off by a factor of h=0.7, but the comparison between the half-mode masses is robust.}, given the higher mean $\Sigma_{\rm{sub}}$ value of $0.055 \rm{kpc^{-2}}$ we infer in this analysis, and the flux uncertainties in the lens sample which are $\sim 6 \%$ on average. 
	
	The overall scaling of the line of sight halo mass function $\delta_{\rm{los}}$ is unconstrained with our sample size and choice of prior. This is likely because the prior on $\delta_{\rm{los}}$ spans a relatively limited range of $\pm 20\%$ around the Sheth-Tormen mass function prediction, and with the current sample size of only eight quads we cannot constrain departures from the Sheth-Tormen prediction at the level of $10-20\%$. 
	
	\subsection{Sources of systematic uncertainties}
	\label{ssec:systematics}
	\subsubsection{The lens macromodels}
	Several works \citep{Gilman++17,Hsueh++18} have investigated the ability of smooth isothermal mass models plus external shear to fit the smooth mass component of galaxy scale strong lenses. These works reach similar conclusions, determining that isothermal models predict image flux ratios to better than $10 \%$ unless a stellar disk is present, in which case explicit modeling of the disk is required \citep[e.g.][]{Hsueh++17,Hsueh++18}. Each of these analysis restricted the smooth lens models to exactly isothermal mass density profiles. 
	
	The deflectors in our sample show no evidence for morphological complexity that would require explicit modeling beyond a power-law ellipsoid model. Specifically, we exclude all lens systems with known stellar disks to avoid any bias they may introduce. To account for remaining uncertainties associated with the lens macromodel, we highlight two features of our lens modeling implemented in an effort to mitigate this source of systematic uncertainty. First, we note that flux ratios are highly localized probes of the surface mass density in the immediate vicinity of the lensed images, and therefore the main requirement for this work is to accurately predict the mass profile in these four small isolated regions. By relaxing the strictly isothermal mass profile assumption and marginalizing over the logarithmic slope of the main deflector mass profile, we allow for the local mass profile in the vicinity of the lensed images to vary. The additional degree of freedom added in the lens macromodel increases our uncertainties, but accounts for deviations from power-law ellipsoids limited to exact $\rho \left(r\right) \propto r^{-2}$ mass profiles. 
	
	Second, we note that smooth power-law models predict a distribution of flux ratios, rather than single values (for example, see Figures A1-A8 in \citet{Nierenberg++19}). Following common practice, \citet{Gilman++17} and \citet{Hsueh++18} identified flux ratio `anomalies' with respect to a single smooth model fit to lensed images, a procedure that does not account for the distribution of flux ratios predicted by smooth lens models that is marginalized over in the full forward modeling analysis we perform. In this work, we also take care to explore the macromodel parameter space and the dark matter hyper-parameter space simultaneously, which accounts for additional covariances that contribute to the model-predicted flux uncertainties. 
	
	\subsubsection{Modeling of the dark matter content}
	We assume specific functional forms for the halo and subhalo mass functions (Equations \ref{eqn:subhalomfunc} and \ref{eqn:losmfunc}), and the mass-concentration-redshift relation (Equation \ref{eqn:cmrelation}). We acknowledge that there are other parameterizations in the literature for both of these quantities \citep[e.g.][]{Schneider++12,Benson++13}, but in this work we implement only one parameterization of WDM effects on the mass function (Equation \ref{eqn:mfuncwdm}) and halo concentrations (Equation  \ref{eqn:cmrelation}), which corresponds to one specific WDM model. We note that additional physics, such as the velocity dispersion of dark matter particles in the early universe, can alter the shape of the mass function, but with the current sample size of lenses it is unlikely we have enough information to constrain these additional features if they were included in the model. 
	
	It is possible that free-streaming effects on the halo mass function near the half-mode mass scale may become more pronounced at high redshifts. This could affect both the location and shape of the turnover in the mass function. However, in the absence of a specific prediction for the evolution of the turnover with redshift, we apply the parameterization in Equation \ref{eqn:mfuncwdm} through the relevant redshift range $z = 0 - 3.5$. We note that since the lensing efficiency of halos decreases approaching source redshift, systematic errors from possible redshift evolution of the WDM turnover will be correspondingly down-weighted. We note that the mass-concentration-redshift relation for WDM calibrated by \citet{Bose++16} that we implement does evolve with redshift, as does the CDM mass-concentration relation from \citep{DiemerJoyce18}. 
	
	\subsection{Implications for WDM models}
	\label{ssec:implications}
	
	Galaxy-galaxy strong lensing provides a useful compliment to the strongest existing probe of the free-streaming length of dark matter from the Lyman-$\alpha$ forest \citep{Viel13,Irsic++17}. Our $2 \sigma$ bound on the thermal relic mass of $m_{\rm{DM}} > 5.2 \rm{keV}$ surpasses than the 3.3 keV constraint from \citet{Viel13} and matches the $5.3 \rm{keV}$ constraint from \citet{Irsic++17}, who invoked additional assumptions regarding the relevant thermodynamics. The key point of this comparison, however, is not so much which method achieves the most precision, but the fact that both methods provide stringent limits and that they are completely independent of each other in observational data and astrophysical assumptions. Independently and in combination, the results from lensing and the Lyman-$\alpha$ forest support the following statement: the halo mass function extends down in a scale-free manner to mass scales of $\sim 10^{8} \msun$, where halos are mostly, if not completely, dark. There appears to be little room left for a viable warm dark matter solution to the small-scale issues of cold dark matter. 
	
	\section*{Acknowledgments}
	We thank the anonymous referee for a careful reading of the paper and thoughtful comments. We are also grateful to Alex Kusenko and Annika Peter for useful discussions throughout the course of this project, and thank David Gilman for useful comments on an early draft version of this work.    
	
	DG, TT, and SB acknowledge support by the US National Science Foundation through grant AST-1714953. DG, TT, SB and AN acknowledge support from HST-GO-15177. AJB and XD acknowledge support from NASA ATP grant 17-ATP17-0120. Support for Program number GO-15177 was provided by NASA through a grant from the Space Telescope Science Institute, which is operated by the Association of Universities for Research in Astronomy, Incorporated, under NASA contract NAS5-26555. TT and AN acknowledge support from HST-GO-13732. AN acknowledges support from the NASA Postdoctoral Program Fellowship, the UC Irvine Chancellor's Fellowship, and the Center for Cosmology and Astroparticle Physics Fellowship. 
	
	This work used computational and storage services associated with the Hoffman2 Shared Cluster provided by the UCLA Institute for Digital Research and Education's Research Technology Group. This work also used computational and storage services associated with the Aurora and Halo super computers. These resources were provided by funding from the JPL Office of the Chief Information Officer. In addition, calculations were performed on the {\tt memex} compute cluster, a resource provided by the Carnegie Institution for Science.
	
	Part of this work is based on observations made with the NASA/ESA Hubble Space Telescope, obtained at the Space Telescope Science Institute, which is operated by the Association of Universities for Research in Astronomy, Inc., under NASA contract NAS 5-26555. These observations are associated with programs $\#13732$ and $\#15177$. Support for programs $\#13732$ and $\#15177$ was provided by NASA through a grant from the Space Telescope Science Institute, which is operated by the Association of Universities for Research in Astronomy, Inc., under NASA contract NAS5-26555.
	
	\bibliographystyle{mnras}
	\bibliography{bibliography}
	
	\appendix
	
	\section{\bf Convergence of the posterior distributions}
	\label{app:A}
	The approximation of the true posterior obtained in Approximate Bayesian Computing (ABC) algorithms converges to the true posterior distribution as the acceptance criterion becomes increasingly more stringent. In our framework, changing the acceptance criterion is equivalent to reducing the number of forward model samples while keeping the number of total accepted realizations fixed. We exploit this property to test for convergence of the posteriors. 
	
\begin{figure*}
		\includegraphics[clip,trim=0cm 0cm 0cm
		0cm,width=.95\textwidth,keepaspectratio]{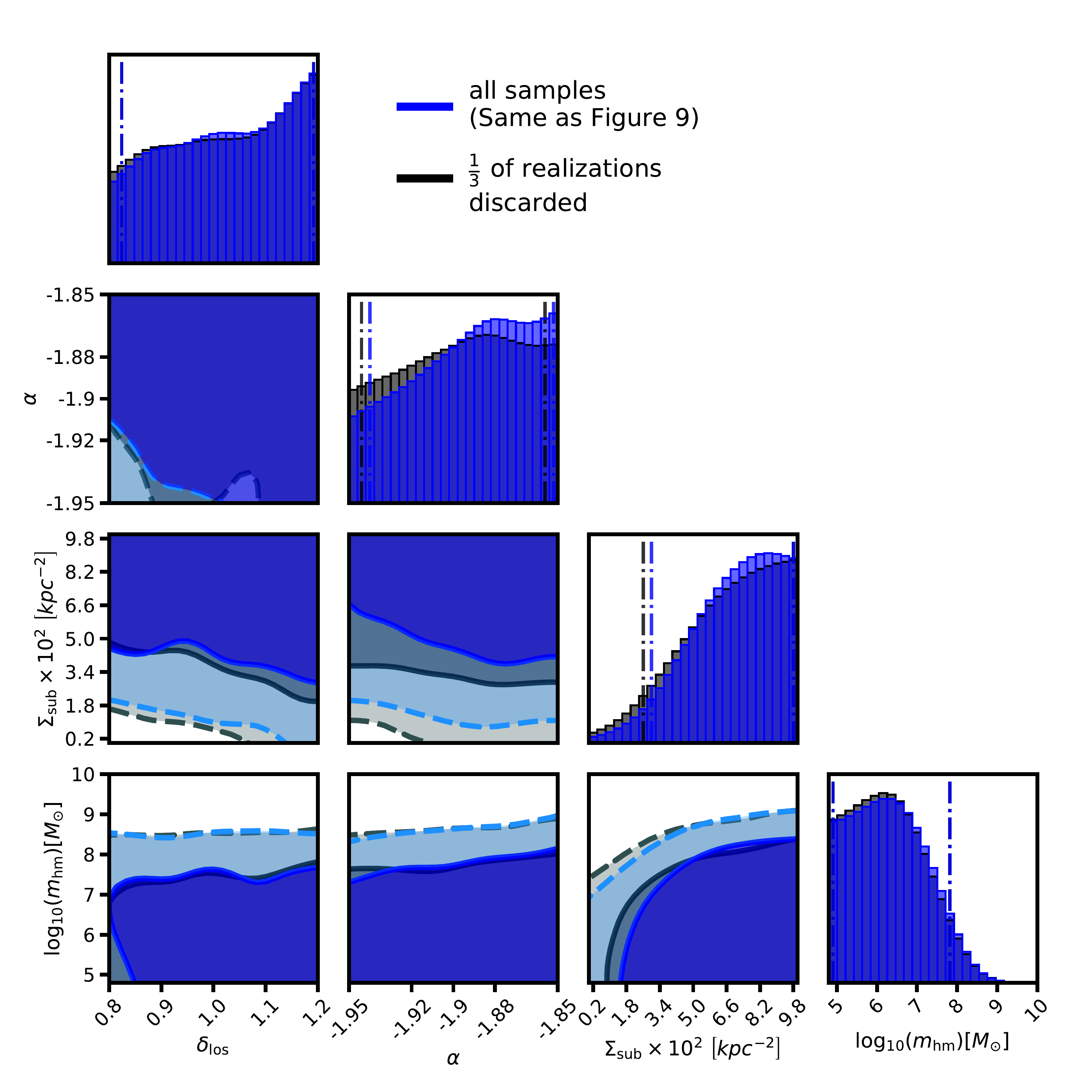}
		\caption{\label{fig:convergence} A convergence test of the posterior distributions. By discarding one-third of the forward model samples and applying the same rejection criterion to those that remain, we verify the inference obtained through the ABC rejection algorithm is robust. }
	\end{figure*}

	In Figure \ref{fig:convergence}, we compare the posterior constructed from the full set of forward model samples (the same as Figure \ref{fig:mainresult}) with a second posterior derived from a depleted set of forward model samples, where we have discarded one-third of the realizations and accepted the same rejection criterion (accept the realizations corresponding to the 800 lowest values of $S_{\rm{lens}}$) to those that remain. The mass of the posterior distributions remains relatively unchanged, and the $1 \sigma$ and $2 \sigma$ contours are nearly identical. We conclude we have generated enough realizations of dark matter structure to reliably construct posterior distributions using the ABC rejection algorithm described in Section \ref{sec:inference}. 
	
	\begin{figure*}
		\includegraphics[clip,trim=0cm 0cm 0cm
		0cm,width=.95\textwidth,keepaspectratio]{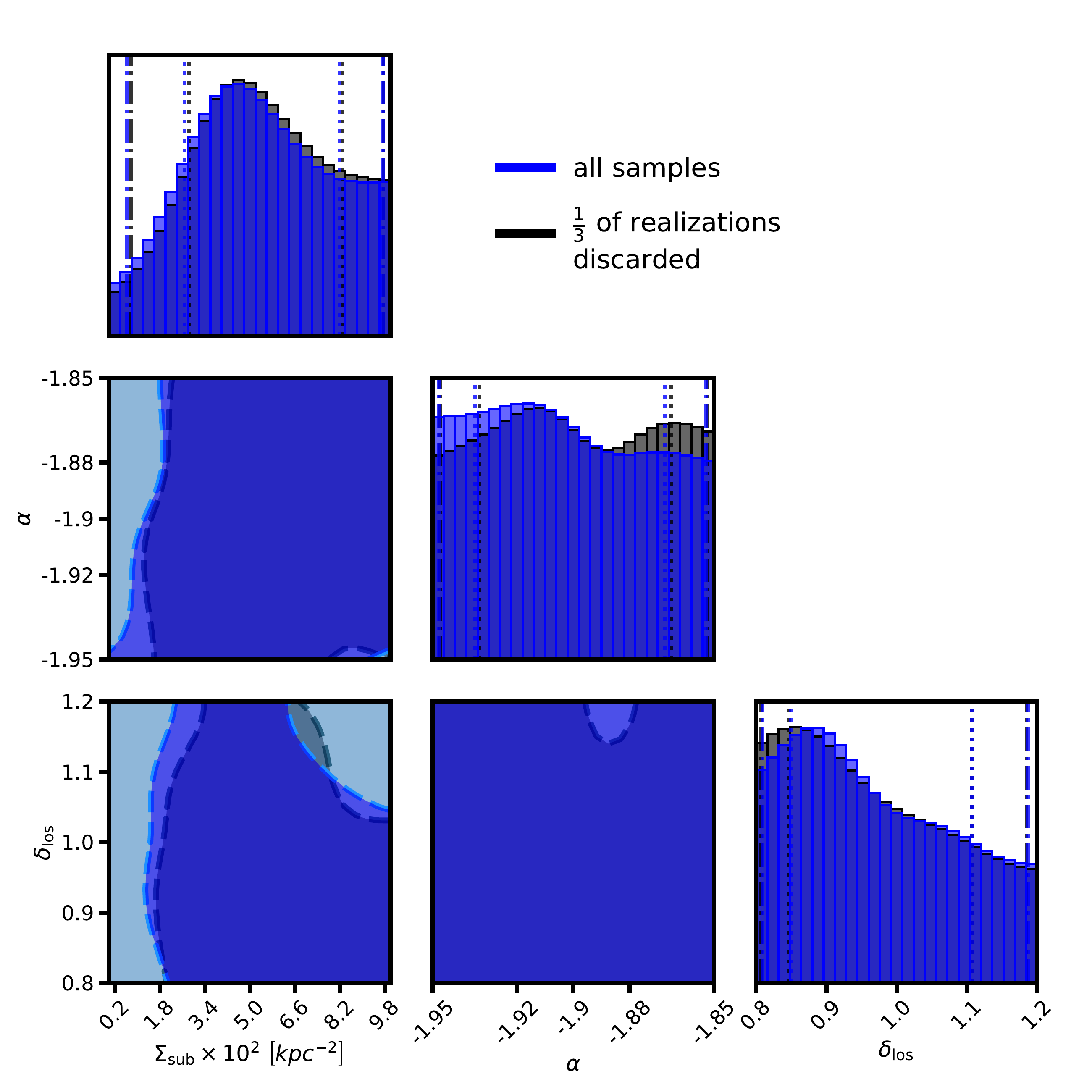}
		\caption{\label{fig:convergence_CDM} A convergence test of the posterior distributions assuming CDM. Like Figure \ref{fig:convergence}, one-third of the samples are discarded and the same number of realizations are accepted into the posterior.}
	\end{figure*}
	
	\section{\bf Obtaining deflector redshifts}
	\label{app:B}
	\begin{figure}
		\includegraphics[clip,trim=0cm 0cm 0cm
		0cm,width=.2\textwidth,keepaspectratio]{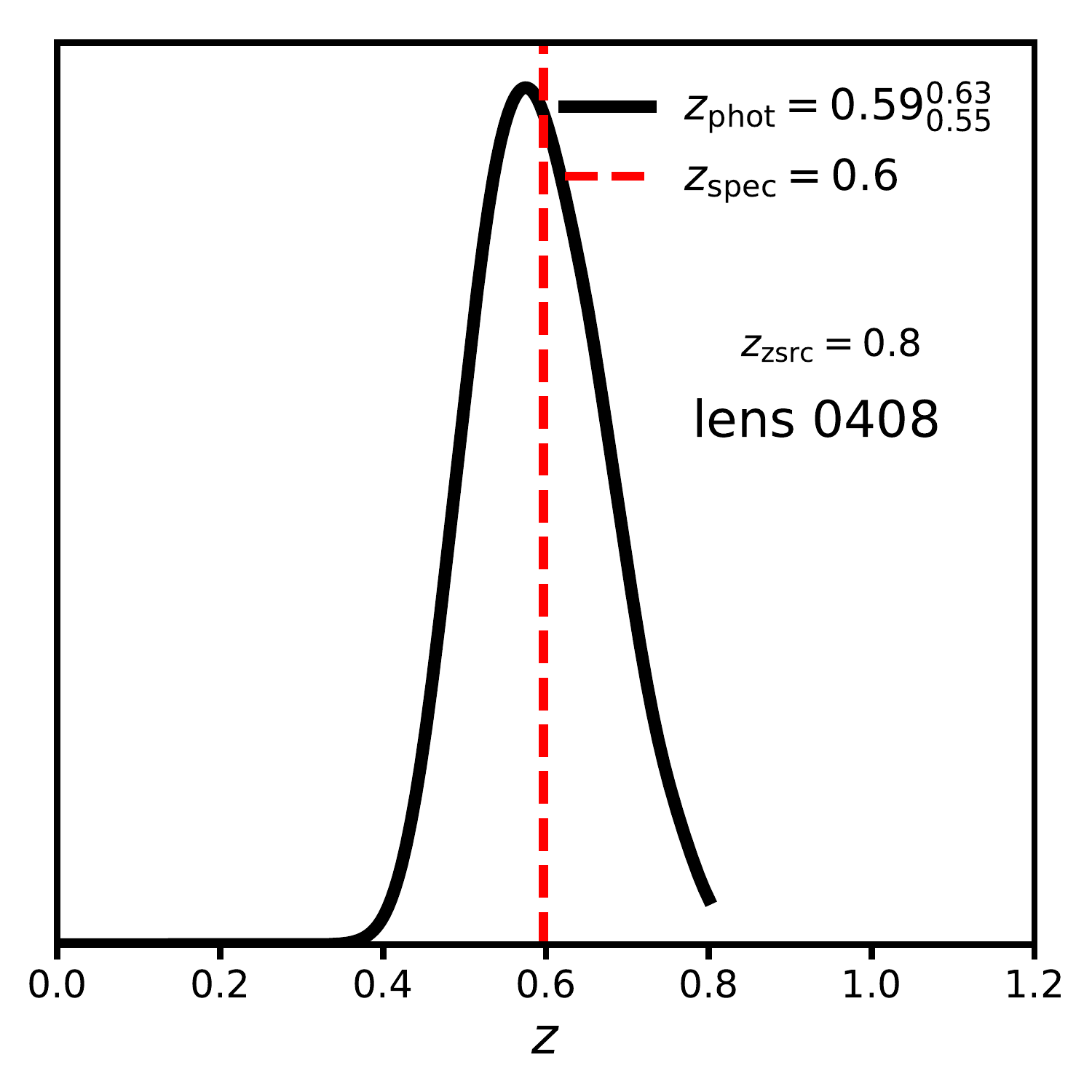}
		\includegraphics[clip,trim=0cm 0cm 0cm
		0cm,width=.2\textwidth,keepaspectratio]{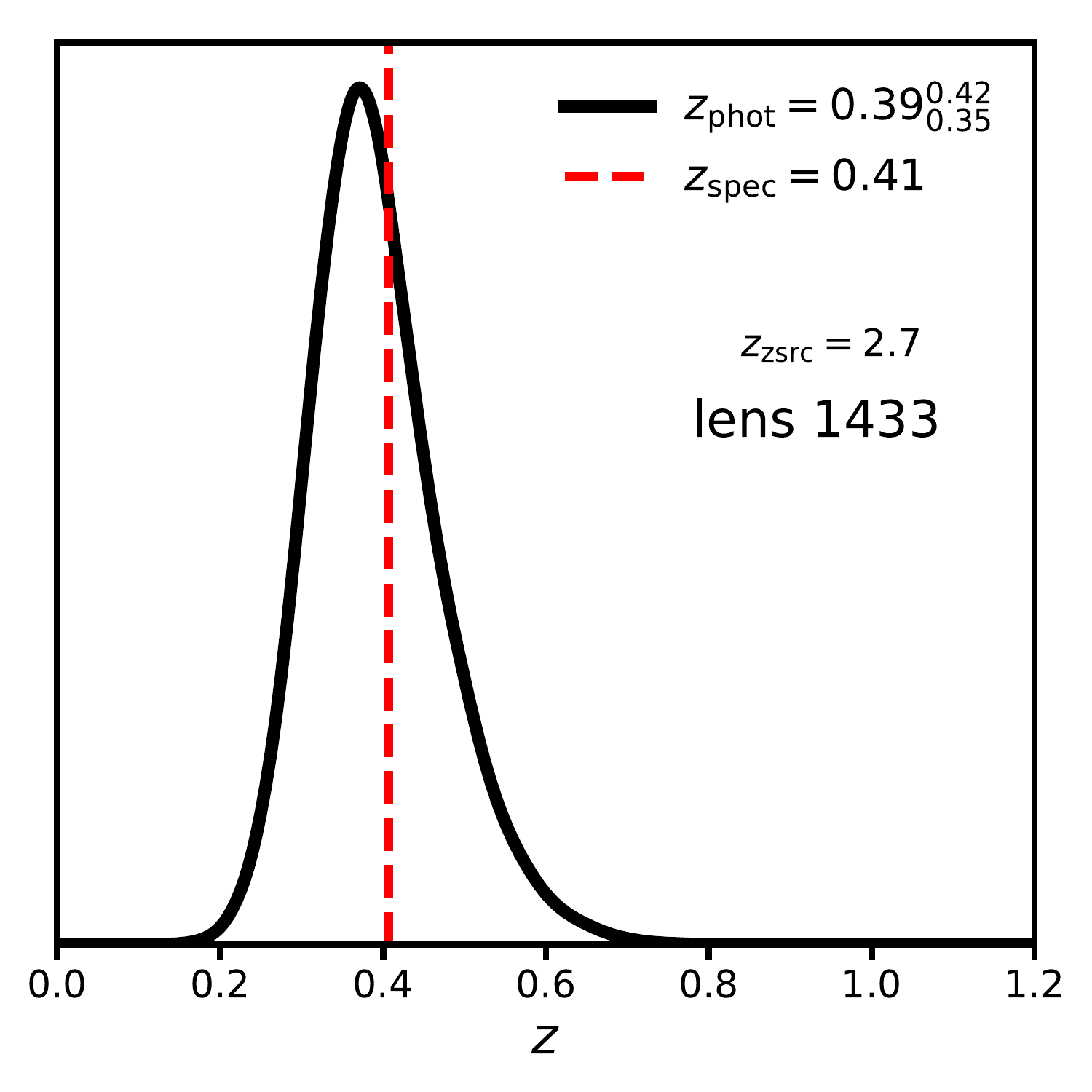}
		\includegraphics[clip,trim=0cm 0cm 0cm
		0cm,width=.2\textwidth,keepaspectratio]{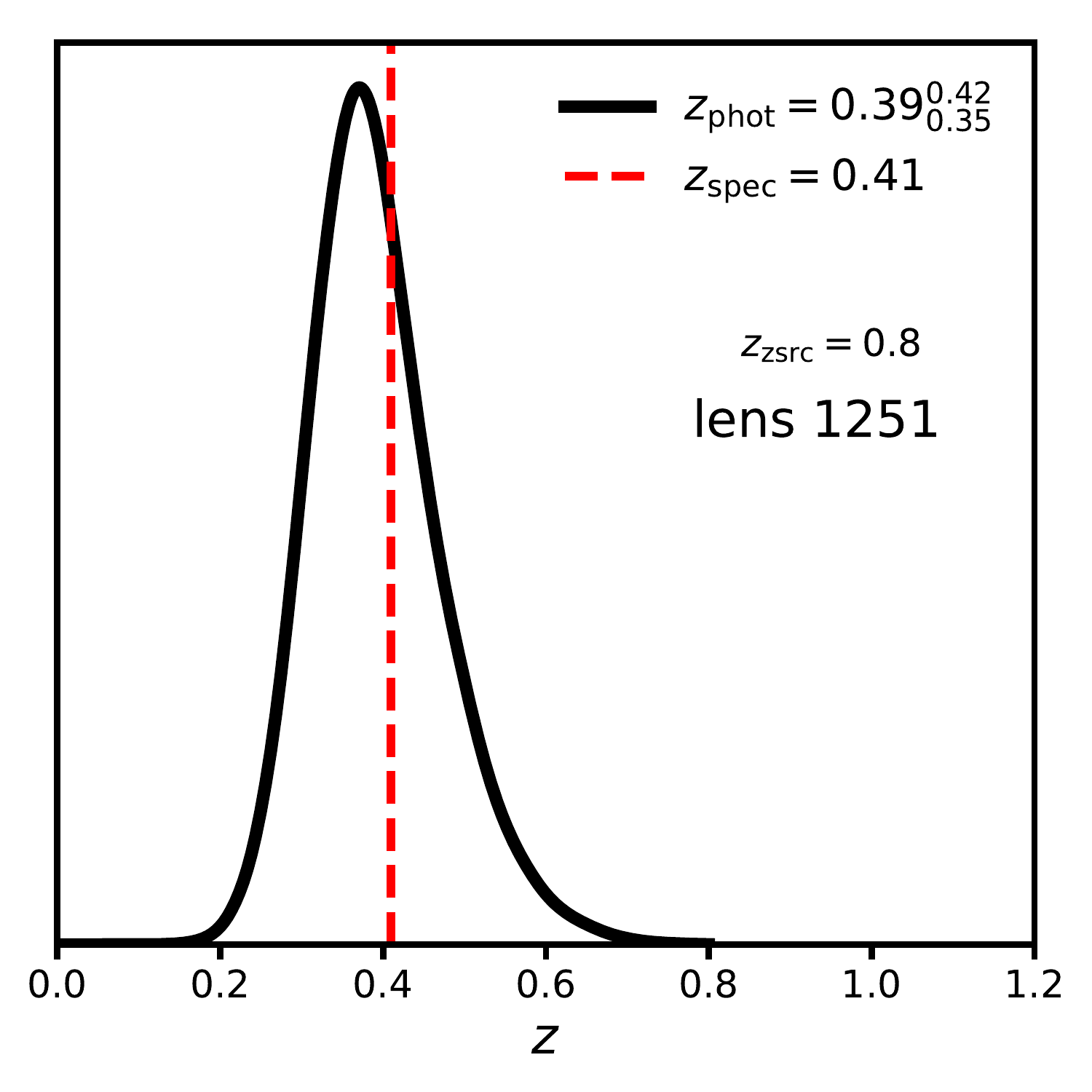}
		\includegraphics[clip,trim=0cm 0cm 0cm
		0cm,width=.2\textwidth,keepaspectratio]{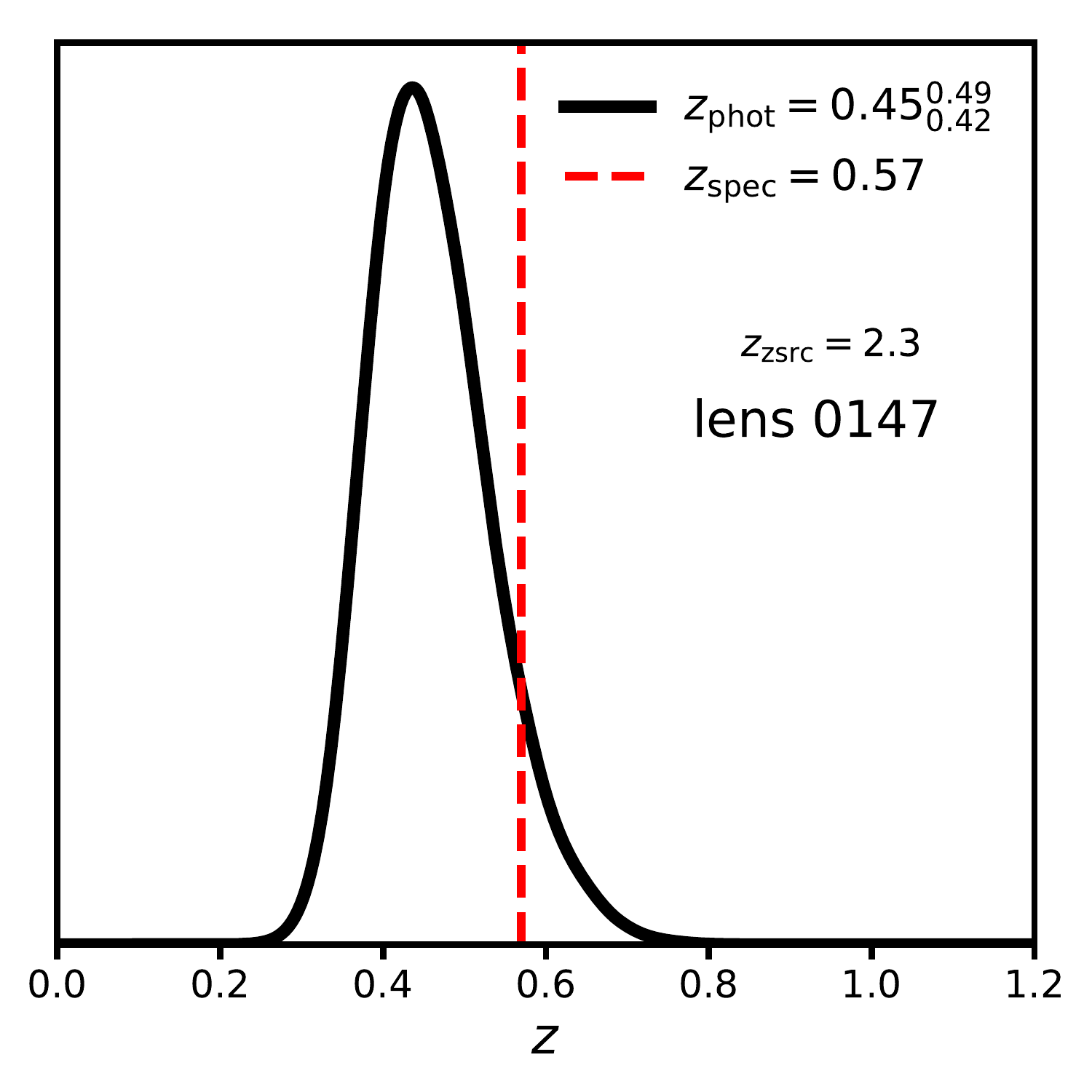}
		\includegraphics[clip,trim=0cm 0cm 0cm
		0cm,width=.2\textwidth,keepaspectratio]{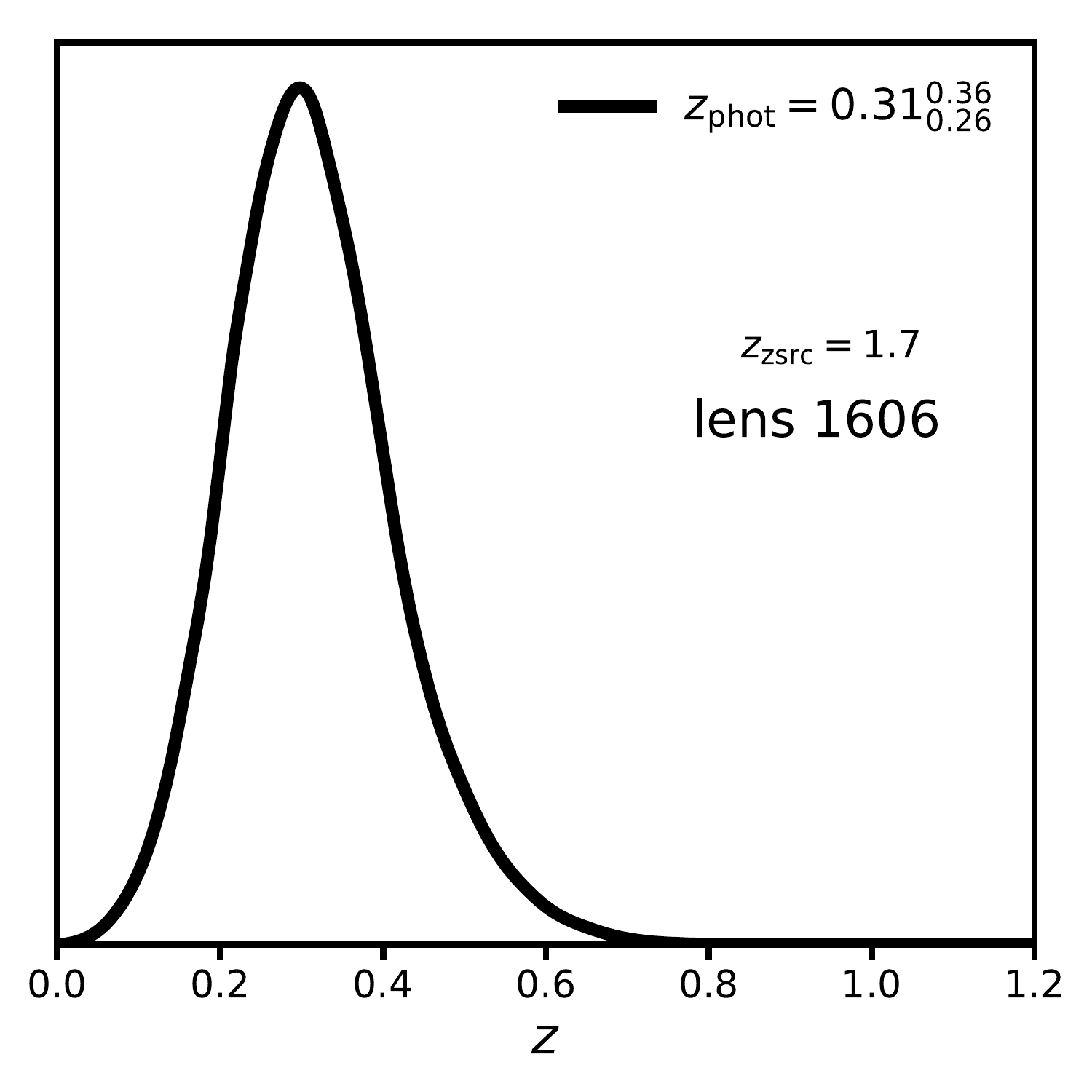}
		\includegraphics[clip,trim=0cm 0cm 0cm
		0cm,width=.2\textwidth,keepaspectratio]{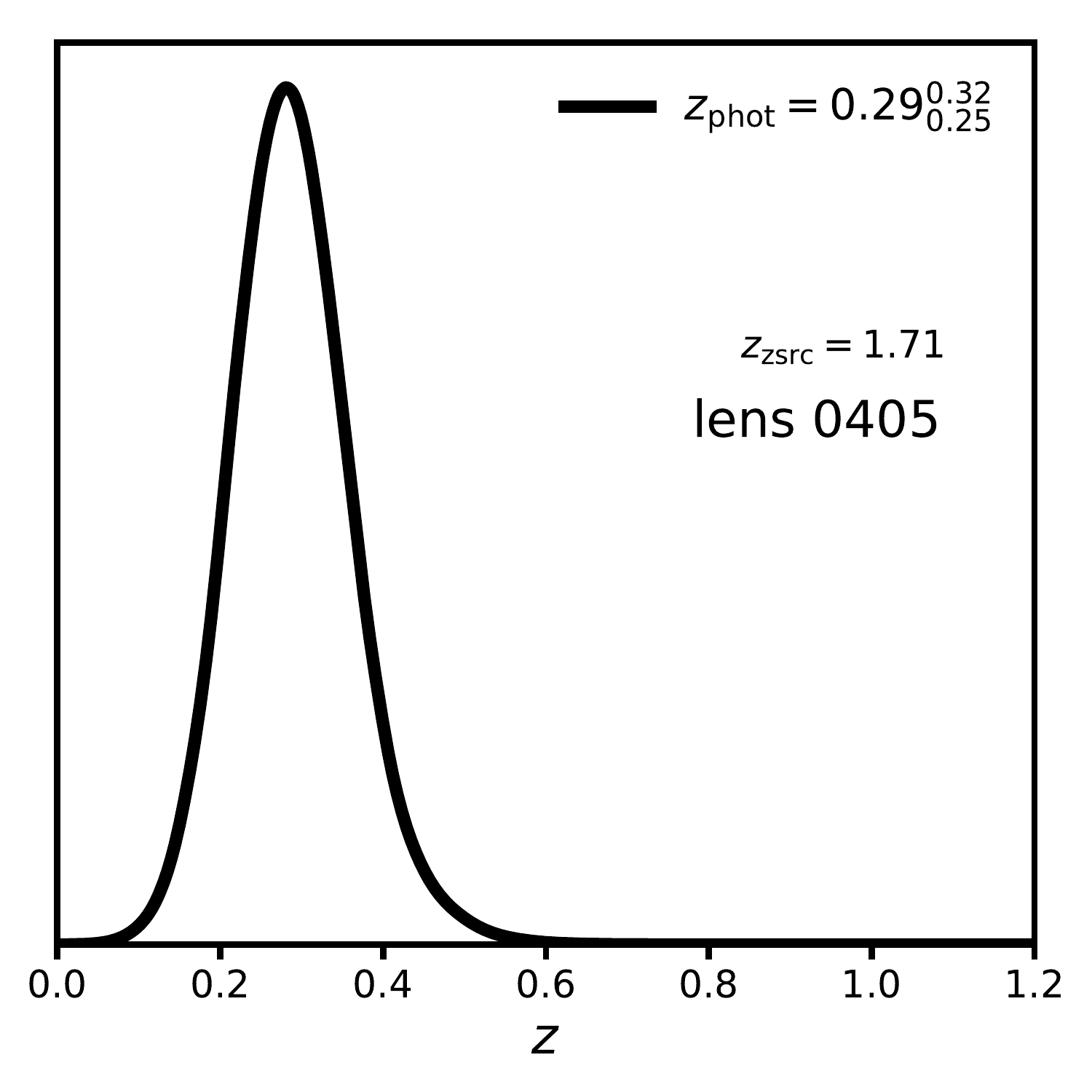}
		\caption{\label{fig:photozs} PDFs for main deflector redshifts computed with the software {\tt{eazy}} and photometry from \citet{Shajib++18}, restricting the photometry templates to those of early-type galaxies. Top rows show four applications of this procedure to quads with measured spectroscopic redshifts (red dotted lines). The bottom row shows the results of this procedure, using the same photometry and template assumptions, applied to the quads PS J1606 and WGD J0405, which do not have spectroscopic redshift measurements.}
	\end{figure}	
	\begin{figure}
		\includegraphics[clip,trim=0cm 0cm 0cm
		0cm,width=.3\textwidth,keepaspectratio]{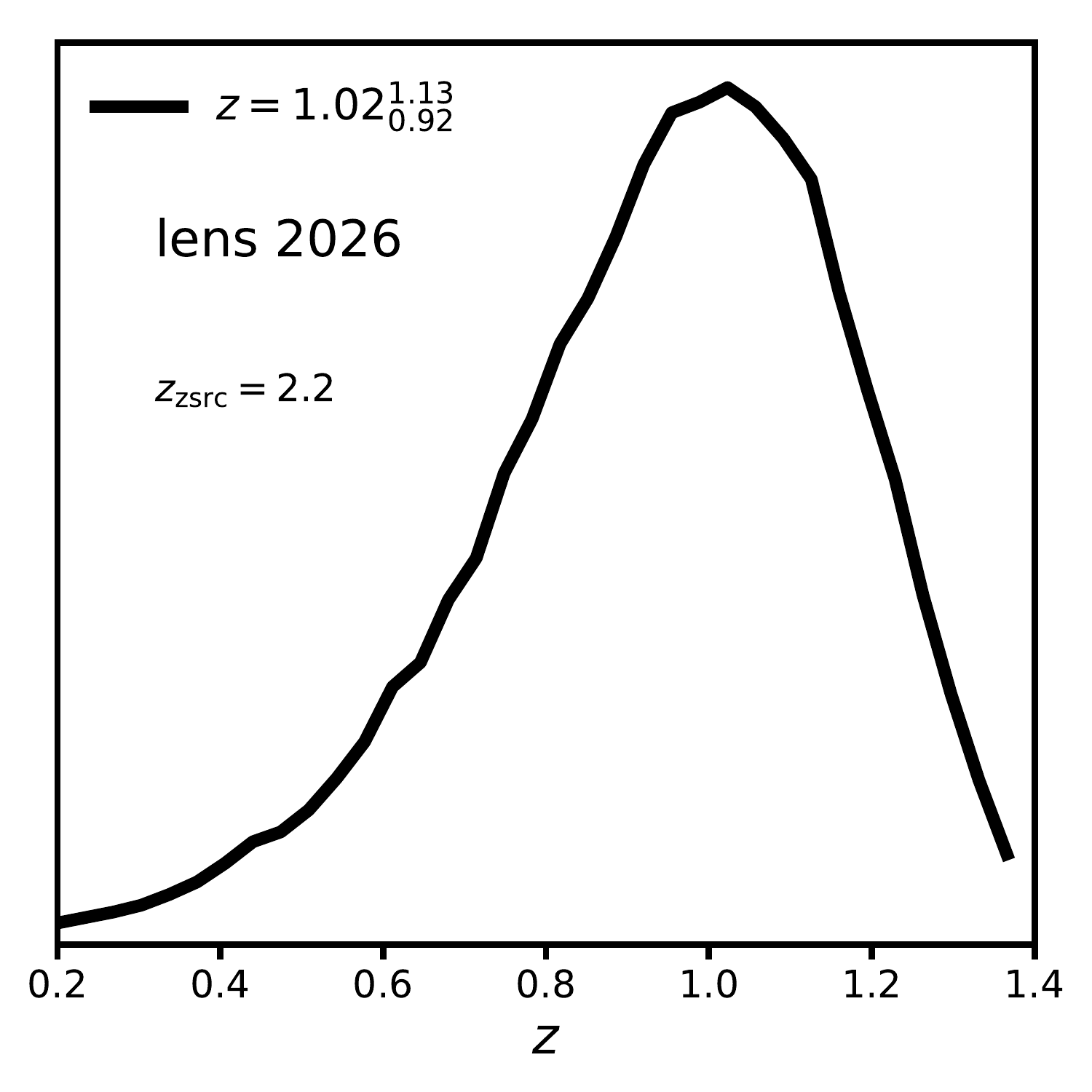}
		\caption{\label{fig:2026z} The PDF for the deflector redshift of WFI 2026 obtained by assuming a velocity dispersion of $240 \pm 30 \ \rm{km} \ \rm{s^{-1}}$ and a roughly isothermal mass profile. }
	\end{figure}
	The quads PS J1606 and WGD J0405 do not have measured spectroscopic redshifts, so we use photometry from \citet{Shajib++18} to obtain photometric redshift estimates. The photometry from \citet{Shajib++18} comes in three bands: F160W, F814W, and F475X with magnitude uncertainties of $0.1-0.3$ dex. We use the software package {\tt{eazy}} \citep{Brammer++08}, and restrict the templates to only consider the SEDs for early-type galaxies, which are $90\%$ of galaxies acting as strong lenses. We verify this procedure is accurate by applying it to other deflectors in sample analyzed by \citep{Shajib++18} that have measured spectroscopic redshifts, and  then proceed to derive PDFs for deflector redshifts in the systems PS J1606 and WGD J0405. 
	
	The results are shown in Figure \ref{fig:photozs}. The top row shows four quads from the sample analyzed in \citet{Shajib++18} with measured spectroscopic redshifts, and the bottom row shows the pdfs output by {\tt{eazy}} for the systems PS J1606 and WGD J0405. 
	
	The system WFI 2026 does not have a photometric redshift, and the photometry available in the literature comes in only one or two bands with larger uncertainties. For this system, we use the equation for isothermal mass profiles relating the Einstein radius $R_{\rm{Ein}} $, source redshift $z_s$, lens redshift $z_d$, velocity dispersion $\sigma$ and speed of light $c$
	\begin{equation}
	\label{eqn:iso}
	R_{\rm{Ein}} = 4 \pi \left(\frac{\sigma}{c}\right)^2 \frac{D_{\rm{ds}}\left(z_d, z_s\right)}{D_{\rm{s}}\left(z_s\right)}
	\end{equation}
	where $D_{\rm{ds}}$ and $D_{\rm{s}}$ are angular diameter distances between the lens and the source, and the observer and the source, respectively. 
	
	We sample a Gaussian distribution of velocity dispersions typical of early-type galaxies $240 \pm 30 \rm{km} \rm{s^{-1}}$, evaluate the right hand side of Equation \ref{eqn:iso}, and numerically solve for the lens redshift that yields the resulting angular diameter distance. The resulting PDF showin the bottom right panel of Figure \ref{fig:2026z} peaks around $z_d=1$, for the measured values $R_{\rm{Ein}}  = 0.67"$, $z_s = 2.2$. We have experimented with placing WFI 2026 at various specific redshifts, but find the posteriors for $\Sigma_{\rm{sub}}$, $\delta_{\rm{los}}$, $\alpha$, and $\mhm$ are unchanged within the uncertainties.
	
	\section{\bf Data}
	\label{app:C}
	We summarize the data used in this analysis, and the references for the astrometry, fluxes or flux ratios, and the corresponding uncertainties, and satellite galaxies or nearby nearby deflectors in Table \ref{tab:datasummary}. 
	
	\begin{table*}
		\centering
		\caption{The data used in this analysis. Letters A-D correspond to the lensed images, while G is the galaxy light centroid. The priors sampled for the satellite galaxies or nearby deflectors are quoted in Table \ref{tab:lenspriors}. Discovery papers are marked with a $^{\dagger}$. }
		\label{tab:datasummary}
		\begin{tabular}{lcccr} 
			\hline
			Lens & Image & dRA & dDec & NL flux \\
			\hline
			WGD J0405-3308 & A & $1.066 \pm 0.003$& $0.323 \pm 0.003$ & $1.00 \pm 0.04$\\
			\citet{Nierenberg++19}& B & $0\pm 0.003$& $0 \pm 0.003$ & $0.65 \pm 0.04$\\
			$^{\dagger}$\citet{Anguita++18}& C & $0.721 \pm 0.003$& $1.159 \pm 0.003$ & $1.25 \pm 0.03$\\
			& D & $-0.157 \pm 0.003$& $1.021 \pm 0.003$ & $1.17 \pm 0.04$ \\
			& G & $0.358 \pm 0.05$ & $0.567 \pm 0.05$ & - \\
			\hline
			HE0435-1223 & A & $2.424 \pm 0.008$& $0.792 \pm 0.008$ & $0.97 \pm 0.05$\\
			\citet{Nierenberg++17}& B & $1.458 \pm 0.008$&$-0.456 \pm 0.008$ & $0.98 \pm 0.049$ \\
			\citet{Wong++17}& C & $0 \pm 0.008$ & $0 \pm 0.008$ &$1 \pm 0.048$ \\
			$^{\dagger}$\citet{Witsotzki++02}& D & $0.768 \pm 0.008$& $1.662 \pm 0.008$ & $0.54 \pm 0.056$\\
			& G & $1.152 \pm 0.05 $&$0.636 \pm 0.05$ &- \\
			\hline
			RX J0911+0551 & A & $0 \pm 0.003$& $0 \pm 0.003$ & $0.56 \pm 0.04$\\
			\citet{Nierenberg++19}& B & $0.258 \pm 0.003$& $0.405 \pm 0.003$ & $1.00 \pm 0.05$\\
			$^{\dagger}$\citet{Bade++97}& C & $-0.016 \pm 0.003$& $0.959 \pm 0.003$ & $0.53 \pm 0.04$\\
			\citet{Blackburne++11}& D  & $-2.971 \pm 0.003$& $0.791 \pm 0.003$ & $0.24 \pm 0.04$\\
			& G  & $-0.688 \pm 0.05$& $0.517 \pm 0.05$ & - \\
			\hline
			B1422+231 & A & $0.387 \pm 0.005$ & $0.315 \pm 0.005$& $0.88 \pm 0.01$\\
			\citet{Nierenberg++14} & B & $0 \pm 0.005$& $0 \pm 0.005$& $1.00 \pm 0.01$\\
			$^{\dagger}$\citet{Patnaik++92}& C & $-0.362 \pm 0.005$ & $-0.728 \pm 0.005$ & $0.474 \pm 0.006$ \\
			& D & $0.941 \pm 0.01$ & $-0.797 \pm 0.01$ & - \\
			& G & $0.734 \pm 0.01$ & $-0.649 \pm 0.01$& - \\ 
			\hline
			PS J1606-2333 & A & $1.622 \pm 0.003$& $0.589 \pm 0.003$ & $1.00 \pm 0.03$\\
			\citet{Nierenberg++19}& B & $0 \pm 0.003$& $0 \pm 0.003$ & $1.00 \pm 0.03$\\
			\citet{Shajib++18}& C & $0.832 \pm 0.003$& $-0.316 \pm 0.003$ & $0.59 \pm 0.02$ \\
			$^{\dagger}$\citet{Lemon++18}& D  & $0.495 \pm 0.003$& $0.739 \pm 0.003$ & $0.79 \pm 0.02$ \\
			& G  & $0.784 \pm 0.05$& $0.211 \pm 0.05$ & - \\
			\hline
			WFI 2026-4536 & A & $0.164 \pm 0.003$& $-1.428 \pm 0.003$ & $1.00 \pm 0.02$ \\
			\citet{Nierenberg++19}& B & $0.417\pm 0.003$& $-1.213 \pm 0.003$ & $0.75 \pm 0.02$ \\
			$^{\dagger}$\citet{Morgan++04}& C & $0\pm 0.003$& $0 \pm 0.003$ & $0.31 \pm 0.02$ \\
			& D & $-0.571 \pm 0.003$& $-1.044 \pm 0.003$ & $0.28 \pm 0.01$ \\
			& G & $-0.023 \pm 0.05$ & $-0.865 \pm 0.05$ & - \\
			\hline
			WFI 2033-4723 & A & $-2.196 \pm 0.003$& $1.260 \pm 0.003$ & $1.00 \pm 0.03$ \\
			\citet{Nierenberg++19}& B & $-1.484 \pm 0.003$& $1.375 \pm 0.003$ & $0.65 \pm 0.03$ \\
			\citet{Vuissoz++08}& C & $0\pm 0.003$& $0 \pm 0.003$ & $0.50 \pm 0.02$ \\
			$^{\dagger}$\citet{Morgan++04}& D & $-2.113 \pm 0.003$& $-0.278 \pm 0.003$ & $0.53 \pm 0.02$ \\
			& G & $-1.445 \pm 0.05$& $2.344 \pm 0.05$ & - \\
			\hline
			WGD 2038-4008 & A & $-2.306 \pm 0.003$& $1.708 \pm 0.003$ & $1.00 \pm 0.01$ \\
			\citet{Nierenberg++19}& B & $0\pm 0.003$& $0 \pm 0.003$ & $1.16 \pm 0.02$ \\
			$^{\dagger}$\citet{Agnello++18}& C & $-1.518 \pm 0.003$& $0.029 \pm 0.003$ & $0.92 \pm 0.02$ \\
			& D & $-0.126 \pm 0.003$& $2.089 \pm 0.003$ & $0.46 \pm 0.01$ \\
			& G & $-0.832 \pm 0.05$ & $1.220 \pm 0.05$ & - \\
			\hline
			
		\end{tabular}
	\end{table*}
	
\end{document}